%% file: main.tex
\documentclass[11pt,a4paper]{article}

\usepackage[english]{babel}
\usepackage[scale=0.7071]{geometry} 
\usepackage[toc, page]{appendix}
\usepackage{cite}
\usepackage[bf]{caption}
\usepackage{subcaption}
\usepackage[bookmarksopen, bookmarksnumbered, bookmarksopenlevel=2, hidelinks]{hyperref}
\usepackage{xparse}

\usepackage{amsmath}
\usepackage{amssymb}
\usepackage{amsthm}
\usepackage{mathtools}
\usepackage{thmtools}
\usepackage{mathrsfs}
\usepackage{dsfont}
\usepackage{nicematrix}
\usepackage{scalerel}
\usepackage{dynkin-diagrams}
\usepackage{bbm}

\numberwithin{equation}{section}
\numberwithin{table}{section}


\newtheorem{theorem}{Theorem}[section]

\newtheorem{proposition}[theorem]{Proposition}

\theoremstyle{definition}

\theoremstyle{definition}

\theoremstyle{remark}

\makeatletter
\def\th@plain{%
  \thm@notefont{}
  \itshape 
}
\def\th@definition{%
  \thm@notefont{}
  \normalfont 
}
\makeatother

\usepackage[capitalise,noabbrev]{cleveref}
\AddToHook{cmd/appendix/before}{
    \crefalias{section}{appendix}%
    \crefalias{subsection}{appendix}
}
\usepackage{graphicx}
\usepackage{xcolor}
\usepackage{tabularray}
\UseTblrLibrary{functional,diagbox} 
\usepackage{tcolorbox}
\tcbset{title filled, arc=0mm}
\usepackage{enumitem}
\usepackage{rotating}
\usepackage[all]{nowidow}
\usepackage{fvextra}
\usepackage{microtype}
\usepackage{titling}

\usepackage[dvipsnames]{xcolor}
\definecolor{gruvbox-dark-red}{HTML}{CC241D}
\definecolor{gruvbox-dark-green}{HTML}{98971A}
\definecolor{gruvbox-dark-blue}{HTML}{458588}

\usepackage{tikz}
\usetikzlibrary{cd,calc,intersections}

\newcommand{\subtitle}[1]{%
    \posttitle{%
        \par\end{center}
        \begin{center}\large#1\end{center}\vskip0.5em}
}

\newcommand{\infdih}{\mathrm{I}_{2}(\infty)}
\newcommand{\deltaone}{\left\langle u'_{1}, d_{\perp} \right\rangle}
\newcommand{\deltatwo}{\left\langle u'_{2}, d_{\perp} \right\rangle}
\newcommand{\etad}{\left\langle n_{1}, d_{\parallel} \right\rangle}
\DeclareMathOperator{\arccosh}{arccosh}
\DeclareMathOperator{\rank}{rank}

\input{figures/Coxeter-diagrams}

\newcommand{\CoxDiagramCell}[1]{%
    \raisebox{.5\depth-.5\height}{#1}%
}
\newcommand{\CoxTextCell}[1]{%
    \raisebox{.5\depth-.5\height}{\strut #1}%
}

\ExplSyntaxOn
\NewDocumentCommand{\NPerLine}{m}{
	\clist_set:Nn \l_tmpa_clist {#1}
	\int_zero:N \l_tmpa_int
	\int_set:Nn \l_tmpb_int { \clist_count:N \l_tmpa_clist }
	\tl_clear:N \l_tmpa_tl

	\clist_map_inline:Nn \l_tmpa_clist {
		\int_incr:N \l_tmpa_int
		\tl_put_right:Nx \l_tmpa_tl { \tl_trim_spaces:n {##1} }

		\int_compare:nNnTF { \l_tmpa_int } = { \l_tmpb_int }
			{ }
			{
                \int_compare:nNnTF { \int_mod:nn { \l_tmpa_int } {8} } = {0}
					{ \tl_put_right:Nn \l_tmpa_tl {,\\} }
					{ \tl_put_right:Nn \l_tmpa_tl {,\space} }
			}
	}

	\CoxTextCell{%
        \shortstack[l]{\strut \tl_use:N \l_tmpa_tl \strut}%
    }
}
\ExplSyntaxOff

\NewDocumentCommand{\CenteredTikzPair}{m m m m m m m m}
	{%
		\begingroup
			\setbox0=\hbox{\input{#3}}%
			\setbox2=\hbox{\input{#6}}%
			\dimen0=\dimexpr\ht0+\dp0\relax
			\dimen2=\dimexpr\ht2+\dp2\relax
			\ifdim\dimen2>\dimen0 \dimen0=\dimen2\fi

			\def\CTP@raise##1{%
				\raisebox{\dimexpr(\dimen0-\height-\depth)/2+\depth\relax}{##1}%
			}%

			\begin{subfigure}[t]{#1}
				\centering
				\CTP@raise{\box0}%
                \vspace{6pt}
				\caption{#4}
				\label{#5}
			\end{subfigure}%
			\hfill
			\begin{subfigure}[t]{#2}
				\centering
				\CTP@raise{\box2}%
                \vspace{5pt}
				\caption{#7}
				\label{#8}
			\end{subfigure}%
		\endgroup
	}%

\input{titlepage-style.tex}
\input{titlepage-data}

\SetupPDFMetadataFromTitleData

\begin{document}

\MakeFrontMatter

\input{sections/introduction}
\input{sections/review-isomorphic-flops}
\input{sections/CICY-Coxeter-database}
\input{sections/dihedral-Coxeter-prepotentials}
\input{sections/harmonic-analysis}
\input{sections/general-Coxeter-prepotentials}
\input{sections/conclusions}
\input{sections/acknowledgements}

\appendix

\input{appendices/dihedral-stabilizer}
\input{appendices/fundamental-Mori-cone-infdih}
\input{appendices/AB-properties}

\clearpage
\bibliography{references}  
\bibliographystyle{JHEP}

\end{document}

%% file: figures/Coxeter-diagrams.tex
\newlength{\CoxEdgeLen}
\newlength{\CoxSepLen}
\pgfkeys{/Dynkin diagram,
    edge length=\CoxEdgeLen,
    separator length=\CoxSepLen
}

\newcommand{\PlaceStarRoot}[5]{%
    \dynkinPlaceRootRelativeTo{#1}{#2}{#3}{#4}{#5}%
    \dynkinRootMark{*}#1%
}

\newcommand{\CoxEdge}[3]{%
    \dynkinDefiniteSingleEdge #1 #2%
    \dynkinEdgeLabel{#1}{#2}{#3}%
}

\newcommand{\CoxEdgeFlipLabel}[3]{%
    \dynkinDefiniteSingleEdge #1 #2%
    \dynkinEdgeLabel{#2}{#1}{#3}%
}

\newcommand{\PTriRight}{\raisebox{0ex}{\ensuremath{\mathrm{P}}}\mkern0.75mu}
\newcommand{\PTriBottom}{\raisebox{-0.4ex}{\ensuremath{\mathrm{P}}}\mkern0mu}
\newcommand{\PTriTop}{\raisebox{-1.5ex}{\ensuremath{\mathrm{P}}}\mkern0mu}
\newcommand{\PTriTopAlt}{\raisebox{0.35ex}{\ensuremath{\mathrm{P}}}\mkern0mu}

\newcommand{\HTriBottom}{\raisebox{-0.4ex}{\ensuremath{\mathrm{H}}}\mkern0mu}

\makeatletter

\newcommand{\CoxDiagramVPH}{%
    \begin{dynkinDiagram}[Coxeter,label=false]A{*}%
        \advance\dynkin@nodes by 2%
        \PlaceStarRoot{2}{1}{east}{right}{right}%
        \PlaceStarRoot{3}{1}{north}{above}{above}%
        \CoxEdge{1}{3}{\PTriRight}%
        \CoxEdgeFlipLabel{1}{2}{\HTriBottom}%
  \end{dynkinDiagram}%
}

\newcommand{\CoxDiagramVPP}{%
    \begin{dynkinDiagram}[Coxeter,label=false]A{*}%
        \advance\dynkin@nodes by 2%
        \PlaceStarRoot{2}{1}{east}{right}{right}%
        \PlaceStarRoot{3}{1}{north}{above}{above}%
        \CoxEdge{1}{3}{\PTriRight}%
        \CoxEdgeFlipLabel{1}{2}{\PTriBottom}%
    \end{dynkinDiagram}%
}

\newcommand{\CoxDiagramTriangleAPP}{%
    \begin{dynkinDiagram}[Coxeter,label=false]A{*}%
        \advance\dynkin@nodes by 2%
        \PlaceStarRoot{2}{1}{east}{right}{right}%
        \PlaceStarRoot{3}{1}{north}{above}{above}%
        \dynkinDefiniteSingleEdge 1 2 \dynkinEdgeLabel{2}{1}{\PTriBottom}%
        \dynkinDefiniteSingleEdge 1 3 \dynkinEdgeLabel{1}{3}{\PTriRight}
        \dynkinDefiniteSingleEdge 2 3 \dynkinEdgeLabel{3}{2}{}%
    \end{dynkinDiagram}%
}

\newcommand{\CoxDiagramTrianglePPH}{%
    \begin{dynkinDiagram}[Coxeter,label=false]A{*}%
        \advance\dynkin@nodes by 2%
        \PlaceStarRoot{2}{1}{east}{right}{right}%
        \PlaceStarRoot{3}{1}{north}{above}{above}%
        \dynkinDefiniteSingleEdge 1 2 \dynkinEdgeLabel{2}{1}{\PTriBottom}%
        \dynkinDefiniteSingleEdge 1 3 \dynkinEdgeLabel{1}{3}{\PTriRight}%
        \dynkinDefiniteSingleEdge 2 3 \dynkinEdgeLabel{3}{2}{\mathrm{H}}%
    \end{dynkinDiagram}%
}

\newcommand{\CoxDiagramTrianglePPP}{%
    \begin{dynkinDiagram}[Coxeter,label=false]A{*}%
        \advance\dynkin@nodes by 2%
        \PlaceStarRoot{2}{1}{east}{right}{right}%
        \PlaceStarRoot{3}{1}{north}{above}{above}%
        \dynkinDefiniteSingleEdge 1 2 \dynkinEdgeLabel{2}{1}{\PTriBottom}%
        \dynkinDefiniteSingleEdge 1 3 \dynkinEdgeLabel{1}{3}{\PTriRight}
        \dynkinDefiniteSingleEdge 2 3 \dynkinEdgeLabel{3}{2}{\mathrm{P}}%
    \end{dynkinDiagram}%
}

\newcommand{\CoxDiagramStarThreeLegP}{%
    \begin{dynkinDiagram}[Coxeter,label=false]A{*}%
        \advance\dynkin@nodes by 3%

        \PlaceStarRoot{2}{1}{north}{above}{above}%
        \begin{scope}[rotate=120]
            \PlaceStarRoot{3}{1}{north}{above}{above}%
        \end{scope}
        \begin{scope}[rotate=-120]
            \PlaceStarRoot{4}{1}{north}{above}{above}%
        \end{scope}

        \CoxEdge{1}{2}{\PTriRight}%
        \CoxEdge{1}{3}{\mathrm{P}}%
        \CoxEdgeFlipLabel{1}{4}{\mathrm{P}}%
  \end{dynkinDiagram}%
}

\newcommand{\CoxDiagramPStarATwoEdge}{%
    \begin{dynkinDiagram}[Coxeter,label=false,rotate=90]A{*}%
        \advance\dynkin@nodes by 3%
        
        \PlaceStarRoot{2}{1}{south}{below}{below}
        \PlaceStarRoot{3}{1}{northwest}{above}{left}
        \PlaceStarRoot{4}{1}{northeast}{above}{right}

        \CoxEdge{1}{2}{\mathrm{P}}%
        \CoxEdge{1}{3}{\PTriBottom}%
        \CoxEdgeFlipLabel{1}{4}{\mathrm{P}}%
        
        \dynkinDefiniteSingleEdge 3 4%
    \end{dynkinDiagram}%
}

\newcommand{\CoxDiagramKFourAllP}{%
    \begin{dynkinDiagram}[Coxeter,label=false]A{*}%
        \advance\dynkin@nodes by 3%

        \PlaceStarRoot{2}{1}{north}{above}{above}%
        \begin{scope}[rotate=120]\PlaceStarRoot{3}{1}{north}{above}{above}\end{scope}%
        \begin{scope}[rotate=-120]\PlaceStarRoot{4}{1}{north}{above}{above}\end{scope}%

        \CoxEdge{1}{2}{\PTriTop}%
        \CoxEdge{1}{3}{\PTriTopAlt}%
        \CoxEdgeFlipLabel{1}{4}{\PTriTopAlt}%

        \CoxEdgeFlipLabel{2}{3}{\mathrm{P}}%
        \CoxEdge{2}{4}{\mathrm{P}}%
        \CoxEdgeFlipLabel{3}{4}{\PTriBottom}%
    \end{dynkinDiagram}%
}

\newcommand{\CoxDiagramStarFourLegFour}{%
    \begin{dynkinDiagram}[Coxeter,label=false,rotate=45]A{*}%
        \advance\dynkin@nodes by 4%
        \PlaceStarRoot{2}{1}{north}{above}{above}%
        \PlaceStarRoot{3}{1}{east}{right}{right}%
        \PlaceStarRoot{4}{1}{south}{below}{below}%
        \PlaceStarRoot{5}{1}{west}{left}{left}%
        \CoxEdgeFlipLabel{1}{2}{4}%
        \CoxEdge{1}{3}{4}%
        \CoxEdgeFlipLabel{1}{4}{4}%
        \CoxEdge{1}{5}{4}%
    \end{dynkinDiagram}%
}

\newcommand{\CoxDiagramKFiveAllThree@pic}{%
    \begingroup
        \pgfmathsetlengthmacro{\CoxR}{\the\CoxEdgeLen/(2*sin(36))}%
        \begin{tikzpicture}[baseline]
            \foreach \i/\ang in {1/90,2/18,3/-54,4/-126,5/162}{%
                \coordinate (K\i) at (\ang:\CoxR);%
            }%

            \begin{pgfonlayer}{Dynkin behind}
                \foreach \i/\j in {1/2,2/3,3/4,4/5,5/1,1/3,1/4,2/4,2/5,3/5}{%
                    \draw[/Dynkin diagram/edge] (K\i) -- (K\j);%
                }%
            \end{pgfonlayer}

            \foreach \i in {1,...,5}{%
                \filldraw[draw=black,fill=black] (K\i) circle[radius=\dynkin@root@radius];%
            }%
        \end{tikzpicture}%
    \endgroup
}

\newcommand{\CoxDiagramKFiveAllThree}{%
    \ifmmode
        \mathord{\vcenter{\hbox{\CoxDiagramKFiveAllThree@pic}}}%
    \else
        \CoxDiagramKFiveAllThree@pic
    \fi
}

\makeatother

%% file: titlepage-data.tex
\Title{Kaleidoscopes, Waves and the Prepotential}
\PDFTitle{Kaleidoscopes, Waves and the Prepotential}

\Author{Rafael \'Alvarez-Garc\'ia}{1,4}
\Author{Fabian Ruehle}{1,2,3}

\Affiliation{1}{Department of Physics, Northeastern University, Boston, MA 02115, USA}
\Affiliation{2}{Department of Mathematics, Northeastern University, Boston, MA 02115, USA}
\Affiliation{3}{The NSF AI Institute for Artificial Intelligence and Fundamental Interactions}
\Affiliation{4}{Jefferson Physical Laboratory, Harvard University, Cambridge, MA 02138, USA}

\Email{ralvarezgarcia@fas.harvard.edu}
\Email{f.ruehle@northeastern.edu}

\Abstract{Isomorphic flops are topology-changing transitions connecting two diffeomorphic families of Calabi-Yau threefolds. They correspond to the generators of certain Coxeter groups acting on the moduli space. As a consequence of these symmetries, the prepotential of 4D $\mathcal{N} = 2$ Type IIA compactifications on such varieties must assemble into Coxeter-invariant functions. We construct a database of all Coxeter symmetries from isomorphic flops in K\"ahler-favorable CICYs. The action of the Coxeter group on the K\"ahler moduli space leaves a symmetric bilinear form invariant, which we interpret as a metric and construct its associated Laplace-Beltrami operator. We argue that the Coxeter-invariant functions featured in the prepotential solve the Helmholtz equation with this Laplacian, and that the prepotential can then be resummed into a decomposition in terms of eigenfunctions of the Laplace-Beltrami operator. The convergence rate of the raw orbit sums of worldsheet instanton contributions and the resummed expressions are complementary, with the latter sharply localizing around the first few terms in the interior of the moduli space.}

%% file: sections/introduction.tex
\section{Introduction and summary}
\label{sec:introduction}

In four-dimensional compactifications of Type IIA string theory on Calabi-Yau threefolds, the vector multiplet sector of the low-energy 4D $\mathcal{N} = 2$ SUGRA is completely described, at the two-derivative level, by the prepotential \cite{Bodner:1990zm}. This central object controls abelian gauge couplings, Yukawa couplings, the central charge of wrapped D-branes and a plethora of other physical quantities. Moreover, it also features prominently in the analysis of flux vacua descending from such compactifications \cite{Grana:2005jc,Douglas:2006es}. Hence, further constraining the structure of the prepotential can lead to many prospective applications.

The prepotential is inextricably linked to the special K\"ahler geometry of the K\"ahler moduli space, which it also determines. The walls of the K\"ahler cone correspond to different types of degenerations of the Calabi-Yau threefold, leading to the contraction of curves, divisors or the variety itself \cite{Wilson1992KahlerCone}. Crossing some of these walls, the Calabi-Yau threefold can undergo topology-changing transitions, which can be understood within string theory \cite{Aspinwall:1993yb,Aspinwall:1993nu,Greene:1996cy}. The lower-dimensional theories on each side of a wall are closely related to each other, and one can exploit this fact to better understand their features.

We will precisely follow this strategy by focusing on a particularly constraining type of topology-changing transition, namely, isomorphic flop transitions. These are crepant small birational modifications such that the two chambers of the extended K\"ahler cone that are connected by the flop correspond to diffeomorphic families of Calabi-Yau threefolds. Hence, either side of the isomorphic flop wall must yield equivalent descriptions of the same lower-dimensional theory and, as a consequence, the existence of such a wall restricts the form of the prepotential of the underlying Calabi-Yau~\cite{Lukas:2022crp}. Isomorphic flop walls may seem a rather special phenomenon, but they are common in the moduli space of CICYs \cite{Brodie:2021toe}.

More explicitly, for an isomorphic flop wall, divide the prepotential into three pieces
\begin{equation}
    \mathcal{F}(T) = \mathcal{F}_{\mathrm{class}}(T) + \mathcal{F}_{\mathrm{flop}}(T) + \mathcal{F}_{\mathrm{non-flop}}(T)\,,
\end{equation}
corresponding to the classical part, the worldsheet instanton corrections associated to curve classes shrinking along the isomorphic flop and the rest of worldsheet instanton corrections, respectively. Only a very small number of curve classes contribute to $\mathcal{F}_{\mathrm{flop}}(T)$, and their interplay with $\mathcal{F}_{\mathrm{class}}(T)$ is well understood~\cite{Candelas:1993dm,Aspinwall:1993nu,Witten:1996qb}. Hence, the paper focuses on the $\mathcal{F}_{\mathrm{non-flop}}(T)$ piece, to which an infinite number of curve classes can contribute. Pairs of curve classes relevant to $\mathcal{F}_{\mathrm{non-flop}}(T)$ identified under the isomorphic flop must share the same genus-zero Gopakumar-Vafa invariants, which can therefore be factored out of the sum of their prepotential contributions. These sums must be invariant under the isomorphic flop $\mathbb{Z}_{2}$-action on the moduli space\,---\,a reflection across a K\"ahler cone wall. This generalizes to a set of isomorphic flops where each acts as a simple reflection, jointly generating a Coxeter group $W$ acting on the moduli space. The contributions to $\mathcal{F}_{\mathrm{non-flop}}(T)$ must therefore organize into Coxeter-invariant functions $\psi_{ld}^{W}(T)$, schematically defined by
\begin{equation}
    \psi_{ld}^{W}(T) \coloneqq \sum_{\substack{W\text{-orbit}\\[0.2em] \text{of } d}} e^{2\pi i l \langle wd, T \rangle}\,.
\end{equation}
This structure was first noticed in \cite{Lukas:2022crp}, but the Coxeter-invariant functions remain largely uncharacterized beyond their definition as a raw orbit sum.

In view of the above facts, two natural questions arise. First, what are the Coxeter groups generated by isomorphic flops that can appear acting on the moduli space of Calabi-Yau threefolds? And second, what are the properties of the Coxeter-invariant functions $\psi_{ld}^{W}(T)$ entering in the construction of the prepotential whenever isomorphic flops are present? These lines of inquiry guide the present work, the results of which we now summarize. Since this is a comprehensive study of the subject, the authors decided to publish an abridged version of this paper, emphasizing the results in \cref{sec:harmonic-analysis}, as a letter in~\cite{Alvarez-Garcia:2026uaq}.

\subsection*{Summary of the results}

We begin with a review of isomorphic flops and their interplay with the prepotential in the context of 4D $\mathcal{N} = 2$ Type IIA compactifications in \cref{sec:review-isomorphic-flops}. In addition to making the paper self-contained, this serves to establish notations and conventions used throughout the text.

In \cref{sec:CICY-Coxeter-Database} we answer the first question above for the 4874 K\"ahler-favorable CICYs \cite{Anderson:2017aux}. For this collection of Calabi-Yau threefolds, we perform an exhaustive classification of their Coxeter symmetries generated by isomorphic flops, finding a total of 19 different Coxeter groups in the database. The count of models exhibiting isomorphic flops is 2182, out of which the 590 with $\rank(W) \geq 2$ Coxeter symmetries are listed in \cref{tab:CICY-Coxeter-database}. The CICY Coxeter Database\,---\,a version of the favorable CICY list of \cite{Anderson:2017aux} enlarged to contain data on isomorphic flops, the Coxeter groups that they generate and the representation through which they act on the moduli space\,---\,is provided as ancillary material with the arXiv submission of this paper and as an online version at \cite{AlvarezGarciaCICYCoxeterDatabase}.

The second question is addressed in \cref{sec:dihedral-Coxeter-prepotentials}, where we study the Coxeter-invariant functions $\psi_{ld}^{\mathrm{I}_{2}(m)}(T)$ for isomorphic flops that generate the dihedral group. The $W = \mathrm{I}_{2}(m)$ choice is motivated due to its abundance within the CICY Coxeter Database (481 models out of the 590 with a $\rank(W) \geq 2$ Coxeter group) and due to the fact that this is the simplest case for which infinite symmetry groups can arise, accounting for 189 of the dihedral examples. The study of this particular scenario illuminates the features that are expected for general Coxeter-invariant functions. Depending on the $\mathrm{I}_{2}(m)$-representation acting on the moduli space, dihedral models can be classified into hyperbolic (H), parabolic (P) and elliptic (E). We find that the raw orbit sums $\psi_{ld}^{\mathrm{I}_{2}(m)}(T) \coloneqq \sum_{\mathrm{I}_{2}(m)} e^{2\pi i l \langle wd, T \rangle}$ can be resummed into the expressions
\begingroup
\allowdisplaybreaks
\begin{align}
    \text{H:}\quad &\psi_{ld}^{\infdih}(T) = e^{2\pi i l \mathcal{C}} \frac{2}{u} \left[ K_{0}(\Omega) + 2\sum_{m=1}^{\infty} K_{\frac{\pi i m}{u}}(\Omega) \cos(m\zeta) \cos(m\eta) \right]\,,\\
    \text{P:}\quad &\psi_{ld}^{\infdih}(T) = e^{2\pi il \left\langle d, T \right\rangle} \left[ \vartheta\left( z_{\theta}, \tau \right) + e^{2\pi i\xi} \vartheta\left( z_{\theta}^{S}, \tau \right) \right]\,,\\
    \text{E:}\quad &\psi_{ld}^{\mathrm{I}_{2}(m)}(T) = e^{2\pi i l \mathcal{C}} 2m \left[ J_{0}(\Omega') + 2\sum_{r=1}^{\infty} i^{mr} J_{mr}(\Omega') \cos(mr\zeta') \cos(mr\eta') \right]\,,
\end{align}
\endgroup
where $K_{\nu}$ and $J_{m}$ are the modified Bessel function of the second kind and the ordinary Bessel function of the first kind, respectively, and $\vartheta$ stands for the Jacobi theta function. The objects $\{ \mathcal{C},\Omega,\zeta,\eta,z_{\vartheta},z_{\vartheta}^{S},\xi,\tau,\Omega',\zeta',\eta' \}$ are functions of the curve class and the complexified K\"ahler moduli, while $\{u,m\}$ are constants determined by the representation of the dihedral group. The appearance of Jacobi theta functions for the parabolic case was already noted in \cite{Lukas:2022crp}. These resummed expressions for $\psi_{ld}^{\mathrm{I}_{2}(m)}$ are more natural from the isomorphic flop perspective in the sense that they are Coxeter-invariant term by term. In addition to their structural interest, their convergence properties are complementary to those of the raw orbit sums, acting as a spectral dual decomposition. The convergence rate of the orbit sum of exponential worldsheet instanton contributions is best in the large volume regime, where the sum localizes around a few leading instanton terms while the rest are very suppressed. The resummed expressions, on the other hand, converge rather slowly in this regime, with their contribution broadly distributed among many Bessel-modes. In the interior of the moduli space, the opposite is true; while the raw orbit sum becomes flatly distributed among many exponential terms and converges at a slow rate, the resummed expressions are strongly peaked around the first Bessel-modes and converge rapidly. \cref{sec:dihedral-stabilizer,sec:fundamental-Mori-cone-infdih,sec:AB-properties} contain the justification to certain claims in \cref{sec:dihedral-Coxeter-prepotentials} and are provided as complementary material.

Motivated by the preceding observation, in \cref{sec:harmonic-analysis} we provide a natural geometric origin for the resummed expressions of \cref{sec:dihedral-Coxeter-prepotentials} by developing a harmonic analysis of the Gromov-Witten expansion. Heuristically speaking, the expansion can be regarded as a superposition of plane waves in the moduli space, with the axions acting as the angular variable and the real K\"ahler moduli controlling the exponential damping effect at large volume. Isomorphic flops act as a set of mirrors in the moduli space\,---\,a kaleidoscope\,---\,and the resummed expressions are a decomposition of the Gromov-Witten waves into harmonics adapted to the Coxeter symmetries. More precisely, the Coxeter action on the real vector space in which the Mori cone lives leads to a real quotient geometry whose metric incorporates the Coxeter symmetries as isometries. The pairing between the curve classes and the complexified K\"ahler moduli defines a K\"ahler-dependent embedding of the real quotient into a complex quotient geometry, with essentially the same metric. From the metric, we can construct a Laplace-Beltrami operator (the canonical second-order differential operator respecting the isometries) in the complex quotient geometry. The eigenfunctions of this operator, i.e., the solutions of the Helmholtz equation in the complex quotient space are the special functions encountered in \cref{sec:dihedral-Coxeter-prepotentials}. Taking the pullback to the K\"ahler moduli space we find the Coxeter-invariant functions $\psi_{ld}^{\mathrm{I}_{2}(m)}$. Although the discussion is illustrated through the dihedral example, the general picture becomes clear: In the presence of isomorphic flops, a natural Laplace-Beltrami operator respecting them as isometries can be defined in the moduli space. The Gromov-Witten expansion can then be resummed\,---\,observed through the kaleidoscope\,---\,into a superposition of eigenfunctions of the differential operator. The resulting expressions act as a spectral dual to the raw orbit sums and have complementary convergence properties.

The obvious next step would be to construct the Laplace-Beltrami operator for the remaining groups appearing in the CICY Coxeter Database and perform the corresponding harmonic analysis; we leave the complete treatment of the general case for future work. Instead of approaching the general problem directly, in \cref{sec:general-Coxeter-prepotentials} we perform a dihedral block decomposition of the Coxeter-invariant functions $\psi_{ld}^{W}(T)$ that allows us to leverage the results of \cref{sec:dihedral-Coxeter-prepotentials}. The first complication with respect to the infinite dihedral case comes from efficiently expressing the raw orbit sums. While the Cayley graph of the infinite dihedral group is a line and its elements can be indexed by the integers in a straightforward manner, general Coxeter groups require more care to ensure that no overcounting occurs while summing over instanton contributions. To address this issue, and inspired by the literature on automatic groups \cite{EpsteinCannonHoltLevyPatersonThurston1992,BrinkHowlett1993FinitenessProperty}, we construct finite-state automata traversing the Cayley graph of a generic Coxeter group without repeating elements. This allows us to express the orbit sum in terms of a resolvent-type formula involving the adjacency matrix of the automaton. As a consequence, the instanton orbit sum of general Coxeter groups is reduced to a problem specified by finite linear algebra objects. Refining the automaton such that it can remember at least the last two steps taken in the Cayley graph, we can distinguish those alternations of letters contained within a dihedral subgroup associated to a pair of simple reflection generators. Summing over these produces special functions similar to those appearing in \cref{sec:dihedral-Coxeter-prepotentials}. The complete $\psi_{ld}^{W}(T)$ is given by a remainder resolvent-type formula that encodes the mixing between dihedral channels after having resummed within them. \cref{sec:general-Coxeter-prepotentials} can be omitted on a first reading without compromising the central narrative of the paper.

We offer a recapitulation of results in \cref{sec:conclusions}, concluding with a brief outlook on avenues for future studies.

%% file: sections/review-isomorphic-flops.tex
\section{Review: Isomorphic flops and Coxeter groups}
\label{sec:review-isomorphic-flops}

Throughout this paper we will study how Coxeter symmetries in the K\"ahler moduli space constrain the non-perturbative prepotential of 4D $\mathcal{N} = 2$ Type IIA compactifications. The Coxeter symmetries are generated by isomorphic flops (iso-flops) of the internal Calabi-Yau threefold. An iso-flop is a flop connecting two chambers in the extended K\"ahler cone corresponding to families of Calabi-Yau threefolds that are not only birational to each other but also diffeomorphic. In this section we review this central notion and establish the notation we will use throughout this paper.

In recent years, a series of papers have discussed iso-flops in the context of string compactifications and the Swampland Program \cite{Vafa:2005ui}. Their relation to symmetries of line bundle cohomology and their interplay with Gromov-Witten invariants were explored in \cite{Brodie:2020fiq}. A study of iso-flops in the context of the Swampland Distance Conjecture \cite{Ooguri:2006in} was undertaken in \cite{Brodie:2021ain}. For M-theory compactified on a Calabi-Yau threefold whose extended K\"ahler cone has an infinite number of chambers, it is naively possible to construct paths of infinite geodesic length along which the spectrum does not vary significantly, contradicting the Swampland Distance Conjecture. Taking the quotient by the Coxeter symmetry generated by iso-flops, a remnant of the gauged eleven-dimensional diffeomorphisms, removes the infinite number of K\"ahler chambers corresponding to isomorphic (in the smooth category) Calabi-Yau threefolds, thereby lifting the contradiction; the non-isomorphic case is taken care of by the Kawamata-Morrison conjecture. Relatedly, an in-depth study of the effective cone of Calabi-Yau threefolds with $h^{1,1}(X) = 2$ was carried out in \cite{Brodie:2021nit}, classifying the types of cone walls and matching geodesic solutions across flops. An analysis of the infinite Coxeter symmetry of the mirror five-parameter Hulek-Verrill family and how it organizes its Gopakumar-Vafa invariants was performed in \cite{Candelas:2021lkc}, with the relation between Coxeter and GLSM symmetries, as well as their use in constraining higher-genus Gopakumar-Vafa invariants via the BCOV holomorphic anomaly recursion \cite{Bershadsky:1993cx}, discussed in \cite{Kuusela:2023vgi}. Iso-flops have also featured in discussions of non-perturbative contributions to the superpotential \cite{Gendler:2022qof}, the algorithmic reconstruction of the extended K\"ahler cone \cite{Gendler:2022ztv} and the counting of its different birational phases \cite{Gendler:2023ujl}.

This work relates most directly to \cite{Brodie:2021toe,Lukas:2022crp}. A characterization of the conditions that must be met for a flop wall to exist in a K\"ahler-favorable family of CICYs, expressed in terms of its configuration matrix properties, was provided in \cite{Brodie:2021toe}. Particular attention was devoted to the iso-flop subcase; this informs our construction of the CICY Coxeter Database in \cref{sec:CICY-Coxeter-Database}. The study of 4D $\mathcal{N} = 2$ Type IIA prepotential constraints from iso-flop Coxeter symmetries was initiated in \cite{Lukas:2022crp}. This is the analysis that we refine and expand throughout the paper. We review the conclusions of both these works in greater detail below.

\subsection{Isomorphic flops}
\label{sec:iso-flops}

A flop is a birational map $f:  X \dashrightarrow X'$ between normal varieties that is an isomorphism in codimension one and factors through a common singular model $X_{\mathrm{sing}}$ by two small contractions \cite{KollarMori1998BirationalGeometry}
\begin{equation}
	\begin{tikzpicture}[baseline=(current bounding box.center)]
		\node (X) at (150:1.5) {$X$};
		\node (Xp) at (30:1.5) {$X'$};
		\node (Xsing) at (-90:1.5) {$X_{\mathrm{sing}}\mathrlap{\,.}$};

		\draw[dashed, ->] (X) -- node[midway, above] {$f$} (Xp);
		\draw[->] (X) -- node[midway, left] {$\pi$} (Xsing);
		\draw[->] (Xp) -- node[midway, right] {$\pi'$} (Xsing);
	\end{tikzpicture}
\end{equation}
Flops are crepant small birational modifications. Focusing on smooth Calabi-Yau threefolds, a flop contracts a finite collection of rational curves $\{ C_{i} \}_{i \in \{1, \dotsc, N\}}$ on $X$ to points and replaces them by a different collection of rational curves $\{ C'_{i} \}_{i \in \{1, \dotsc, N\}}$ on $X'$ while preserving the Calabi-Yau condition. For the birational map to be a flop, there must exist a divisor\footnote{Throughout the text, we will often not distinguish between divisors, divisor classes and line bundles.} $D \in \mathrm{Pic}(X)$ and, through the Picard group identification induced by $X \setminus \bigcup_{i=1}^{N} C_{i} \cong X' \setminus \bigcup_{i=1}^{N} C'_{i}$, a corresponding divisor $D'\in \mathrm{Pic}(X')$ such that $D \cdot C_{i} > 0$ and $D'\cdot C'_{i} < 0$. The Calabi-Yau threefolds $X$ and $X'$ share the same Hodge numbers, but their triple intersection form and second Chern class transform under the natural codimension-one identification. More concretely, for any Type~I contraction \cite{Wilson1992KahlerCone} with $\eta \in H_{2}(X,\mathbb{Z})$ the primitive curve class generating the flopping ray, we have that \cite{Wilson1999FlopsTypeIII}
\begin{equation}
    \left( D' \right)^{3} = \left( D \right)^{3} - (D \cdot \eta)^{3} \sum_{k \in \mathbb{Z}_{>0}} n_{k\eta} k^{3}
\end{equation}
and
\begin{equation}
    c_{2} \left( X' \right) \cdot D' = c_2 \left( X \right) \cdot D + 2 (D \cdot \eta) \sum_{k \in \mathbb{Z}_{>0}} n_{k\eta} k\,,
\label{eq:c2-flop-change}
\end{equation}
where $n_{k\eta}$ are the genus-zero Gopakumar-Vafa invariants of the flopping classes. Hence, $X$ and $X'$ are birational to each other but not necessarily diffeomorphic.

In what follows, we will center our attention on flops for which $X$ and $X'$ are, in fact, diffeomorphic. Since this is an isomorphism in the smooth category, we will refer to these as isomorphic flops or iso-flops for short. According to Wall's theorem \cite{Wall1966ClassificationV}, the diffeomorphism class of a simply connected Calabi-Yau threefold with torsion-free homology is completely determined by its Hodge numbers $h^{1,1}(X)$ and $h^{2,1}(X)$, its second Chern class $c_{2}(X)$ and its triple intersection numbers $\kappa_{ijk} \coloneqq D_{i} \cdot D_{j} \cdot D_{k}$, where $\{ D_{i}\}_{i \in \{ 1, \dotsc, h^{1,1}(X) \}}$ is a basis of $\mathrm{Pic}(X) \cong H^{2}(X,\mathbb{Z}) \cong \mathbb{Z}^{h^{1,1}(X)}$. Since the Hodge numbers remain unchanged under flops, only the latter two objects need to be checked in order to certify whether a candidate flop is of the isomorphic kind or not. In \cref{sec:CICY-Coxeter-Database} we construct a database of all iso-flops for K\"ahler-favorable CICYs following this natural strategy: we first enumerate all possible flops and then check whether the corresponding integral divisor lattice isomorphism $\phi: H^{2}(X,\mathbb{Z}) \rightarrow H^{2}(X',\mathbb{Z})$ is such that $\kappa_{X} = \phi^{*}(\kappa_{X'})$ and $c_{2}(X) = \phi^{*}(c_{2}(X'))$, where $\kappa_{X}$ is the triple intersection form defined by $\kappa_{X}(D_{i},D_{j},D_{k}) = \kappa_{ijk}$.\footnote{In the second pullback we have employed the pairing $H^{4}(X,\mathbb{Z}) \times H^{2}(X,\mathbb{Z}) \rightarrow \mathbb{Z}$, which is perfect for torsion-free homology, to regard $c_{2}(X)$ as a linear form in $H^{2}(X,\mathbb{Z})$, and likewise for $X'$.} K\"ahler-favorable CICYs enjoy a series of properties that simplify this process, which we now review. Below, we use the compact notation $h \coloneqq h^{1,1}(X)$.

\subsection{Iso-flops in K\"ahler-favorable CICYs}
\label{sec:iso-flops-Kahler-favorable}

While the conclusions of the paper are more general, the explicitly worked-out examples and the CICY Coxeter Database focus on K\"ahler-favorable CICYs, a subset of CICYs \cite{Candelas:1987kf} highlighted in the favorable CICY list \cite{Anderson:2017aux} for their nice properties.

For a CICY $X$, let us denote its ambient space by $\mathcal{A}$. We say that $X$ has a favorable embedding if the divisors of $X$ descend from those of $\mathcal{A}$ in the sense that $\mathrm{Pic}(\mathcal{A})$ surjects onto $\mathrm{Pic}(X)$, implying that $h = h^{1,1}(X) \leq \rank(\mathrm{Pic}(\mathcal{A}))$. An example of a favorable CICY with a non-injective descent of the Picard group is the Schoen manifold; see the discussion in \cite{Anderson:2017aux}. Note that some of the favorable embeddings found in \cite{Anderson:2017aux} require taking the ambient space to be a product of almost del Pezzo surfaces, rather than projective spaces. Under the restriction map $\rho: H^{2}(\mathcal{A},\mathbb{Z}) \otimes \mathbb{R} \rightarrow H^{2}(X,\mathbb{Z}) \otimes \mathbb{R}$, all K\"ahler classes of the ambient space descend to K\"ahler classes of the CICY, meaning that $\rho(\mathcal{K}_{\mathcal{A}}) \subseteq \mathcal{K}_{X}$. For those families admitting a favorable embedding in a product of projective spaces $\mathcal{A} = \mathbb{P}^{n_{1}} \times \cdots \times \mathbb{P}^{n_{F}}$, we have that $\mathrm{Pic(\mathcal{A})} = \langle H_{1}, \dotsc, H_{F} \rangle_{\mathbb{Z}}$ and the K\"ahler cone $\mathcal{K}_{\mathcal{A}}$ of the ambient space is simply the positive orthant in $H^{2}(\mathcal{A},\mathbb{Z}) \otimes \mathbb{R}$. As a consequence, the favorable property implies that the Kähler cone $\mathcal{K}_{X}$ of $X$ is at least as big as the positive orthant in $H^{2}(X,\mathbb{Z}) \otimes \mathbb{R}$. All models of this type in the favorable CICY list have $h = h^{1,1}(X) = \rank(\mathrm{Pic}(\mathcal{A}))$, with $\{ D_{i} \}_{i \in \{ 1, \dotsc, F \}}$ giving a basis of $\mathrm{Pic}(X)$, where $D_{i} \coloneqq \left. H_{i} \right|_{X}$ are the restrictions of the hyperplane classes.

We say that $X$ is K\"ahler-favorable if, in addition, the K\"ahler cone $\mathcal{K}_{X}$ of $X$ descends from that of $\mathcal{A}$, i.e., if the restriction map is such that $\mathcal{K}_{X} = \rho(\mathcal{K}_{\mathcal{A}})$. This means that every K\"ahler class in $X$ is the restriction of a K\"ahler class in $\mathcal{A}$. When the ambient space is $\mathcal{A} = \mathbb{P}^{n_{1}} \times \cdots \times \mathbb{P}^{n_{F}}$ it then follows that $\mathcal{K}_{X}$ is simply the positive orthant $\langle D_{1}, \dotsc, D_{h} \rangle_{\mathbb{R}_{>0}} \subset H^{2}(X,\mathbb{Z}) \otimes \mathbb{R}$. This makes K\"ahler-favorable CICYs a particularly computable class of Calabi-Yau threefolds. Throughout the rest of the paper, we will focus on such Calabi-Yau threefolds constructed in products of projective spaces.

A concise way of describing a family of CICYs is through its configuration matrix. For K\"ahler-favorable CICYs, each row of the configuration matrix corresponds to a generator, and therefore also a wall, of the K\"ahler cone. The flop walls of this class of Calabi-Yau threefolds were characterized in \cite{Brodie:2021toe} in terms of certain configuration matrix row patterns. Below we will make use of the following proposition.

\begin{proposition}
    Let $X$ be a K\"ahler-favorable CICY threefold with generic defining equations. The configuration matrix of $X$ indicates the presence (or absence) of flops and, in particular, that flops occur only in the following two cases:
    \begin{itemize}
        \item \textbf{Type 1:} The configuration matrix is of the form
        \begin{equation}
            \begingroup
            \delimiterfactor=1000
            \begin{bNiceArray}{c|ccccccc}
                \mathbb{P}^{n} & 1 & 1 & \cdots & 1 & 0 & \cdots & 0\\
                \vec{\mathbb{P}} & \vec{q}_{1} & \vec{q}_{2} & \cdots & \vec{q}_{n+1} & \vec{q}_{n+2} & \cdots & \vec{q}_{K}\\
            \end{bNiceArray}
            \endgroup
            \,.
        \end{equation}
        In this case, the CICY threefold can be flopped along the $\mathbb{P}^{n}$ direction
        \begin{enumerate}[label=(\alph*)]
            \item to a diffeomorphic Calabi-Yau threefold, if $\vec{q}_{1} = \vec{q}_{2} = \cdots = \vec{q}_{n+1}$; or
            \item to a typically, but not necessarily, non-diffeomorphic Calabi-Yau threefold otherwise.
        \end{enumerate}

        \item \textbf{Type 2:} The configuration matrix is of the form
        \begin{equation}
            \begingroup
            \delimiterfactor=1000
            \begin{bNiceArray}{c|ccccccc}
                \mathbb{P}^{n} & 2 & 1 & \cdots & 1 & 0 & \cdots & 0\\
                \vec{\mathbb{P}} & \vec{p}_{1} & \vec{p}_{2} & \cdots & \vec{p}_{n} & \vec{p}_{n+1} & \cdots & \vec{p}_{K}\\
            \end{bNiceArray}
            \endgroup
            \,,
        \end{equation}
        including the case with no unit entries in the first row. In this case, the CICY can be flopped along the $\mathbb{P}^{n}$ direction to a diffeomorphic Calabi-Yau threefold.
    \end{itemize}
\label{prop:iso-flop-row-types}
\end{proposition}
We will refer to these types of configuration matrix rows as Type~1a, Type~1b and Type 2 rows, respectively. While constructing the CICY Coxeter Database in \cref{sec:CICY-Coxeter-Database}, we will exploit this proposition to determine which candidate integral involutions of the divisor lattice can directly be regarded as iso-flops and which ones require additional scrutiny via Wall's theorem.

In what follows, we will rarely make reference to the ambient space $\mathcal{A}$ and therefore use the abbreviated notation $\mathcal{K} \coloneqq \mathcal{K}_{X}$ to refer to the K\"ahler cone of $X$.

\subsection{Mori and K\"ahler representations of the iso-flop involutions}
\label{sec:iso-flop-representations}

To study how the iso-flops constrain the prepotential, it will be useful to determine their action on the K\"ahler moduli and the curve classes. Crossing an iso-flop wall brings us to a chamber of the extended K\"ahler cone representing a diffeomorphic family of Calabi-Yau threefolds. This K\"ahler cone chamber is glued to the initial one by the iso-flop wall. Hence, if we cross the same iso-flop wall from the new chamber, we return to the initial K\"ahler cone chamber. This implies that each iso-flop is an involution $W = \mathbb{Z}_{2}$ whose representation we want to explicitly determine and study.

Let us denote the curve classes by $d \in \mathcal{M}$ and the complexified K\"ahler moduli by $T^{i} = b^{i} + it^{i}$, for $i \in \{ 1, \dotsc, h \}$, where the $\{ t^{i} \}_{i \in \{ 1, \dotsc, h \}}$ appearing in the decomposition of the K\"ahler form $J = t^{i}D_{i} \in H^{2}(X,\mathbb{R})$ are the real K\"ahler moduli and the $\{ b^{i} \}_{i \in \{ 1, \dotsc, h \}}$ appearing in the decomposition of the Kalb–Ramond field $B = b^{i}D_{i} \in H^{2}(X,\mathbb{R})/H^{2}(X,\mathbb{Z})$ are their axionic superpartners. Consider the vector space $V \cong \mathbb{R}^{h} \supset \mathcal{M}$ and its dual $\hat{V} \cong \mathbb{R}^{h} \supset \mathcal{K}$. The action of $W$ on curve classes is naturally understood in terms of a real representation $\rho: W \rightarrow\mathrm{GL}(V)$, that we will call the Mori representation. Due to supersymmetry, the real and imaginary parts of the complexified K\"ahler moduli must transform in the same way under the action of $W$. Hence, we can focus on the dual real representation $\hat{\rho}: W \rightarrow \mathrm{GL}(\hat{V})$ acting on the real K\"ahler moduli, which we will call the K\"ahler representation, and complexify it whenever we act on the complexified K\"ahler moduli. The Mori and K\"ahler representations of a group element $w \in W$ are related by $\rho(w) = \hat{\rho}\big(w^{-1}\big)^{T}$. Since iso-flops lead to integral involutions of the divisor and curve lattices, the matrices representing $W$ will have integral entries and iso-flops will generate crystallographic groups.

Focusing on Calabi-Yau threefolds with $h \geq 2$ and a simplicial K\"ahler cone, a necessary condition for an iso-flop to exist along the $\{ t^{i} = 0 \}$ K\"ahler cone wall is the existence of an integral vector $u_{i}$ with components $u_{i}^{a}$, for $a \in \{ 1, \dotsc, h \}$, satisfying \cite{Lukas:2022crp}
\begin{equation}
    u_{i}^{i} = 2\,,\qquad u_{i}^{a} \leq 0\,,\quad a \neq i\,,\qquad \kappa_{abc}u_{i}^{a} = 0\,,\quad b,c \neq i\,,\qquad u_{i}^{a} \in \mathbb{Z}\,.
\label{eq:uvec-definition}
\end{equation}
The K\"ahler representation of the corresponding candidate iso-flop involution is then given by the matrix
\begin{equation}
    \hat{\mathcal{M}}_{i} = \mathds{1}_{h} - \left( \vec{0}, \dotsc, \vec{0}, u_{i}, \vec{0}, \dotsc, \vec{0} \right)\,,
\label{eq:Kahler-rep-u-vector}
\end{equation}
with the dual Mori representation obtained by transposition $\mathcal{M}_{i} = \hat{\mathcal{M}}_{i}^{T}$. To certify that this candidate involution is indeed an iso-flop, one needs to compute the relevant topological data across the flop transition and check whether Wall's theorem confirms that the manifolds are diffeomorphic. In the context of K\"ahler-favorable CICYs, we can sometimes skip this check, depending on the configuration matrix row to which the $u_{i}$-vector is associated with, by invoking \cref{prop:iso-flop-row-types}.

\subsection{Iso-flop Coxeter groups}
\label{sec:iso-flop-Coxeter-groups}

Consider a Calabi-Yau threefold $X$ whose K\"ahler cone has iso-flop walls corresponding to the involutions $\{s_{i}\}_{i \in I}$. In the K\"ahler representation $\hat{\mathcal{M}}_{i} = \hat{\rho}(s_{i})$ it is easy to directly check that $\hat{\mathcal{M}}_{i}^{2} = \mathds{1}_{h}$ with $(h-1)$ unit eigenvalues and a single $-1$ eigenvalue. Hence, these act as reflections in the moduli space and serve as the generators of a Coxeter group \cite{Vinberg1971DiscreteLinearGroups}.

A Coxeter system $(W,S)$ consists of a group $W$ generated by involutions $S \coloneqq \{s_{i}\}_{i \in I}$ with the presentation
\begin{equation}
    W = \left\langle s_{1}, \dotsc, s_{n} \mid (s_{i}s_{j})^{c_{ij}} = 1 \right\rangle\,,
\end{equation}
where the relations are defined in terms of the $c_{ij} \in \mathbb{Z} \cup \{ \infty \}$ satisfying
\begin{equation}
    c_{ii} = 1\qquad \text{and}\qquad c_{ij} = c_{ji} \in \{2, 3, \dotsc, \infty\}\quad(i \neq j)\,.
\label{eq:Coxeter-matrix-def}
\end{equation}
Whenever $c_{ij} = \infty$ no relation is to be imposed among $s_{i}$ and $s_{j}$. The elements of $S$ are known as simple reflections and they are not uniquely determined by the group $W$ in general. It is common to present the integers $c_{ij}$ as the entries of the Coxeter matrix $C$. This matrix can be depicted graphically in terms of a Coxeter diagram, associating a node to each simple reflection in $S$ and joining two nodes with an edge if $c_{ij} \geq 3$. The edges carry labels $c_{ij}$ except for the $c_{ij} = 3$ case, for which the label is conventionally omitted. A useful auxiliary object is the Schl\"afli matrix $B$ with entries $b_{ij} \coloneqq -2 \cos(\pi/c_{ij})$. Its eigenvalues determine the type of Coxeter group at hand, namely of finite type if the Schl\"afli matrix is positive definite, of affine type if its eigenvalues are non-negative with at least one being zero and of indefinite type otherwise. Coxeter groups are an old subject and more information can be found in a plethora of textbooks, see, for example, \cite{Humphreys1990ReflectionGroups,Davis2008GeometryTopologyCoxeter,Kane2001ReflectionGroups}.

In our setting, the moduli space geometry selects not only a concrete Coxeter system $(W,S)$ for the Coxeter group but also a particular pair of K\"ahler and Mori representations, which may differ from the standard geometric representation of $W$. These were discussed in \cref{sec:iso-flop-representations} for a simple reflection, with the general group element obtained from products of the (transposes of the) matrices \eqref{eq:Kahler-rep-u-vector} as
\begin{equation}
    w = s_{i_{1}} \cdots s_{i_{k}} \implies \rho(w) = \mathcal{M}_{i_{1}} \cdots \mathcal{M}_{i_{k}}\,.
\end{equation}
Here and in most of the text we will use the Mori representation, with this choice being arbitrary and leading to the same final expressions as the K\"ahler representation.

It is convenient to introduce some additional notation that will be used in later sections. Consider two iso-flop walls, which we take without loss of generality along the K\"ahler cone boundaries $\{ t^{1} = 0\}$ and $\{ t^{2} = 0\}$. Define the integers $m_{1} \coloneqq -u_{1}^{2}$ and $m_{2} \coloneqq -u_{2}^{1}$. We can split the $u$-vectors for this pair of simple reflections into
\begin{subequations}
\begin{align}
    u_{1} &= \left( 2,-m_{1}, -(u'_{1})^{1}, \dotsc, -(u'_{1})^{h-2} \right)\,,\\
    u_{2} &= \left( -m_{2},2, -(u'_{2})^{1}, \dotsc, -(u'_{2})^{h-2} \right)\,,
\end{align}
\label{eq:uprimevec-definition}%
\end{subequations}
where the $(u'_{i})^{a'}$, $i \in \{1,2\}$, $a' \in \{1, \dotsc, h-2\}$ are non-negative and form the components of a reduced $u'$-vector. Defining the matrices
\begin{equation}
    M_{1} =
    \begin{pNiceArray}{cc}[cell-space-limits=2.5pt]
        -1 & m_{1}\\
        0 & 1
    \end{pNiceArray}
    \,,\quad
    M_{2} =
    \begin{pNiceArray}{cc}[cell-space-limits=2.5pt]
        1 & 0\\
        m_{2} & -1
    \end{pNiceArray}
    \,,\quad
    U_{1} =
    \begin{pNiceArray}{c}[cell-space-limits=2.5pt]
        u'_{1}\\
        \vec{0}^{\;T}
    \end{pNiceArray}
    \,,\quad U_{2} =
    \begin{pNiceArray}{c}[cell-space-limits=2.5pt]
        \vec{0}^{\;T}\\
        u'_{2}
    \end{pNiceArray}
    \,,
\end{equation}
we see that the matrices representing the two simple reflections under consideration have the product structure
\begin{equation}
	\mathcal{M}_{1} =
	\begin{pNiceArray}{c|l}[cell-space-limits=5pt]
		M_{1}  & U_{1}\\
		\hline
		0 & \mathds{1}_{h-2}
	\end{pNiceArray}
	\,,\qquad
	\mathcal{M}_{2} =
	\begin{pNiceArray}{c|l}[cell-space-limits=5pt]
		M_{2} & U_{2}\\
		\hline
		0 & \mathds{1}_{h-2}
	\end{pNiceArray}
	\,.
\label{eq:block-structure-M}
\end{equation}
The rotation obtained from the product of two reflections will also be referenced frequently, using the notation $\mathcal{Q}_{ij} \coloneqq \mathcal{M}_{i}\mathcal{M}_{j}$ and $\mathcal{S} \coloneqq \mathcal{M}_{i}$. For the selected pair at hand, this leads to the block structure
\begin{equation}
	\mathcal{Q} =
	\begin{pNiceArray}{c|l}[cell-space-limits=5pt]
		Q  & U\\
		\hline
		0 & \mathds{1}_{h-2}
	\end{pNiceArray}
	\,,\qquad
	\mathcal{S} \coloneqq \mathcal{M}_{1}
	\,,
\label{eq:block-structure-QS}
\end{equation}
where
\begin{equation}
    Q \coloneqq M_{1} M_{2}\,,\qquad U = M_{1}U_{2} + U_{1} =
    \begin{pNiceArray}{c}[cell-space-limits=2.5pt]
        u'_{1} + m_{1}u'_{2}\\
        u'_{2}
    \end{pNiceArray}
    \,.
\end{equation}

In addition to the iso-flop Coxeter group $W$ discussed above, the configuration matrix of many K\"ahler-favorable CICYs is invariant under the permutation of certain rows. Since for this class of Calabi-Yau threefolds there is a correspondence between configuration matrix rows and K\"ahler moduli, this results in an additional $S_{n}$ symmetry that can permute the corresponding iso-flop walls. In such cases, the full duality group is $G = W \rtimes S_{n}$. While many of our conclusions hold for $G$, we will focus on $W$ throughout most of the text.

\subsection{Coxeter prepotentials}
\label{sec:coxeter-prepotentials}

Having discussed the conditions under which iso-flop walls appear in the K\"ahler moduli space and how they act as the simple reflections generating a Coxeter group $W$, we are equipped to discuss the effects of this symmetry on the 4D $\mathcal{N} = 2$ Type IIA prepotential, following the study initiated in \cite{Lukas:2022crp}.

The prepotential can be divided into a classical and an instanton-corrected piece
\begin{equation}
    \mathcal{F}(T) = \mathcal{F}_{\mathrm{class}}(T) + \mathcal{F}_{\mathrm{inst}}(T)\,,\qquad \mathcal{F}_{\mathrm{inst}}(T) \coloneqq \sum_{d \in \mathcal{M}} n_{d} \mathrm{Li}_{3} \left( e^{2\pi i \langle d, T \rangle} \right)\,,
\end{equation}
where the $n_{d}$ are the genus-zero Gopakumar-Vafa (GV) invariants. Since the worldsheet instanton contributions are obtained by summing over the effective curve classes, and the Coxeter group directly acts on these, it is natural to expect some form of Coxeter-invariant structure to arise for the prepotential. Not all curves behave the same under the flop; hence, it is convenient to split the Mori cone into different curve class types.

Start by considering the initial K\"ahler moduli space chamber $\mathcal{K}$. By crossing the iso-flop walls we land on adjacent chambers, glued to the previous ones by the fixed locus of the reflections. The union of all these chambers yields a subcone of the effective cone $\mathcal{K}_{\mathrm{eff}}$, which we call the iso-effective cone
\begin{equation}
    \mathcal{K}_{\mathrm{iso}} \coloneqq \bigcup_{w \in W} \mathcal{K}_{w} \subseteq \mathcal{K}_{\mathrm{eff}}\,.
\end{equation}
For each iso-flop K\"ahler chamber $\mathcal{K}_{w}$, take the dual Mori cone $\mathcal{M}_{w} = w\mathcal{M} = \mathcal{K}_{w}^{\vee}$. We can then define the restricted Mori cone
\begin{equation}
    \mathcal{M}_{\mathrm{restr}} \coloneqq \bigcap_{w \in W} \mathcal{M}_{w} = \mathcal{K}_{\mathrm{iso}}^{\vee} \subset \mathcal{M}\,.
\end{equation}
With this definition, it is clear that the restricted Mori cone is invariant under the action of the Coxeter group. It will also be useful to define the fundamental Mori cone
\begin{equation}
    \mathcal{M}_{f} \subset \mathcal{M}_{\mathrm{restr}} \subset \mathcal{M}
\end{equation}
by taking the fundamental domain of the $W$-action on $\mathcal{M}_{\mathrm{restr}}$. Note that, although the fundamental domain for the $W$-action on the iso-effective cone can be taken to be $\mathcal{K}_{f} =\mathcal{K}$, we always have the strict inclusion $\mathcal{M}_{f} \subset \mathcal{K}^{\vee}$, i.e., the two fundamental cones are not dual to each other. This is clear from the fact that the iso-flop walls in $\mathcal{K}$ are dual to flopping rays in $\mathcal{M}$, meaning that they lie in $\mathcal{M} \setminus \mathcal{M}_{\mathrm{restr}}$.

Consider now how the genus-zero GV invariants organize into orbits of the Coxeter group action.\footnote{Here, we focus on genus-zero GV invariants due to our interest in the 4D $\mathcal{N} = 2$ Type IIA prepotential, but higher-genus GV invariants also appear in orbits of the Coxeter group.} First, we have to distinguish flopping curve classes from non-flopping curve classes in the Mori cone $\mathcal{M}$. The image under the $W$-action of a flopping curve class is another flopping curve class, possibly for one of the diffeomorphic Calabi-Yau families rather than the original one. Denoting the set of flopping curve classes of $X$ by $\mathcal{B} \subseteq \mathcal{M}\setminus\mathcal{M}_{\mathrm{restr}}$, take a subset $\mathcal{B}_{f} \subseteq \mathcal{B}$ covering all of $\mathcal{B}$ under the $W$-action without repeating curve classes. The prepotential should then sum over all $wd \in \mathcal{M}$ for $d \in \mathcal{B}_{f}$ and $w \in W$, which share the same genus-zero GV invariants. The images $wd \notin \mathcal{M}$ correspond to flopping curves for a diffeomorphic family and are not summed over; they will in general have different genus-zero GV invariants (when computed in the diffeomorphic family). Second, note that for a non-flopping class $d$, the genus-zero GV invariants are constant along $W$-orbits, i.e., $n_{wd} = n_{d}$ for $w \in W$. In fact, they are invariant across $G$-orbits, where $G = W \rtimes S_{n}$ includes the configuration matrix row permutations, as explained earlier. This forces non-flopping classes outside the restricted Mori cone to have vanishing genus-zero GV invariants. Altogether, the instanton prepotential can be written as
\begin{equation}
    \mathcal{F}_{\mathrm{inst}}(T) = \mathcal{F}_{\mathrm{flop}}(T) + \mathcal{F}_{\mathrm{non-flop}}(T)\,,
\end{equation}
where
\begin{subequations}
\begin{align}
    \mathcal{F}_{\mathrm{flop}}(T) &\coloneqq \sum_{d \in \mathcal{B}_{f}} n_{d} \sum_{\substack{w \in W\\wd \in \mathcal{M}}} \mathrm{Li}_{3} \left( e^{2\pi i \langle wd, T \rangle} \right)\,,\\
    \mathcal{F}_{\mathrm{non-flop}}(T) &\coloneqq \sum_{d \in \mathcal{M}_{f}} n_{d} \Psi_{d}^{W}(T)\,.
\end{align}
\end{subequations}
Here we have introduced the $W$-invariant functions $\Psi_{d}^{W}(T)$, defined by averaging over $W$-orbits as
\begin{equation}
    \Psi_{d}^{W}(T) \coloneqq \sum_{w \in W/\mathrm{Stab}_{W}(d)} \mathrm{Li}_{3}\left( e^{2\pi i \langle wd, T \rangle} \right)\,.
\end{equation}
Taking into account the stabilizer of the curve class under consideration is crucial to not obtain expressions that overcount contributions or are directly divergent.

Going to the Gromov-Witten expansion of the prepotential, we can introduce the $W$-invariant functions $\psi_{ld}^{W}(T)$, defined by
\begin{equation}
    \psi_{ld}^{W}(T) \coloneqq \sum_{w \in W/\mathrm{Stab}_{W}(d)} e^{2\pi i l \langle wd, T \rangle}
\end{equation}
and related to the ones above by the usual multi-covering formula
\begin{equation}
    \Psi_{d}^{W}(T) = \sum_{l \in \mathbb{Z}_{>0}} \frac{1}{l^{3}} \psi_{ld}^{W}(T)\,.
\end{equation}
The properties of $\Psi_{d}^{W}(T)$ and $\psi_{ld}^{W}(T)$ remain thus far largely unexplored. In this paper, we determine resummed expressions for them with favorable convergence properties in the interior of the K\"ahler cone and interpret the results in terms of the geometry of the moduli space.

%% file: sections/CICY-Coxeter-Database.tex
\section{The CICY Coxeter Database}
\label{sec:CICY-Coxeter-Database}

The set of Calabi-Yau threefolds obtained as complete intersections in products of projective spaces \cite{Green:1986ck} remains one of the standard testbeds in string geometry. The complete list of 7890 CICYs was first compiled in the seminal paper \cite{Candelas:1987kf} and later refined and expanded with the addition of Hodge number \cite{Green:1987cr}, free quotient \cite{Braun:2010vc,Candelas:2010ve,Candelas:2015amz,Constantin:2016xlj} and non-freely acting symmetries \cite{Lukas:2017vqp} data.

A major modernization of the CICY dataset was carried out in \cite{Anderson:2017aux}, producing favorable presentations for all but 48 CICY threefolds (that can nonetheless be favorably described in products of almost del Pezzo surfaces), the determination of the full topological data for the list and an in-depth study of fibration structures. The novel favorable presentations permitted a new search for freely acting symmetries \cite{Gray:2021kax}. Most importantly for us, the favorable CICY list \cite{Anderson:2017aux} highlights a subset of 4874 K\"ahler-favorable CICYs with a product of projective spaces for ambient space, meaning that their K\"ahler and Mori cones directly descend from the latter, as reviewed in \cref{sec:iso-flops-Kahler-favorable}. There are an additional 83 CICYs that are K\"ahler-favorable when embedded in a product of almost del Pezzo surfaces, but we will ignore these in what follows.

Our study of iso-flop walls requires us to explicitly determine the K\"ahler and Mori cones of the Calabi-Yau threefolds under consideration. In general, this is very complicated, since the cones can be non-simplicial or even non-polyhedral (e.g., for the Schoen manifold).\footnote{The Kawamata-Morrison \cite{Morrison1993CompactificationsModuli,Kawamata1997ConeDivisorsCYFiber} conjecture claims that, in these cases, the action of the automorphism group of the variety on the K\"ahler cone has a rational polyhedral fundamental domain. This has been confirmed, e.g., for the Schoen manifold \cite{GrassiMorrison1993AutomorphismsKahlerCone}.} Hence, the 4874 K\"ahler-favorable CICYs offer an excellent starting point for a classification of iso-flop walls and the Coxeter symmetries of the moduli space that they lead to. We build on the favorable CICY list \cite{Anderson:2017aux} and add Coxeter data for all K\"ahler-favorable models.

Employing the triple intersection data of the CICYs, we test whether a $u$-vector satisfying~\eqref{eq:uvec-definition} exists for each row of the configuration matrices of the K\"ahler-favorable models; this is a necessary condition for an iso-flop to exist. The $u$-vector scan yields a total of 2921 iso-flop row candidates, split into 399 Type~1a rows, 250 Type~1b rows and 2272 Type~2 rows. As reviewed in \cref{prop:iso-flop-row-types}, Type~1a and Type~2 rows are guaranteed to yield iso-flop walls \cite{Brodie:2021toe}. Type~1b rows, of which there are a total 28132 among the K\"ahler-favorable models, result in flop walls that are typically non-isomorphic. The $u$-vector test ensures the existence of an integral involution, represented by the matrix $\hat{\mathcal{M}}$ given in \eqref{eq:Kahler-rep-u-vector}, identifying the divisor lattices of the two chambers and under which the cubic triple intersection form is invariant. In other words, once the existence of a flop $f: X \dashrightarrow X'$ is established, the pre-flop cubic triple intersection form is the pullback of the post-flop one. Moreover, the Hodge numbers $h^{1,1}(X)$ and $h^{2,1}(X)$ remain invariant under flops. According to Wall's theorem \cite{Wall1966ClassificationV}, checking that the involution is such that the pre-flop $c_{2}(X)$ is the pullback of the post-flop $c_{2}(X')$ would certify the isomorphism in the smooth category. Indeed, for all 250 Type~1b row candidates one can check using \eqref{eq:c2-flop-change} that\footnote{Strictly speaking, we are checking that the linear functional defined by $\ell_{X}(D) \coloneqq \int_{X} c_{2}(X) \smile D$ satisfies $\ell_{X'}(\phi(D)) = \ell_{X}(D)$ for all $D \in H^{2}(X,\mathbb{Z})$, where $\phi: H^{2}(X,\mathbb{Z}) \rightarrow H^{2}(X',\mathbb{Z})$ is the identification of divisor lattices stemming from the isomorphism in codimension-one associated with the birational map \mbox{$f: X \dashrightarrow X'$}. This implies that $c_{2}(X) = \phi^{*}(c_{2}(X'))$ in $H^{4}(X,\mathbb{Z})$ when $H^{4}(X,\mathbb{Z}) \times H^{2}(X,\mathbb{Z}) \rightarrow \mathbb{Z}$ is perfect, which is the case for a closed and oriented manifold $X$ with torsion free homology.}
\begin{equation}
    c_{2}\left( X' \right) \cdot D'_{i} = \left( \hat{\mathcal{M}}_{ij}^{-1} \right)^{T} \left( c_{2}(X) \cdot D_{j} \right) = \mathcal{M}_{ij} \left( c_{2}(X) \cdot D_{j} \right)\,,\qquad i,j \in \{1, \dotsc, h\}\,,
\end{equation}
see \cref{sec:elliptic-example-I2(4)} for an illustrative example. In this way, we obtain a complete list of all iso-flop configuration matrix rows and the matrices that represent the associated involutions acting on the K\"ahler/Mori cone of the K\"ahler-favorable CICYs.

The aforementioned involutions correspond to the simple reflections generating the Coxeter groups that we are interested in. Given a concrete K\"ahler-favorable CICY and its iso-flop walls, we can determine the Coxeter matrix \eqref{eq:Coxeter-matrix-def} of the group that they lead to by computing the order of $\mathcal{Q}_{ij} = \mathcal{M}_{i}\mathcal{M}_{j}$ for all pairs of associated involution matrices.\footnote{An efficient way of determining the order of $\mathcal{Q}_{ij}$ is the following. Factor its characteristic polynomial $\chi_{\mathcal{Q}_{ij}} = \prod_{\alpha \in A} f_{\alpha}$ over $\mathbb{Q}$ and, for each irreducible factor $f_{\alpha}$ of $\deg(f_{\alpha}) = d_{\alpha}$, certify cyclotomicity by a single divisibility check $f_{\alpha} \mid x^{M(d_{\alpha})} - 1$, where $M(d_{\alpha}) = \prod_{p-1 \leq d_{\alpha}} p^{E_{p}(d_{\alpha})}$ and $E_{p}(d_{\alpha})$ is maximal such that $p^{E_{p}(d_{\alpha})-1}(p-1) \leq d_{\alpha}$ from Euler's totient on prime powers. If $f_{\alpha}$ passes the test, obtain $n_{\alpha} \coloneqq \min\{ k \geq 1 : f_{\alpha} \mid x^{k}-1 \}$ by successively dividing out prime factors off $M(d_{\alpha})$. If any irreducible factor fails the divisibility check, assign $\mathrm{ord}(\mathcal{Q}_{ij}) = \infty$. Otherwise, compute $\mathrm{lcm}_{\alpha \in A}(n_{\alpha})$ and check whether the corresponding power of $\mathcal{Q}_{ij}$ is the identity. If it is not, assign $\mathrm{ord}(\mathcal{Q}_{ij}) = \infty$. If the check is successful, reduce $\mathrm{lcm}_{\alpha \in A}(n_{\alpha})$ to $\mathrm{ord}(\mathcal{Q}_{ij})$ by dividing out appropriate prime factors. Compute the powers $x^{k} \mod f_{\alpha}$ and $\mathcal{Q}_{ij}^{k}$ using binary exponentiation to increase the efficiency of the algorithm.} Out of the 4874 K\"ahler-favorable CICYs, 2182 enjoy a Coxeter action on their moduli space. The most abundant case, corresponding to 1592 models, is that in which a single iso-flop wall appears, leading to $W = \mathbb{Z}_{2}$. The 590 remaining cases, with $\rank(W) \geq 2$, feature a total of 18 different Coxeter diagrams which, according to the eigenvalues of their Schl\"afli matrix, can be classified into 9 finite Coxeter groups
\begin{equation}
\begin{aligned}
    &\makebox[0.8\linewidth][s]{%
        \dynkins{[Coxeter]A1|[Coxeter]A1}\hfil
        \dynkins{[Coxeter]A1|[Coxeter]A1|[Coxeter]A1}\hfil
        \dynkins{[Coxeter]A1|[Coxeter]A1|[Coxeter]A2}
    }\\[8pt]
    &\makebox[0.8\linewidth][s]{%
        \dynkins{[Coxeter]A2}\hfil
        \dynkins{[Coxeter]A1|[Coxeter]A2}\hfil
        \dynkins{[Coxeter]A2|[Coxeter]A2}
    }\\[8pt]
    &\makebox[0.8\linewidth][s]{%
        \dynkins{[Coxeter]B2}\hfil
        \dynkins{[Coxeter]A1|[Coxeter]B2}\hfil
        \dynkins{[Coxeter]B2|[Coxeter]B2}\rlap{\,,}
    }
\end{aligned}
\end{equation}
1 affine Coxeter group
\begin{equation}
\begin{aligned}
    \makebox[0.5\linewidth][s]{
        \begin{dynkinDiagram}[Coxeter,label=false]A{**}
            \dynkinEdgeLabel{1}{2}{\mathrm{H}}
        \end{dynkinDiagram}
    }\\[8pt]
    \makebox[0.5\linewidth][s]{
        \begin{dynkinDiagram}[Coxeter,label=false]A{**}
            \dynkinEdgeLabel{1}{2}{\mathrm{P}}
        \end{dynkinDiagram}\rlap{\,,}
    }
\end{aligned}
\end{equation}
and 8 indefinite Coxeter groups
{\allowdisplaybreaks%
\begin{align}
    &\makebox[0.8\linewidth][s]{%
        \CoxDiagramVPH\hfil
        \CoxDiagramStarThreeLegP\hfil
        \CoxDiagramStarFourLegFour
    }\nonumber\\[8pt]
    &\makebox[0.8\linewidth][s]{%
        \CoxDiagramVPP\hfil
        \CoxDiagramPStarATwoEdge\hfil
        \CoxDiagramKFiveAllThree
    }\nonumber\\[8pt]
    &\makebox[0.8\linewidth][s]{%
        \CoxDiagramTriangleAPP\hfil
        \CoxDiagramKFourAllP\hfil
        \phantom{\CoxDiagramKFiveAllThree}
    }\\[8pt]
    &\makebox[0.8\linewidth][s]{%
        \CoxDiagramTrianglePPH\hfil
        \phantom{\CoxDiagramKFourAllP}\hfil
        \phantom{\CoxDiagramKFiveAllThree}
    }\nonumber\\[8pt]
    &\makebox[0.8\linewidth][s]{%
        \CoxDiagramTrianglePPP\rlap{\,.}\hfil
        \phantom{\CoxDiagramKFourAllP}\hfil
        \phantom{\CoxDiagramKFiveAllThree}
    }\nonumber
\end{align}%
}%
Here, the edge labels P and H are to be interpreted as $\infty$. The additional split they provide into 21 diagrams refers to the properties of their Mori/K\"ahler representation, as will be explained in \cref{sec:dihedral-Mori-Kahler-representations}, while the abstract Coxeter group remains identical. The number of models presenting a given $\rank(W)$ and $h^{1,1}(X)$ are collected in \cref{tab:CICY-Coxeter-database-counts},
\begin{table}[t]
    \centering
    \resizebox{\linewidth}{!}{
    \begin{tblr}[evaluate=\fileInput]{
        width  = \linewidth,
        colsep = 3pt,
        column{1-2} = {colsep=3pt},
        colspec = {c c *{10}{c c} c c},
        cells   = {c, valign=m},
        hlines = {solid},
        vlines = {solid}
    }
    \SetCell[c=2]{c}{\diagbox[dir=NW]{rank}{\raisebox{-0.75ex}{\(h^{1,1}(X)\)}}} & & \SetCell[c=2]{c}{\(2\)}  & & \SetCell[c=2]{c}{\(3\)}  & & \SetCell[c=2]{c}{\(4\)}  & & \SetCell[c=2]{c}{\(5\)}  & & \SetCell[c=2]{c}{\(6\)}  & & \SetCell[c=2]{c}{\(7\)}  & & \SetCell[c=2]{c}{\(8\)}  & & \SetCell[c=2]{c}{\(9\)}  & & \SetCell[c=2]{c}{\(10\)} & & \SetCell[c=2]{c}{\(11\)} & & \SetCell[c=2]{c}{Total}  & \\
    \fileInput{data/CICY-Coxeter-Database-counts-summary}
    \end{tblr}
    }
    \caption{Number of K\"ahler-favorable CICYs by $\rank(W)$ and $h^{1,1}(X)$.}
    \label{tab:CICY-Coxeter-database-counts}
\end{table}
while the concrete K\"ahler-favorable CICYs displaying a particular Coxeter group with $\rank(W) \geq 2$ are presented in \cref{tab:CICY-Coxeter-database}, grouped by $h^{1,1}(X)$. One can observe that the more constraining infinite order Coxeter symmetries appear for a total of 251 models.

Note that all Coxeter groups found in the scan are either finite, affine, or indefinite crystallographic Coxeter groups, see \cref{sec:iso-flop-representations}. As such, the range of, e.g., dihedral groups $\mathrm{I}_{2}(m)$ that can be realized could have been predicted to be $m \in \{ 2,3,4,6, \infty \}$ a priori. Interestingly, the $\mathrm{I}_{2}(6)$ case is not realized among the K\"ahler-favorable CICYs. In our exhaustive search, we do find models with $W = \mathrm{I}_{2}(4)$, which did not appear in the $\rank(W) = 2$ (subgroups) scan of \cite{Lukas:2022crp}. The latter focused on finding Type 1a and Type 2 iso-flop row pairs and missed the 250 Type~1b rows that do lead to iso-flops; these appear in all pairs of CICY configuration matrix rows associated with $\mathrm{I}_{2}(4)$ (sub)groups generated by a pair of simple reflections that are collected in the database.

From this overview of the CICY Coxeter Database we can extract a few observations. First, although the K\"ahler-favorable CICYs have $h^{1,1}(X) \leq 12$, Coxeter groups of infinite order are only realized for $h^{1,1}(X) \leq 5$. It is for $h^{1,1}(X) = 5$ that the maximum value of $\rank(W) = 5$ is achieved as well. Naively, one could have expected higher $h^{1,1}(X)$ to provide more room for iso-flop walls and complicated Coxeter dynamics; whether this concentration at low $h^{1,1}(X)$ and $\rank(W)$ is a feature of K\"ahler-favorable CICYs or more general would be interesting to study by extending the analysis to, e.g., the Kreuzer-Skarke database \cite{Kreuzer:2000xy} or to even more general constructions of Calabi-Yau threefolds. The most prevalent infinite order Coxeter group is $\mathrm{I}_{2}(\infty)$, appearing a total of 189 times, which is why we choose the dihedral group $\mathrm{I}_{2}(m)$ as the example on which we build the general discussion in \cref{sec:dihedral-Coxeter-prepotentials,sec:harmonic-analysis}. Moreover, any pair of iso-flop walls generates an $\mathrm{I}_{2}(m)$ subgroup of $W$, see \cref{tab:dihedral-subgroups-count}, meaning that the general case can be approached from a dihedral building blocks perspective as we do in \cref{sec:general-Coxeter-prepotentials}.
\begin{table}[t]
    \centering
    \begin{tblr}{
        columns = {c},
        column{1-Z} = {mode=math},
        hlines = {solid},
        vlines = {solid}
    }
        \mathrm{I}_{2}(2) & \mathrm{I}_{2}(3) & \mathrm{I}_{2}(4) & \mathrm{I}_{2}(\infty) \text{ (P)} & \mathrm{I}_{2}(\infty) \text{ (H)} \\
        474 & 69 & 68 & 285 & 39
    \end{tblr}
    \caption{Number of $\mathrm{I}_{2}(m)$ subgroups of $W$ generated by pairs of iso-flop walls among the K\"ahler-favorable CICYs.}
    \label{tab:dihedral-subgroups-count}
\end{table}

\begin{sidewaystable}[p!]
    \centering
    \resizebox{\linewidth}{!}{
    \begin{tblr}[evaluate=\fileInput]{
        columns = {c},
        cells = {valign=m},
        row{1} = {abovesep=3pt,  belowsep=3pt},
        row{2-Z} = {abovesep=7.5pt,belowsep=7.5pt},
        cell{2-Z}{1} = {cmd=\CoxDiagramCell},
        cell{2-Z}{2} = {cmd=\CoxTextCell},
        cell{2-Z}{3} = {cmd=\CoxTextCell},
        column{4-Z} = {cmd=\NPerLine},
        hlines = {solid},
        vlines = {solid}
    }
        Coxeter Matrix & Type & Count & $h^{1,1}(X)$ = 2 & $h^{1,1}(X)$ = 3 & $h^{1,1}(X)$ = 4 & $h^{1,1}(X)$ = 5 & $h^{1,1}(X)$ = 6 & $h^{1,1}(X)$ = 7 & $h^{1,1}(X)$ = 8 & $h^{1,1}(X)$ = 9 & $h^{1,1}(X)$ = 10\\
        \fileInput{data/CICY-Coxeter-Database-summary-diagrams} 
    \end{tblr}
    }
    \caption{K\"ahler-favorable CICYs with Coxeter groups of $\rank(W) \geq 2$.}
    \label{tab:CICY-Coxeter-database}
\end{sidewaystable}

The complete CICY Coxeter Database is provided as ancillary material with the arXiv submission of this paper. An online version is available at \cite{AlvarezGarciaCICYCoxeterDatabase}. As an example, let us look at the CICY 6771 entry.
\begin{center}
\noindent\begin{minipage}{\linewidth}
\begin{Verbatim}[fontsize=\small, breaklines=true, frame=single]
    {Num -> 6771,
     H11 -> 3,
     H21 -> 35,
     C2 -> {24, 44, 44},
     Conf -> {{1, 1, 0, 0}, {2, 0, 1, 1}, {0, 2, 1, 1}},
     Favour -> True,
     KahlerPos -> True,
     IsProduct -> False,
     IsoFlopRows -> {{2, "Type 2"}, {3, "Type 2"}},
     KahlerRefGens -> {{{1, 1, 0}, {0, -1, 0}, {0, 4, 1}},
                       {{1, 0, 1}, {0, 1, 4}, {0, 0, -1}}},
     CoxeterMat -> {{1, H}, {H, 1}}}
\end{Verbatim}
\end{minipage}
\end{center}
It contains the same substitution rules as in \cite{Anderson:2017aux} with the addition of the Coxeter data \texttt{IsoFlopRows}, \texttt{KahlerRefGens} and \texttt{CoxeterMat} (which read \texttt{NonKahlerPos} for those models that are non-K\"ahler-favorable). The field \texttt{IsoFlopRows} establishes, in the concrete presentation of the configuration matrix used in the database, which rows lead to \mbox{iso-flops}. The matrices corresponding to the associated simple reflections in the K\"ahler representation are stored in \texttt{KahlerRefGens}. Finally, \texttt{CoxeterMat} prints the relations between simple reflection pairs in the order established by \texttt{IsoFlopRows}. This concrete model is used as an example in \cref{sec:hyperbolic-dihedral-example}.

%% file: sections/dihedral-Coxeter-prepotentials.tex
\section{Dihedral Coxeter Prepotentials}
\label{sec:dihedral-Coxeter-prepotentials}

The dihedral group is the most abundant $\rank(W) \geq 2$ case in the CICY Coxeter Database and the simplest among them. Despite being generated by just two iso-flops, its action on the moduli space leads to strong structural consequences for the form of the instanton prepotential in 4D $\mathcal{N} = 2$ Type IIA compactifications, especially for the infinite dihedral group subcase.

In the present section we focus on this Coxeter group and, building on the analysis initiated in \cite{Lukas:2022crp}, fully characterize the Coxeter-invariant functions appearing in the prepotential through direct computation. The study of dihedral models can be split into three subcases: hyperbolic, parabolic and elliptic. The first two correspond to the infinite dihedral group and the last one agglomerates the finite dihedral group models. Starting from the definition of the Coxeter-invariant functions as orbit averages, we obtain resummed expressions that are invariant term by term and more natural from the point of view of the Coxeter-symmetric moduli space. The hyperbolic case leads to a superposition of modified Bessel functions of the second kind, while the parabolic case results in objects involving Jacobi theta functions. The elliptic case is finite and not as rich, but can nonetheless be rewritten in terms of ordinary Bessel functions of the first kind to put it on the same footing as the other two. This   presentation of the elliptic case is natural from invariant theory, since it yields functions whose arguments are written in terms of the generators of the invariant polynomial algebra of the dihedral group.

The resummed expressions can be seen to localize in the opposite region of the moduli space than the raw orbit sums that define the Coxeter-invariant functions, as was mentioned in the introduction and is explained below. This fact inspires the harmonic analysis of the Gromov-Witten expansion performed in \cref{sec:harmonic-analysis}, which leads to a natural explanation for the special functions found through direct computation here.

Every pair of nodes in a Coxeter diagram corresponds to a dihedral subgroup generated by a pair of simple reflections. Hence, the types of functions that we encounter below in this section will also feature in the dihedral decomposition approach to the general case that we take in \cref{sec:general-Coxeter-prepotentials}.

\subsection{Mori/K\"ahler representations of the dihedral group}
\label{sec:dihedral-Mori-Kahler-representations}

The dihedral group $\mathrm{I}_{2}(m) \cong D_{m} \cong \mathbb{Z}_{m} \rtimes \mathbb{Z}_{2}$ is succinctly described by the Coxeter diagram
\begin{equation}
    \dynkin[Coxeter,gonality=m]{I}{}\,,
\end{equation}
corresponding to the Coxeter presentation
\begin{equation}
    \mathrm{I}_{2}(m) = \left\langle s_{1}, s_{2} \mid s_{1}^{2} = s_{2}^{2} = 1,\, (s_{1}s_{2})^{m}=1 \right\rangle\,.
\end{equation}
Defining $r \coloneqq s_{1} s_{2}$, we will also frequently make use of the non-Coxeter presentations
\begin{equation}
    \mathrm{I}_{2}(m) = \left\langle r, s_{i} \mid s_{i}^{2} = 1, s_{i} r s_{i} = r^{-1},\, r^{m} = 1 \right\rangle\,,\quad i = 1, 2\,.
\label{eq:infdih-non-Coxeter-presentation}
\end{equation}

Following the conventions introduced in \cref{sec:review-isomorphic-flops}, consider the Mori representation
\begin{equation}
    \mathcal{Q} \coloneqq \mathcal{Q}_{12} = \rho(r)\,,\qquad \mathcal{S}\coloneqq \mathcal{S}_{1} = \rho(s_{1})\,,
\end{equation}
with analogous $\hat{\mathcal{Q}}$ and $\hat{\mathcal{S}}$ for the K\"ahler representation. Recall from \eqref{eq:block-structure-M} and \eqref{eq:block-structure-QS} that, possibly after a reordering of the K\"ahler basis, $\mathcal{Q}$ and $\mathcal{S}$ have the block structure
\begin{equation}
	\mathcal{Q} =
	\begin{pNiceArray}{c|l}[cell-space-limits=5pt]
		Q  & U\\
		\hline
		0 & \mathds{1}_{h-2}
	\end{pNiceArray}
	\,,\qquad
	\mathcal{S} =
	\begin{pNiceArray}{c|l}[cell-space-limits=5pt]
		M_{1} & U_{1}\\
		\hline
		0 & \mathds{1}_{h-2}
	\end{pNiceArray}
	\,.
\end{equation}
Defining
\begin{equation}
    V_{0} \coloneqq \left\{ (x,0,\ldots,0) \mid x \in \mathbb{R}^{2} \right\} \subset \mathbb{R}^{h} \cong V\,,
\end{equation}
it is apparent that we have the short exact sequence of $\mathrm{I}_{2}(m)$-modules
\begin{equation}
    0 \longrightarrow \big(V_{0}, \left. \rho \right|_{V_{0}} \big) \longrightarrow \big(V, \rho\big) \longrightarrow \big(\mathbb{R}^{h-2}, \mathrm{Id}\big) \longrightarrow 0\,.
\label{eq:dihedral-rep-short-exact-sequence}
\end{equation}
Given the block structure of $\mathcal{Q}$ and $\mathcal{S}$, it will be useful to split the complexified K\"ahler moduli vector $T$ and curve class vector $d$ accordingly,
\begin{equation}
    T_{\parallel} \coloneqq
    \begin{pmatrix}
        T_{1}\\
        T_{2}
    \end{pmatrix}
    \,,\quad
    T_{\perp} \coloneqq
    \begin{pmatrix}
        T_{3}\\
        \vdots\\
        T_{h}
    \end{pmatrix}
    \,,\qquad
    d_{\parallel} \coloneqq
    \begin{pmatrix}
        d_{1}\\
        d_{2}
    \end{pmatrix}
    \,,\quad
    d_{\perp} \coloneqq
    \begin{pmatrix}
        d_{3}\\
        \vdots\\
        d_{h}
    \end{pmatrix}
    \,.
\end{equation}

The matrix $\mathcal{S}$ represents the reflection generator of $\mathrm{I}_{2}(m)$ and, therefore, all possible infinite-order behavior lies in $\mathcal{Q}$, which represents the rotation generator. Hence, we will subdivide the $\mathrm{I}_{2}(m)$ case according to the properties of $\mathcal{Q}$, which, in turn, mostly depend on $\left. \rho \right|_{V_{0}}(r) = Q$ due to \eqref{eq:dihedral-rep-short-exact-sequence}. Since $Q \in \mathrm{PSL(2,\mathbb{Z})}$ is a M\"obius transformation, it is traditionally classified according to its number of fixed points when regarded as an orientation-preserving isometry of the (closure of the) upper half-plane
\begin{equation}
    \overline{\mathbb{H}} \coloneqq \{ z \in \mathbb{C} \mid \mathrm{Im(z) \geq 0}\} = \mathbb{H} \cup \partial \mathbb{H}
\end{equation}
into the following types \cite{Katok1992FuchsianGroups}:
\begin{itemize}
    \item Elliptic $\Leftrightarrow$ $|\mathrm{Tr}(Q)| < 2 \Rightarrow m_{1} m_{2} < 4$, $m_{1}, m_{2} \geq 0$. It has a single fixed point in $\mathbb{H}$ and represents a rotation around this fixed point. The matrix $\mathcal{Q}$ has then $\mathrm{ord}(\mathcal{Q}) = m < \infty$ and it leads to a representation of $\mathrm{I}_{2}(m)$.

    \item Parabolic $\Leftrightarrow$ $|\mathrm{Tr}(Q)| = 2 \Rightarrow m_{1} m_{2} = 4$, $m_{1}, m_{2} \geq 0$. It has one fixed point on $\partial \mathbb{H}$ and preserves a family of horocycles based at this boundary point. The matrix $\mathcal{Q}$ has $\mathrm{ord}(\mathcal{Q}) = \infty$ and leads to a representation of $\infdih$. From the fact that $Q$ is unipotent with $Q = \mathds{1}_{2} + N$, $N^{2} = 0$, it is immediate that $\mathcal{Q} = \mathds{1}_{h} + \mathcal{N}$ with $\mathcal{N}^{3} = 0$, meaning that it is unipotent of index at most 3. One can check that
    \begin{equation}
        \mathcal{N}^{2} = 0 \iff NU = 0 \iff u'_{1} = -\frac{m_{1}}{2} u'_{2}\,,
    \label{eq:parabolic-unipotent-index-2}
    \end{equation}
    where $u'_{1}$ and $u'_{2}$ are defined as in \eqref{eq:uprimevec-definition}. Since $m_{1} m_{2} = 4$ and $(u'_{i})^{a'} \geq 0$ for all $i \in \{1,2\}$, $a' \in \{1, \dotsc, h-2\}$, the above relation can only be satisfied when $u'_{1} = u'_{2} = 0$. For iso-flops in K\"ahler favorable CICYs this never occurs. Hence, for the parabolic case we always encounter $\mathcal{Q}$ unipotent of index exactly 3.

    \item Hyperbolic $\Leftrightarrow$ $|\mathrm{Tr}(Q)| > 2 \Rightarrow m_{1} m_{2} > 4$, $m_{1}, m_{2} \geq 0$. It has two fixed points on $\partial \mathbb{H}$ and represents a hyperbolic translation along the geodesic that joins them, with translation length $\ell(Q) = 2 \log \lambda_{+}$, where $\lambda_{+}$ is the bigger eigenvalue of $Q$. The matrix $\mathcal{Q}$ has $\mathrm{ord}(\mathcal{Q}) = \infty$ and leads to a representation of $\infdih$. Its spectral radius is $\rho(\mathcal{Q}) > 1$, meaning that it is not unipotent.
\end{itemize}

Finding a change of basis $\rho'(g) = P^{-1} \rho(g) P$ with
\begin{equation}
    P =
    \begin{pNiceArray}{c|l}[cell-space-limits=5pt]
		\mathds{1}_{2}  & K\\
		\hline
		0 & \mathds{1}_{h-2}
	\end{pNiceArray}
    \,,\qquad
    K =
    \begin{pmatrix}
        k_{1}\\
        k_{2}
    \end{pmatrix}
    \,,
\end{equation}
that manifestly turns $\mathcal{Q}$ and $\mathcal{S}$ (alternatively $\mathcal{M}_{1}$ and $\mathcal{M}_{2}$) into block diagonal matrices is equivalent to solving the linear system of equations
\begin{equation}
    \begin{pNiceArray}{cc}[cell-space-limits=2.5pt]
        2 & -m_{1}\\
        -m_{2} & 2
    \end{pNiceArray}
    \begin{pNiceArray}{c}[cell-space-limits=2.5pt]
        k_{1}\\
        k_{2}
    \end{pNiceArray}
    =
    \begin{pNiceArray}{c}[cell-space-limits=2.5pt]
        u'_{1}\\
        u'_{2}
    \end{pNiceArray}
    \,,
\end{equation}
which is non-singular for the elliptic and hyperbolic cases. The singular parabolic case has solutions whenever \eqref{eq:parabolic-unipotent-index-2} is satisfied, since a block diagonal $\mathcal{Q}$ would automatically be unipotent of index 2. Hence, for K\"ahler favorable CICYs the extension \eqref{eq:dihedral-rep-short-exact-sequence} is always split in the elliptic and hyperbolic cases and non-split in the parabolic one.

\subsection[Infinite dihedral group \texorpdfstring{$\infdih$}{I2(Inf)}: hyperbolic representation]{Infinite dihedral group \texorpdfstring{$\boldsymbol{\infdih}$}{I2(Inf)}: hyperbolic representation}
\label{sec:dihedral-hyperbolic-representations}

Let us begin our analysis by treating the hyperbolic case in detail. Starting from the definition of the $\infdih$-invariant functions as orbit averages
\begin{equation}
    \psi_{ld}^{\infdih}(T) \coloneqq \sum_{w\, \in\, \infdih/\mathrm{Stab}_{\infdih}(d)} e^{2\pi i l\langle wd, T \rangle}\,,
\label{eq:infdih-raw-orbit-sum}
\end{equation}
the aim is to obtain resummed expressions that are natural from the point of view of the Coxeter-symmetric moduli space. To avoid overcounting contributions, it is important to split dihedral computations into three subcases according to the stabilizer of the curve class under consideration:
\begin{itemize}
    \item $d \notin \mathrm{Fix}(\mathcal{M}_{1}) \cup \mathrm{Fix}(\mathcal{M}_{2})$, for which $\mathrm{Stab}_{\infdih}(d) = 1$;
    \item $d \in ( \mathrm{Fix}(\mathcal{M}_{1}) \cup \mathrm{Fix}(\mathcal{M}_{2}) ) \setminus ( \mathrm{Fix}(\mathcal{M}_{1}) \cap \mathrm{Fix}(\mathcal{M}_{2}) )$, for which $\mathrm{Stab}_{\infdih}(d) = \mathbb{Z}_{2}$; and
    \item $d \in \mathrm{Fix}(\mathcal{M}_{1}) \cap \mathrm{Fix}(\mathcal{M}_{2})$, for which $\mathrm{Stab}_{\infdih}(d) = \infdih$.
\end{itemize}
The last subcase is never realized in hyperbolic models and can be ignored for now, see \cref{sec:dihedral-stabilizer,sec:fundamental-Mori-cone-infdih} for more details. Note that all classes $d \in \mathrm{int}(\mathcal{M}_{f})$ have trivial stabilizer; the converse statement is not true.

We start by focusing on the case $d \notin \mathrm{Fix}(\mathcal{M}_{1}) \cup \mathrm{Fix}(\mathcal{M}_{2})$ with trivial stabilizer. The raw orbit sum can then be indexed by the integers as
\begin{equation}
    \psi_{ld}^{\infdih}(T) = \sum_{k \in \mathbb{Z}} \left[ e^{2\pi i l \left\langle \mathcal{Q}^{k} d, T \right\rangle} + e^{2\pi i l \left\langle \mathcal{Q}^{k} \mathcal{S} d, T \right\rangle} \right]\,.
\label{eq:hyperbolic-infdih-raw-sum}
\end{equation}
To compute $\psi_{ld}^{\infdih}(T)$ it will be useful to obtain a closed form for $\mathcal{Q}^{k}$ and $\mathcal{Q}^{k}\mathcal{S}$, with $k \in \mathbb{Z}$. Through direct multiplication, one finds
\begin{equation}
    \mathcal{Q}^{k} =
    \begin{pNiceArray}{c|l}[cell-space-limits=5pt]
        Q^{k} & R_{k} U\\
        \hline
        0 & \mathds{1}_{h-2}
    \end{pNiceArray}
    \,,\qquad
    \mathcal{Q}^{k}\mathcal{S} =
    \begin{pNiceArray}{c|l}[cell-space-limits=5pt]
        Q^{k}S & Q^{k}U_{1} + R_{k} U\\
        \hline
        0 & \mathds{1}_{h-2}
    \end{pNiceArray}
    \,,
\end{equation}
where
\begin{equation}
    R_{k} \coloneqq
    \begin{cases}
        \phantom{-}\sum_{j=0}^{k-1} Q^{j}\,,\quad k\geq 0\,,\\[2.5pt]
        -\sum_{j=k}^{-1} Q^{j}\,,\quad k < 0\,.
    \end{cases}
\end{equation}
As noted in \cite{Lukas:2022crp}, this means that (at least some of) the entries of $\mathcal{Q}^{k}$ and $\mathcal{Q}^{k}\mathcal{S}$ grow as powers of $k$, leading to complicated, non-modular, $\infdih$-invariant functions that were left largely unexplored. We can reduce the problem to finding a closed form for $Q^{k}$ through the identity
\begin{equation}
    R_{k} = \big( \mathds{1}_{2} -Q^{k} \big) \big( \mathds{1}_{2} -Q \big)^{-1}\,,\quad k \in \mathbb{Z}\,.
\label{eq:hyperbolic-closed-Rk}
\end{equation}
This suggests the definition of a shift matrix
\begin{equation}
    B \coloneqq \big( \mathds{1}_{2} - Q \big)^{-1} U = \frac{1}{4-m_{1}m_{2}}
    \begin{pNiceArray}{c}[cell-space-limits=2.5pt]
        2u'_{1} + m_{1}u'_{2}\\
        2u'_{2} + m_{2}u'_{1}
    \end{pNiceArray}
    \,,
\end{equation}
which encodes the extension data $u'_{1}$ and $u'_{2}$ of the representation. We are then led to
\begin{equation}
    \mathcal{Q}^{k} =
    \begin{pNiceArray}{c|l}[cell-space-limits=5pt]
        Q^{k} & \big( \mathds{1}_{2} -Q^{k} \big) B\\
        \hline
        0 & \mathds{1}_{h-2}
    \end{pNiceArray}
    \,,\qquad
    \mathcal{Q}^{k}\mathcal{S} =
    \begin{pNiceArray}{c|l}[cell-space-limits=5pt]
        Q^{k}S & Q^{k}U_{1} + \big( \mathds{1}_{2} -Q^{k} \big) B\\
        \hline
        0 & \mathds{1}_{h-2}
    \end{pNiceArray}
    \,.
\label{eq:Qk-QkS-hyperbolic}
\end{equation}

Let us now focus on the two-dimensional hyperbolic block $\big( V_{0},\left. \rho \right|_{V_{0}} \big)$, where we have the geometric representation of $\infdih$ acting on the real span of its simple roots \cite{Humphreys1990ReflectionGroups}. There exists a symmetric bilinear form\footnote{The standard geometric representation of $\infdih$ is obtained for the parabolic case with $m_{1} = m_{2} = 2$, for which $\Sigma$ becomes degenerate. The other geometric representations we consider differ from that one by a non-standard choice of $\infdih$-invariant bilinear form.}
\begin{equation}
    \Sigma =
    \begin{pNiceArray}{cc}[cell-space-limits=2.5pt]
        m_{2} & -\frac{m_{1}m_{2}}{2}\\
        -\frac{m_{1}m_{2}}{2} & m_{1}
    \end{pNiceArray}
\label{eq:invariant-bilinear-form}
\end{equation}
which is invariant under the Coxeter group action. For the hyperbolic case, $\Sigma$ is a \mbox{non-degenerate} Lorentzian bilinear form with $\mathrm{sig}(\Sigma) = (1,1)$ and two null directions spanned by $v_{+}$ and $v_{-}$, which are defined by
\begin{equation}
    \mu_{\pm} \coloneqq \frac{1}{2} \left( m_{1} \pm \sqrt{\frac{m_{1}}{m_{2}}(m_{1} m_{2}-4)} \right)\,,\qquad v_{+} \coloneqq
    \begin{pmatrix}
        \mu_{+}\\
        1
    \end{pmatrix}
    \,,\qquad
    v_{-} \coloneqq
    \begin{pmatrix}
        \mu_{-}\\
        1
    \end{pmatrix}
    \,.
\label{eq:hyperbolic-Q-eigenvectors}
\end{equation}
The normalization of $v_{+}$ and $v_{-}$ has been chosen such that $\mathcal{S}v_{\pm} = v_{\mp}$. These are also the eigenvectors of $Q$, with the corresponding eigenvalues given by
\begin{equation}
    \lambda_{\pm} \coloneqq \frac{(m_{1}m_{2}-2) \pm \sqrt{m_{1}m_{2}(m_{1}m_{2}-4)}}{2}\,.
\label{eq:Q-eigenvalues}
\end{equation}
It will be useful to exploit the geometric interpretation of $Q$ as a hyperbolic translation by defining a new coordinate
\begin{equation}
    u \coloneqq \frac{\ell(Q)}{4} \Rightarrow \lambda_{\pm} = e^{\pm 2u}\,,\quad \cosh(2u) = \frac{m_{1}m_{2}-2}{2}\,,\quad \sinh(2u) = \frac{\sqrt{m_{1}m_{2}(m_{1}m_{2}-4)}}{2}\,.
\end{equation}
One then finds that $Q^{k}$ can be written in closed form as
\begin{equation}
    Q^{k} = \frac{1}{\sinh(u)}
    \begin{pNiceArray}{cc}[cell-space-limits=2.5pt]
        \sinh((2k+1)u) & -\frac{m_{1} \sinh(2ku)}{2 \cosh(u)}\\
        \frac{2 \cosh(u) \sinh(2ku)}{m_{1}} & -\sinh((2k-1)u)
    \end{pNiceArray}
    \,.
\end{equation}
We also define the extension shift vector 
\begin{equation}
    b(d_{\perp}) \coloneqq B d_{\perp} =
    \frac{1}{4-m_{1}m_{2}}
    \begin{pNiceArray}{c}[cell-space-limits=2.5pt]
        2 \deltaone + m_{1} \deltatwo\\
        2 \deltatwo + m_{2} \deltaone
    \end{pNiceArray}
    \,,
\label{eq:extension-shift-vector}
\end{equation}
the orbit sum seeds
\begin{subequations}
\begin{align}
    y(d) &\coloneqq d_{\parallel} - b(d_{\perp})\,,\\
    y'(d) &\coloneqq M_{1} d_{\parallel} + U_{1} d_{\perp} - b(d_{\perp}) = Sy(d)\,,
\end{align}
\label{eq:hyperbolic-orbit-sum-seeds}%
\end{subequations}
and the extension shift scalar
\begin{equation}
    \mathcal{C}(d_{\perp},T) \coloneqq \left\langle b(d_{\perp}), T_{\parallel} \right\rangle + \left\langle d_{\perp}, T_{\perp} \right\rangle\,.
\label{eq:hyperbolic-extension-shift-scalar}
\end{equation}
The orbit sum seeds act as untwisted coordinates for the action of $\infdih$ on the curve classes with respect to the extension short exact sequence \eqref{eq:dihedral-rep-short-exact-sequence}, exploiting the fact that it is a split extension for the hyperbolic case. The inner products featured in $\psi_{ld}^{\infdih}(T)$ then lead to
\begin{subequations}
\begin{align}
    \big\langle \mathcal{Q}^{k}d,T \big\rangle &= \big\langle Q^{k}y(d), T_{\parallel} \big\rangle + \mathcal{C}(d_{\perp},T)\,,\\
    \big\langle \mathcal{Q}^{k}\mathcal{S}d,T \big\rangle &= \big\langle Q^{k}y'(d), T_{\parallel} \big\rangle + \mathcal{C}(d_{\perp},T)\,.
\end{align}
\end{subequations}
Hence, we have to compute
\begin{equation}
    \psi_{ld}^{\infdih}(T) = e^{2\pi i l \mathcal{C}(d_{\perp},T)} \sum_{k \in \mathbb{Z}} \left[ e^{2\pi i l \left\langle Q^{k}y(d), T_{\parallel} \right\rangle} + e^{2\pi i l \left\langle Q^{k}y'(d), T_{\parallel} \right\rangle} \right]\,.
\end{equation}

Both terms above have the same structure but with different orbit sum seeds, namely
\begin{equation}
    S\left( l,z,T_{\parallel}\right) \coloneqq \sum_{k \in \mathbb{Z}} e^{2\pi i l \left\langle Q^{k} z, T_{\parallel} \right\rangle}\,.
\label{eq:hyperbolic-one-seed-orbit-sums}
\end{equation}
In terms of the auxiliary quantities
\begin{subequations}
\begin{align}
    a_{0}\left(z,T_{\parallel}\right) &\coloneqq \left\langle z, T_{\parallel} \right\rangle\,,\\
    a_{1}\left(z,T_{\parallel}\right) &\coloneqq \left\langle Q z, T_{\parallel} \right\rangle\,,
\end{align}
\end{subequations}
the exponents can be written as
\begin{equation}
    \big\langle Q^{k} z, T_{\parallel} \big\rangle = \mathcal{A}\left(z,T_{\parallel}\right) e^{2ku} + \mathcal{B}\left(z,T_{\parallel}\right) e^{-2ku}\,,
\label{eq:QkzTparallel}
\end{equation}
where
\begin{subequations}
\begin{align}
    \mathcal{A}\left(z,T_{\parallel}\right) &\coloneqq \frac{a_{1}\left(z,T_{\parallel}\right) - \lambda_{-} a_{0}\left(z,T_{\parallel}\right)}{\sqrt{m_{1}m_{2}(m_{1}m_{2}-4)}}\,,\\
    \mathcal{B}\left(z,T_{\parallel}\right) &\coloneqq \frac{-a_{1}\left(z,T_{\parallel}\right) + \lambda_{+} a_{0}\left(z,T_{\parallel}\right)}{\sqrt{m_{1}m_{2}(m_{1}m_{2}-4)}}\,.
\end{align}
\label{eq:AB-definitions}%
\end{subequations}
Note that $e^{2ku}$ appears in the exponent of $S$, making its $k$-dependence double-exponential. An important property of $\mathcal{A}\left(z,T_{\parallel}\right)$ and $\mathcal{B}\left(z,T_{\parallel}\right)$ is that
\begin{equation}
    \begin{split}
        &\mathrm{Im}\left(\mathcal{A}\left(z,T_{\parallel}\right)\right) > 0\,,\\
        &\mathrm{Im}\left(\mathcal{B}\left(z,T_{\parallel}\right)\right) > 0\,,
    \end{split}
    \qquad \textrm{for}\qquad z \in C^{+}\,,\quad T_{\parallel} \in \mathrm{int} \left( \hat{\pi}_{0} \left( \mathcal{K} \right) \right)\,,
\end{equation}
where $C^{+} \coloneqq \langle v_{+}, v_{-} \rangle_{\mathbb{R}_{>0}}$. This covers all values of interest, since $y(d),\, y'(d) \in C^{+}$ for $d \in \mathcal{M}_{f}$ with $\mathrm{Stab}_{\infdih}(d) = 1$. We derive this fact in \cref{sec:AB-properties}. Consider now the continuous function
\begin{equation}
    f \left( l,z,T_{\parallel}  \,\middle|\, x \right) \coloneqq e^{2\pi i l \left( \mathcal{A} \left( z,T_{\parallel} \right)x + \mathcal{B}\left( z,T_{\parallel} \right)x^{-1} \right)}\,,\quad x \in \mathbb{R}_{>0}\,.
\label{eq:hyperbolic-kernel}
\end{equation}
This allows us to regard $S\left(l,z,T_{\parallel}\right)$ as the (multiplicative) lattice evaluation of $f$, i.e.,
\begin{equation}
    S\left(l,z,T_{\parallel}\right) = \sum_{k \in \mathbb{Z}} f(q^{k})\,,\qquad q\coloneqq e^{2u}=\lambda_{+}\,.
\end{equation}

In the introduction to this section, we advanced that the parabolic case leads to the appearance of Jacobi theta functions in $\psi_{ld}^{\infdih}(T)$. This was previously noticed in \cite{Lukas:2022crp}, see also the comments in \cref{sec:dihedral-parabolic-representations}. It is worth establishing a brief comparison between the hyperbolic and parabolic cases at this point already.

For the parabolic case, the exponential weighting of worldsheet instantons paired with the unipotency of index 3 of $\mathcal{Q}$ leads to a Gaussian kernel. The parabolic $\infdih$-action results in an orbit sum corresponding to the periodization of this Gaussian kernel over an additive lattice, yielding Jacobi theta functions. In this context, the natural integral transform providing the spectral decomposition is the Fourier transform, which maps kernels in a Gaussian family to other members of the same family. This leads to modular properties for the lattice sum, namely the invariance of Jacobi theta functions under $S \in \mathrm{PSL}(2,\mathbb{Z})$ via Poisson summation. The appearance of modular objects is geometrically natural, given the relation between the parabolic case and genus-one fibered Calabi-Yau threefolds, see \cref{sec:parabolic-dihedral-example}.

In the hyperbolic case, the form of the kernel is $e^{ax+bx^{-1}}$, rather than Gaussian, and the hyperbolic $\infdih$-action results in a periodization over a multiplicative lattice $q^{\mathbb{Z}}$ instead. The natural integral transform in this multiplicative context is the Mellin transform
\begin{equation}
    \mathcal{M}[f](s) \coloneqq \int_{0}^{\infty} x^{s-1} f(x)\, dx
\end{equation}
or, equivalently, the Fourier transform in log-variables, since the multiplicative lattice results from the additive one by exponentiation. The kernel no longer maps to members of the same family under the integral transform, meaning that the modular properties found for the parabolic case are broken. The latter can be formally regarded as a degenerate limit of the hyperbolic case in which the spacing of the multiplicative lattice is rescaled until it becomes additive and the modular transformations are restored. We briefly comment on this in \cref{sec:parabolic-degeneration}.

The spectral decomposition of the hyperbolic case is obtained by Mellin-Poisson summation \cite{Butzer1998FiniteMellin,Bardaro2018QuadratureMellin}, given for some $\sigma > 0$ by
\begin{equation}
    \sum_{k \in \mathbb{Z}} f(e^{k/\sigma}) e^{kc/\sigma} = \sigma \sum_{m \in \mathbb{Z}} \mathcal{M}[f](c + 2\pi i \sigma m)\,.
\end{equation}
For our purposes, we can set $c=0$ and define $q := e^{1/\sigma}$, which leads to
\begin{equation}
    \sum_{k \in \mathbb{Z}} f(q^{k}) = \frac{1}{\log(q)} \sum_{m \in \mathbb{Z}} \mathcal{M}[f] \left( \frac{2\pi i m}{\log(q)} \right)\,.
\label{eq:Mellin-Poisson-summation}
\end{equation}
We can safely apply this to $f\left(z,T_{\parallel} \,\middle|\, x\right)$ due to its super-exponential, uniform absolute convergence ensured by $\mathrm{Im}\left(\mathcal{A}\left(z,T_{\parallel}\right)\right),\, \mathrm{Im}\left(\mathcal{B}\left(z,T_{\parallel}\right)\right) > 0$. The Mellin transform of $f$ follows from the identity \cite[Eq.\ 3.471.9]{Gradshteyn2014TableIntegrals}
\begin{equation}
    \int_{0}^{\infty} x^{\nu -1} e^{-\alpha x - \beta/x} dx = 2 \left( \frac{\beta}{\alpha} \right)^{\frac{\nu}{2}} K_{\nu}(2\sqrt{\alpha\beta})\,,\quad \mathrm{Re}(\alpha),\, \mathrm{Re}(\beta) > 0\,,
\label{eq:gradshteyn-identity}
\end{equation}
where $K_{\nu}$ is the modified Bessel function of the second kind. Substituting
\begin{equation}
    \begin{split}
        \alpha &= -2\pi i l \mathcal{A}\left(z,T_{\parallel}\right)\,,\\
        \beta &= -2\pi i l \mathcal{B}\left(z,T_{\parallel}\right)\,,
    \end{split}
    \qquad \mathrm{Im}\left(\mathcal{A}\left(z,T_{\parallel}\right)\right),\, \mathrm{Im}\left(\mathcal{B}\left(z,T_{\parallel}\right)\right) > 0
\end{equation}
and using \eqref{eq:Mellin-Poisson-summation} we find
\begin{equation}
\begin{split}
    S \left( l,z,T_{\parallel} \right) &= \frac{1}{u} \sum_{m \in \mathbb{Z}} K_{\frac{\pi i m}{u}}\left( \Omega \left( l,z,T_{\parallel} \right) \right) e^{im\theta\left(z,T_{\parallel}\right)}\\ &= \frac{1}{u} \left[ K_{0}\left(\Omega\left(l,z,T_{\parallel}\right)\right) + 2\sum_{m=1}^{\infty} K_{\frac{\pi i m}{u}}\left(\Omega\left(l,z,T_{\parallel}\right)\right) \cos\left(m\theta\left(z,T_{\parallel}\right)\right) \right]\,,
\end{split}
\label{eq:hyperbolic-S-sum-Bessel}%
\end{equation}
where we have defined
\begin{subequations}
\begin{align}
    \Omega \left( l,z,T_{\parallel}\right) &\coloneqq 4\pi l\sqrt{-\mathcal{A}\left(z,T_{\parallel}\right) \mathcal{B}\left(z,T_{\parallel}\right)}\,,\\
    \theta \left( z,T_{\parallel} \right) &\coloneqq \frac{\pi}{2u} \log \left( \frac{\mathcal{B}\left(z,T_{\parallel}\right)}{\mathcal{A}\left(z,T_{\parallel}\right)} \right)\,,
\end{align}
\label{eq:hyperbolic-Delta-theta-definition}%
\end{subequations}
taking the logarithm on the principal branch.

To obtain the final result, first note that
\begin{equation}
    \mathcal{A}\left(y(d),T_{\parallel}\right) \mathcal{B}\left(y(d),T_{\parallel}\right) = \mathcal{A}\left(y'(d),T_{\parallel}\right) \mathcal{B}\left(y'(d),T_{\parallel}\right)\,.
\end{equation}
This allows us to write, with $\Omega$ standing for $\Omega \left( l,z,T_{\parallel}\right)$ and similarly for $\mathcal{C}$, $\zeta$ and $\eta$,
\begin{equation}
    \psi_{ld}^{\infdih}(T) = e^{2\pi i l \mathcal{C}} \frac{2}{u} \left[ K_{0}(\Omega) + 2\sum_{m=1}^{\infty} K_{\frac{\pi i m}{u}}(\Omega) \cos(m\zeta) \cos(m\eta) \right]\,,
\label{eq:hyperbolic-infdih-Bessel-sum}
\end{equation}
where
\begin{equation}
    \zeta\left(T_{\parallel}\right) \coloneqq \frac{\theta\left(y(d),T_{\parallel}\right) + \theta\left(y'(d),T_{\parallel}\right)}{2}\,,\qquad \eta(d) \coloneqq \frac{\theta\left(y(d),T_{\parallel}\right) - \theta\left(y'(d),T_{\parallel}\right)}{2}\,.
\label{eq:phi-psi-definition}
\end{equation}
The fact that the dependence on $T_{\parallel}$ and $d$ splits between the two objects becomes clear from the simplified expressions
\begin{equation}
    \zeta\left( T_{\parallel} \right) = \frac{\pi}{2u} \log\left( \frac{\left\langle v_{-}, T_{\parallel} \right\rangle}{\left\langle v_{+}, T_{\parallel} \right\rangle} \right)\,,\qquad \eta(d) = \frac{\pi}{2u} \log\left( \frac{\beta_{y(d)}}{\alpha_{y(d)}} \right)\,,
\end{equation}
where $\alpha_{y(d)}$ and $\beta_{y(d)}$ are defined through the decomposition $y(d) = \alpha_{y(d)}v_{+} + \beta_{y(d)}v_{-}$. By construction, \eqref{eq:hyperbolic-infdih-Bessel-sum} is manifestly invariant under the hyperbolic $\infdih$-action. Indeed, from the above definitions, one can check that for $w \in \infdih$, the effect of acting with the Mori representation $\rho(w)$ is
\begin{subequations}
\begin{align}
    \mathcal{C}\left(d_{\perp},T_{}\right) &\longmapsto \mathcal{C}\left(d'_{\perp},T\right)\,,\\
    \Omega\left(l,y(d),T_{\parallel}\right) &\longmapsto \Omega\left(l,y(d'),T_{\parallel}\right)\,,\\
    \zeta\left(T_{\parallel}\right) &\longmapsto \zeta\left(T_{\parallel}\right)\,,\\
    \eta\left(d\right) &\longmapsto \eta\left(d'\right) + 2\pi n\,,\quad n \in \mathbb{Z}\,,
\end{align}
\end{subequations}
\textit{mutatis mutandis} for the K\"ahler representation.

For curve classes $d \in ( \mathrm{Fix}(\mathcal{M}_{1}) \cup \mathrm{Fix}(\mathcal{M}_{2}) ) \setminus ( \mathrm{Fix}(\mathcal{M}_{1}) \cap \mathrm{Fix}(\mathcal{M}_{2}) )$, i.e., for those sitting on a mirror hyperplane, \eqref{eq:hyperbolic-infdih-Bessel-sum} overcounts the contributions. The correct result is obtained by taking a single orbit sum seed
\begin{equation}
    \psi_{ld}^{\infdih}(T) = e^{2\pi i l \mathcal{C}} \frac{1}{u} \left[ K_{0}\left(\Omega\right) + 2\sum_{m=1}^{\infty} K_{\frac{\pi i m}{u}}\left(\Omega\right) \cos\left(m\theta\right) \right]\,.
\label{eq:hyperbolic-infdih-Bessel-sum-single-seed}
\end{equation}

Beyond the conceptual and structural interest of \eqref{eq:hyperbolic-infdih-Bessel-sum} and \eqref{eq:hyperbolic-infdih-Bessel-sum-single-seed}, we may also ask whether they offer some practical advantages with respect to the raw Gromov–Witten orbit sum \eqref{eq:infdih-raw-orbit-sum}. The two expressions turn out to be most useful in complementary regimes. In the large volume region, \eqref{eq:infdih-raw-orbit-sum} is dominated by a few leading instanton contributions. This sharply localized distribution translates on the spectral dual side to the need for many Bessel-mode contributions to achieve equivalent convergence. In the deep interior of the moduli space, however, the Gromov-Witten terms decay very slowly and the convergence of \eqref{eq:hyperbolic-infdih-raw-sum} is poor. This broad distribution localizes very efficiently on the spectral dual side around the first few Bessel-modes of \eqref{eq:hyperbolic-infdih-Bessel-sum} and \eqref{eq:hyperbolic-infdih-Bessel-sum-single-seed}, which converge notably fast. The phenomenon is familiar from the Fourier analysis of waves, which inspires the harmonic analysis of the Gromov-Witten expansion in \cref{sec:harmonic-analysis}.

\subsubsection{Three-parameter example}
\label{sec:hyperbolic-dihedral-example}

In the CICY Coxeter Database, we find 26 hyperbolic $\infdih$ models, out of which 6 have $h^{1,1}(X) = 2$ and 20 have $h^{1,1}(X) = 3$. To illustrate the preceding discussion, let us focus on the CICY 6771 family, which has $\left(h^{1,1},h^{2,1}\right) = (3,35)$ and is described by the configuration matrix
\begin{equation}
    X_{6771} \in
    \begingroup
    \delimiterfactor=1000
    \begin{bNiceArray}{c|cccc}
        \mathbb{P}^{3} & 2 & 0 & 1 & 1\\
        \mathbb{P}^{3} & 0 & 2 & 1 & 1\\
        \mathbb{P}^{1} & 1 & 1 & 0 & 0
    \end{bNiceArray}
    ^{3,35}_{-64\rlap{\,.}}
    \endgroup
\end{equation}
Since the embedding of $X_{6771}$ is K\"ahler-favorable, we can identify the iso-flops employing the methods reviewed in \cref{sec:review-isomorphic-flops}. Using the triple intersection numbers
\begin{equation}
\begin{split}
    &\kappa_{111} = \kappa_{222} = 2\,,\quad \kappa_{112} = \kappa_{122} = 6\,,\quad \kappa_{113} = \kappa_{223} = 4\,,\quad \kappa_{123} = 8\,,\\
    &\kappa_{133} = \kappa_{233} = \kappa_{333} = 0\,,
\end{split}
\end{equation}
and the fact that the first two rows are of Type~2, we find the reflection generators
\begin{equation}
    \mathcal{M}_{1} =
    \begin{pNiceArray}{cc|c}
        -1 & 4 & 1 \\
        0 & 1 & 0 \\
        \hline
        0 & 0 & 1
    \end{pNiceArray}
    \,,\qquad
    \mathcal{M}_{2} =
    \begin{pNiceArray}{cc|c}
        1 & 0 & 0 \\
        4 & -1 & 1 \\
        \hline
        0 & 0 & 1
    \end{pNiceArray}
    \,,
\end{equation}
which furnish a hyperbolic representation of $\infdih$. The third row is of Type~1b, but fails the iso-flop necessary condition \eqref{eq:uvec-definition}, meaning that it leads to a non-isomorphic flop wall. Given the above matrices, we get
\begin{equation}
    m_{1} = m_{2} = 4,\qquad u'_{1} = u'_{2} = (1)\,,\qquad \lambda_{\pm} = 7 \pm 4\sqrt{3}\,,\qquad u = \frac{\arccosh{(7)}}{2}\,.
\end{equation}
This leads to
\begingroup
\allowdisplaybreaks
\begin{subequations}
\begin{align}
    \mathcal{C}\left(d_{\perp},T\right) &= -\frac{1}{2} d_{3} \left(T_{1}+T_{2}-2 T_{3}\right)\,,\\
    \Omega\left(l,y(d),T_{\parallel}\right) &= \pi  l \sqrt{\frac{2}{3}\left(T_{1}^2+4 T_{1} T_{2}+T_{2}^2\right) \left(2 \left(d_{1}^2-4 d_{1} d_{2}+d_{2}^2\right)-2 d_{3} \left(d_{1}+d_{2}\right)-d_{3}^2\right)}\,,\\
    \zeta\left(T_{\parallel}\right) &= \frac{\pi  \log \left(\frac{\left(2-\sqrt{3}\right) T_{1}+T_{2}}{\left(2+\sqrt{3}\right) T_{1}+T_{2}}\right)}{\arccosh(7)}\,,\\
    \eta\left(d\right) &= \frac{\pi  \log \left(\frac{6 (2 d_{2}+d_{3})}{2 \sqrt{3} d_{1}+\left(6-4 \sqrt{3}\right) d_{2}-\left(\sqrt{3}-3\right) d_{3}}-1\right)}{\arccosh(7)}\,,
\end{align}
\end{subequations}
\endgroup
from which $\psi_{ld}^{\infdih}(T)$ can be calculated using \eqref{eq:hyperbolic-infdih-Bessel-sum} for $d \notin \mathrm{Fix}(\mathcal{M}_{1}) \cup \mathrm{Fix}(\mathcal{M}_{2})$ or \eqref{eq:hyperbolic-infdih-Bessel-sum-single-seed} for $d \in ( \mathrm{Fix}(\mathcal{M}_{1}) \cup \mathrm{Fix}(\mathcal{M}_{2}) ) \setminus ( \mathrm{Fix}(\mathcal{M}_{1}) \cap \mathrm{Fix}(\mathcal{M}_{2}) )$.

The fundamental Mori cone is
\begin{equation}
    \mathcal{M}_{f} = \left\{ d \in \mathbb{R}^{h}_{\geq 0} \mathrel{\Big|} 2d_{1} - 4d_{2} \leq d_{3}\,,\quad 2d_{2} - 4d_{1} \leq d_{3} \right\}\,,
\end{equation}
which can also be described as
\begin{equation}
    \mathcal{M}_{f} = \left\langle w_{1},w_{2},e_{1},p_{1},q_{1} \right\rangle_{\mathbb{R}_{\geq 0}}\,,
\end{equation}
with
\begin{equation}
    w_{1} =
    \begin{pmatrix}
        4\\
        2\\
        0
    \end{pmatrix}
    \,,\quad
    w_{2} =
    \begin{pmatrix}
        2\\
        4\\
        0
    \end{pmatrix}
    \,,\quad
    e_{1} =
    \begin{pmatrix}
        0\\
        0\\
        1
    \end{pmatrix}
    \,,\quad
    p_{1} =
    \begin{pmatrix}
        1\\
        0\\
        2
    \end{pmatrix}
    \,,\quad
    q_{1} =
    \begin{pmatrix}
        0\\
        1\\
        2
    \end{pmatrix}
    \,,
\end{equation}
see \cref{sec:fundamental-Mori-cone-infdih} for further discussion. The walls of the fundamental Mori cone contain non-trivial classes with $\mathrm{Stab}_{\mathrm{\infdih}}(d) = \mathbb{Z}_{2}$, e.g., $d = (2,1,0)$ with $n_{d} = 224$. It is also possible to find classes $d \in \partial\mathcal{M}_{f}$ with trivial stabilizer, e.g., $d = (3,3,0)$ with $n_{d} = 11840$.

Notice that all the above quantities are symmetric under the simultaneous exchange of $d_{1} \leftrightarrow d_{2}$ and $T_{1} \leftrightarrow T_{2}$. This stems from the permutation symmetry of the first two rows of the configuration matrix, which means the full symmetry group is $G = \infdih \rtimes \mathbb{Z}_{2}$.

The CICY 6771 model above has a K3 fibration $\pi_{\mathrm{K3}}: X \rightarrow \mathbb{P}^{1}$, where the K3 fiber is given by the first two rows in the configuration matrix and the base by the last one. Since the fiber is given by the iso-flop rows, the Coxeter action preserves the fibration and acts fiberwise by isometries of the Picard lattice of the K3 fiber. This occurs for all $h^{1,1}(X) = 3$ hyperbolic dihedral models. The $h^{1,1}(X) = 2$ hyperbolic models do not exhibit any fibration structure; although they can be connected to K3 fibered configurations via a one-step split of their configuration matrix \cite{Candelas:1987kf}, their infinite dihedral symmetry is not seen to descend from the split models in any obvious way.

\subsection[Infinite dihedral group \texorpdfstring{${\infdih}$}{I2(Inf)}: parabolic representation]{Infinite dihedral group \texorpdfstring{$\boldsymbol{\infdih}$}{I2(Inf)}: parabolic representation}
\label{sec:dihedral-parabolic-representations}

While the hyperbolic and elliptic cases are very similar, the parabolic case must be treated slightly differently due to the non-split nature of the extension \eqref{eq:dihedral-rep-short-exact-sequence}. The latter was analyzed in \cite{Lukas:2022crp} already; we refine and expand the analysis below.

When computing $\psi_{ld}^{\infdih}(T)$ for parabolic models, we have to once again distinguish between three types of curve classes:
\begin{itemize}
    \item $d \notin \mathrm{Fix}(\mathcal{M}_{1}) \cup \mathrm{Fix}(\mathcal{M}_{2})$, for which $\mathrm{Stab}_{\infdih}(d) = 1$;
    \item $d \in ( \mathrm{Fix}(\mathcal{M}_{1}) \cup \mathrm{Fix}(\mathcal{M}_{2}) ) \setminus ( \mathrm{Fix}(\mathcal{M}_{1}) \cap \mathrm{Fix}(\mathcal{M}_{2}) )$, for which $\mathrm{Stab}_{\infdih}(d) = \mathbb{Z}_{2}$; and
    \item $d \in \mathrm{Fix}(\mathcal{M}_{1}) \cap \mathrm{Fix}(\mathcal{M}_{2})$, for which $\mathrm{Stab}_{\infdih}(d) = \infdih$.
\end{itemize}
Unlike for the hyperbolic case, it is now possible to find non-trivial curve classes that are stabilized by the entire group, see \cref{sec:dihedral-stabilizer,sec:fundamental-Mori-cone-infdih} for more details.

Let us start by discussing the case $d \notin \mathrm{Fix}(\mathcal{M}_{1}) \cup \mathrm{Fix}(\mathcal{M}_{2})$ with trivial stabilizer. The raw orbit sum is the same as in \eqref{eq:hyperbolic-infdih-raw-sum}. At a computational level, the differences arise due to the unipotent nature of $\mathcal{Q}$ with index 3, related to the non-split nature of the parabolic representations that was mentioned above. The latter manifests as the non-invertibility of $\mathds{1}_{2} - Q$, which prevents us from writing coordinates untwisted by a $b(d_{\perp})$ extension shift vector and invalidates the expression \eqref{eq:hyperbolic-closed-Rk} for $R_{k}$. Using the nilpotent piece of $Q$, we can write it instead as
\begin{equation}
    R_{k} = k \mathds{1}_{2} + \binom{k}{2} N\,,\quad k \in \mathbb{Z}\,.
\end{equation}
This leads to
\begin{subequations}
\begin{align}
    \big( \mathcal{Q}^{k} d \big)_{\parallel} &= d_{\parallel} + k\alpha(d_{\parallel},d_{\perp}) + \binom{k}{2} \beta(d_{\perp})\,,\quad k \in \mathbb{Z}\,,\\
    \big( \mathcal{Q}^{k} \mathcal{S} d \big)_{\parallel} &= d_{\parallel}^{S} + k\alpha(d_{\parallel}^{S},d_{\perp}) + \binom{k}{2} \beta(d_{\perp})\,,
\end{align}
\label{eq:parabolic-Qkd-QkSd}%
\end{subequations}
where we have defined
\begin{equation}
    d_{\parallel}^{S} \coloneqq Sd_{\parallel} + U_{1}d_{\perp}\,,\qquad \alpha\left(z,d_{\perp}\right) \coloneqq N z + U d_{\perp}\,,\qquad \beta\left(d_{\perp}\right) \coloneqq N U d_{\perp}\,.
\end{equation}
We can then rewrite the $\infdih$-invariant function in terms of the lattice sum
\begin{equation}
    S \left( l,z,d_{\perp},T_{\parallel} \right) \coloneqq \sum_{k \in \mathbb{Z}} e^{\pi i k^{2} l \left\langle \beta\left(d_{\perp}\right), T_{\parallel} \right\rangle + 2\pi i k l \left\langle \alpha\left(z,d_{\perp}\right) - \frac{\beta\left(d_{\perp}\right)}{2}, T_{\parallel} \right\rangle}
\end{equation}
as
\begin{equation}
    \psi_{ld}^{\infdih}(T) = e^{2\pi il \left\langle d_{\perp}, T_{\perp} \right\rangle} \left[ e^{2\pi il \left\langle d_{\parallel}, T_{\parallel} \right\rangle} S\big( d_{\parallel},d_{\perp},T_{\parallel} \big) + e^{2\pi il \left\langle d_{\parallel}^{S}, T_{\parallel} \right\rangle} S\big( d_{\parallel}^{S},d_{\perp},T_{\parallel} \big) \right]\,.
\end{equation}

The lattice sum $S\left( l,z,d_{\perp},T_{\parallel} \right)$ can be identified with the Jacobi theta function
\begin{equation}
    \vartheta\left(z_{\vartheta}, \tau\right) \coloneqq \sum_{k \in \mathbb{Z}} e^{\pi i k^{2} \tau + 2\pi i k z_{\vartheta}}\,,\quad z_{\vartheta} \in \mathbb{C}\,,\quad \mathrm{Im}(\tau) > 0
\label{eq:parabolic-z-tau-definition}
\end{equation}
with
\begin{equation}
    z_{\vartheta}\left( l,z,d_{\perp}, T_{\parallel} \right) = l \left\langle \alpha(z,d_{\perp}) - \frac{\beta(d_{\perp})}{2}, T_{\parallel} \right\rangle\,,\qquad \tau\left( l,d_{\perp}, T_{\parallel} \right) = l \left\langle \beta(d_{\perp}), T_{\parallel} \right\rangle
\label{eq:parabolic-ztheta-tau-definition}
\end{equation}
as long as the convergence requirement is met. To see that this is indeed the case for $T \in \mathrm{int} \left( \mathcal{K} \right)$, let us analyze the behavior of $\left\langle \beta(d_{\perp}), \mathrm{Im}\left( T_{\parallel} \right) \right\rangle$. We are performing the orbit sum for $d \in \mathcal{M}_{\mathrm{restr}}$, which implies that
\begin{equation}
    \big\langle \mathcal{Q}^{k} d, \mathrm{Im}(T) \big\rangle > 0\,,\qquad \big\langle \mathcal{Q}^{k}\mathcal{S} d, \mathrm{Im}(T) \big\rangle > 0\,,\qquad \forall k \in \mathbb{Z}\,.
\label{eq:parabolic-Qkd-QkSd-T-inequalities}
\end{equation}
Since $\mathcal{Q}^{k}d \in \mathcal{M}$ for all $k \in \mathbb{Z}$ and $\mathcal{M}$ is a closed cone, $\lim_{|k| \rightarrow \infty} \frac{2}{k^{2}} \left( \mathcal{Q}^{k} d \right) = (\beta(d_{\perp}),0)$ lies in $\mathcal{M}$. As a consequence of this, we immediately obtain $\left\langle \beta(d_{\perp}), \mathrm{Im}\left( T_{\parallel} \right) \right\rangle \geq 0$. Assume now that $\left\langle \beta(d_{\perp}), \mathrm{Im}\left( T_{\parallel} \right) \right\rangle = 0$. Given that $T \in \mathrm{int} \left( \mathcal{K} \right)$, this can only occur for $\beta(d_{\perp}) = 0$. When $\beta(d_{\perp})$ vanishes, using that $\ker(N) = \mathrm{im}(N) = \big\langle (m_{1},2)^{T} \big\rangle_{\mathbb{R}} \eqqcolon \delta$ (see \cref{sec:dihedral-stabilizer}) and that $\big\langle v, \mathrm{Im}\big(T_{\parallel}\big) \big\rangle \neq 0$ for any $0 \neq v \in \delta$ implies that the linear terms in \eqref{eq:parabolic-Qkd-QkSd} become infinitely negative for at least one $k \rightarrow \pm \infty$ direction, leading to a contradiction with \eqref{eq:parabolic-Qkd-QkSd-T-inequalities}, unless $\alpha\left( d_{\parallel}, d_{\perp} \right) = \alpha\left( d_{\parallel}^{S}, d_{\perp} \right) = 0$. But if both the quadratic and linear terms in \eqref{eq:parabolic-Qkd-QkSd} vanish, we conclude that $\mathrm{Stab}_{\infdih}(d) = \infdih$, contradicting our initial trivial stabilizer assumption. Hence, for the case with trivial stabilizer we can indeed rewrite the $\infdih$-invariant functions in terms of Jacobi theta functions as
\begin{equation}
    \psi_{ld}^{\infdih}(T) = e^{2\pi il \left\langle d, T \right\rangle} \left[ \vartheta\left( z_{\theta}, \tau \right) + e^{2\pi i\xi} \vartheta\left( z_{\theta}^{S}, \tau \right) \right]\,,
\label{eq:parabolic-Jacobi-theta-expression}
\end{equation}
where we have again suppressed the explicit dependencies of the appearing quantities on the curve class and K\"ahler data, abused notation by using $z_{\vartheta} = z_{\theta}\left( l,d_{\parallel},d_{\perp}, T_{\parallel} \right)$ and $z_{\vartheta}^{S} = z_{\theta}\left( l,d_{\parallel}^{S},d_{\perp}, T_{\parallel} \right)$, and defined
\begin{equation}
    \xi\left( l, d, T_{1} \right) \coloneqq l \left( \etad + \deltaone \right) T_{1}\,.
\end{equation}

The orbit sum $S \left( l,z,d_{\perp},T_{\parallel} \right)$ corresponds to the periodization of a Gaussian kernel over an additive lattice. The natural integral transform in this context is the Fourier transform and Poisson summation then leads to the well-known $S \in \mathrm{PSL}(2,\mathbb{Z})$ modularity property of the Jacobi theta function
\begin{equation}
    \vartheta(z_{\vartheta},\tau) = \frac{1}{\sqrt{-i\tau}} e^{-\frac{\pi i z_{\vartheta}^{2}}{\tau}} \vartheta\left( \frac{{z_{\vartheta}}}{\tau}, -\frac{1}{\tau} \right) = \frac{1}{\sqrt{-i \tau}} \sum_{k \in \mathbb{Z}} e^{-\frac{\pi i}{\tau}(k + z_{\vartheta})^{2}}\,.
\end{equation}
The expanded form above quickly localizes on a few terms in the small $\tau\left( d_{\perp}, T_{\parallel} \right)$ region of the moduli space, while the original raw sum does so in the opposite regime, cf.\ the hyperbolic case discussion in \cref{sec:dihedral-hyperbolic-representations}.

For curve classes \mbox{$d \in ( \mathrm{Fix}(\mathcal{M}_{1}) \cup \mathrm{Fix}(\mathcal{M}_{2}) ) \setminus ( \mathrm{Fix}(\mathcal{M}_{1}) \cap \mathrm{Fix}(\mathcal{M}_{2}) )$}, i.e., sitting on top of a single mirror hyperplane, the expression \eqref{eq:parabolic-Jacobi-theta-expression} overcounts the contributions to the orbit sum. Note that the exponent in $\xi\left( l, d, T_{1} \right)$ contains the mirror hyperplane equation associated to the simple reflection $\mathcal{S} = \mathcal{M}_{1}$, see \eqref{eq:parabolic-fundamental-Mori-cone}. Taking into account that $\mathrm{Stab}_{\infdih}(d) = \mathbb{Z}_{2}$ leads to
\begin{equation}
    \psi_{ld}^{\infdih}(T) = e^{2\pi il \left\langle d, T \right\rangle} \vartheta\left( z_{\theta}, \tau \right)\,.
\label{eq:parabolic-Jacobi-theta-expression-Z2-stabilizer}
\end{equation}
When $d \in \mathrm{Fix}(\mathcal{M}_{1}) \cap \mathrm{Fix}(\mathcal{M}_{2})$ the entire group stabilizes the curve class, making the $\infdih$-invariant function simply
\begin{equation}
    \psi_{ld}^{\infdih}(T) = e^{2\pi i l \left\langle d, T \right\rangle}\,.
\end{equation}
This occurs for curve classes aligned with the cusp direction in the $d_{\parallel}$-plane, as discussed in \cref{sec:dihedral-stabilizer}.

The resulting expressions are manifestly invariant under the $\infdih$-action, even if their individual pieces are not. Indeed, one can check that, under the $Q$-action
\begin{equation}
    Q: (z_{\vartheta},\tau) \longmapsto (z_{\vartheta} + \tau,\tau)\,,
\end{equation}
the Jacobi theta function is only quasiperiodic, transforming like
\begin{equation}
    Q: \vartheta\left( z_{\vartheta}, \tau \right) \longmapsto e^{-\pi i \tau-2\pi i z_{\vartheta}} \vartheta\left( z_{\vartheta}, \tau \right)\,.
\end{equation}
This is the well-known fact that Jacobi theta functions are holomorphic sections of a line bundle over the universal elliptic curve, see, for example, \cite{BirkenhakeLange2004ComplexAbelianVarieties}. The $e^{2\pi il \left\langle d, T \right\rangle}$ factor transforms in the opposite way,
\begin{equation}
    Q: e^{2\pi i l \left\langle Qz, T_{\parallel} \right\rangle} \longmapsto e^{2\pi i l \left\langle z, T_{\parallel} \right\rangle} e^{2\pi i l \left\langle \alpha\left( z,d_{\perp} \right), T_{\parallel} \right\rangle} = e^{2\pi i l \left\langle z, T_{\parallel} \right\rangle} e^{2\pi i z_{\vartheta}} e^{\pi i \tau}\,,
\end{equation}
making $e^{2\pi i l \left\langle z, T_{\parallel} \right\rangle} S \left( l,z,d_{\perp},T_{\parallel} \right)$ invariant. The $S$-action, on the other hand, exchanges the orbit sums corresponding to each orbit sum seed.

\subsubsection{Three-parameter example}
\label{sec:parabolic-dihedral-example}

The CICY Coxeter Database contains 163 parabolic $\infdih$ models: 17 with $h^{1,1}(X) = 3$, 91 with $h^{1,1}(X) = 4$ and 55 with $h^{1,1}(X) = 5$. As a concrete example, let us explore the CICY 6971 family, with $(h^{1,1},h^{2,1}) = (3,36)$. It is described by the configuration matrix
\begin{equation}
    X_{6971} \in
    \begingroup
    \delimiterfactor=1000
    \begin{bNiceArray}{c|ccc}
        \mathbb{P}^{2} & 2 & 1 & 0\\
        \mathbb{P}^{2} & 0 & 1 & 2\\
        \mathbb{P}^{2} & 1 & 1 & 1
    \end{bNiceArray}
    ^{3,36}_{-66\rlap{\,,}}
    \endgroup
\end{equation}
leading to the triple intersection numbers
\begin{equation}
\begin{split}
    &\kappa_{112} = \kappa_{122} = 3\,,\quad
    \kappa_{113} = \kappa_{223} = 2\,,\quad
    \kappa_{133} = \kappa_{233} = 4\,,\quad
    \kappa_{123} = 8\,,\\
    &\kappa_{111} = \kappa_{222} = \kappa_{333} = 0\,.
\end{split}
\end{equation}
The first two rows are of Type 2 and yield the reflection generators\footnote{The Mori/K\"ahler representations of the parabolic infinite dihedral groups that we encounter are non-semisimple. This property is expected to not hold once a representation of the full duality group is considered \cite{Delgado:2024skw}, of which the Coxeter action is only a subgroup.}
\begin{equation}
    \mathcal{M}_{1} =
    \begin{pNiceArray}{cc|c}
        -1 & 2 & 3 \\
        0 & 1 & 0 \\
        \hline
        0 & 0 & 1
    \end{pNiceArray}
    \,,\qquad
    \mathcal{M}_{2} =
    \begin{pNiceArray}{cc|c}
        1 & 0 & 0 \\
        2 & -1 & 3 \\
        \hline
        0 & 0 & 1
    \end{pNiceArray}
    \,,
\end{equation}
furnishing a parabolic representation of $\infdih$. While the third row is of Type 1b, it leads to a non-isomorphic flop wall due to the failure of the necessary condition \eqref{eq:uvec-definition}. The model corresponds to the data
\begin{equation}
    m_{1} = m_{2} = 2\,,\qquad u'_{1} = u'_{2} = (3)\,,\qquad \delta_{1} = \delta_{2} =
    \begin{pmatrix}
        2\\
        2
    \end{pmatrix}
    \,,\qquad
    n_{1} =
    \begin{pmatrix}
        -2\\
        2
    \end{pmatrix}
    \,.
\end{equation}
For $d \notin \mathrm{Fix}(\mathcal{M}_{1}) \cup \mathrm{Fix}(\mathcal{M}_{2})$ and $d \in ( \mathrm{Fix}(\mathcal{M}_{1}) \cup \mathrm{Fix}(\mathcal{M}_{2}) ) \setminus ( \mathrm{Fix}(\mathcal{M}_{1}) \cap \mathrm{Fix}(\mathcal{M}_{2}) )$ the $\infdih$-invariant functions $\psi_{ld}^{\infdih}$ can be computed using \eqref{eq:parabolic-Jacobi-theta-expression} and \eqref{eq:parabolic-Jacobi-theta-expression-Z2-stabilizer}, respectively, with
\begingroup
\allowdisplaybreaks
\begin{subequations}
\begin{align}
    \tau\left( l,d_{\perp},T_{\parallel} \right) &= 12 l d_{3} (T_{1}+T_{2})\,,\\
    z_{\vartheta} = z_{\vartheta}\left( l,d_{\parallel},d_{\perp},T_{\parallel} \right) &= 3 l d_{3} (T_{1} - T_{2}) + 2 l d_{1} (T_{1} + T_{2}) - 2 l d_{2} (T_{1} + T_{2})\,,\\
    z_{\vartheta}^{S} = z_{\vartheta}\big( l,d_{\parallel}^{S},d_{\perp},T_{\parallel} \big) &= -2 l d_{1} (T_{1} + T_{2}) + 2 l d_{2} (T_{1} + T_{2}) + 3 l d_{3} (3 T_{1} + T_{2})\,,\\
    \xi\left( l,d,T_{1} \right) &= l (-2 d_{1} + 2 d_{2} + 3 d_{3}) T_{1}\,.
\end{align}
\end{subequations}
\endgroup

The fundamental Mori cone is
\begin{equation}
    \mathcal{M}_{f} = \left\{ d \in \mathbb{R}^{h}_{\geq 0} \mathrel{\Big|} -2 d_{1} + 2 d_{2} + 3 d_{3} \geq 0\,,\quad -2 d_{1} + 2 d_{2} - 3d_{3} \leq 0 \right\}\,,
\end{equation}
written in terms of generators as
\begin{equation}
    \mathcal{M}_{f} = \left\langle w_{1},e_{1},p_{1},q_{1} \right\rangle_{\mathbb{R}_{\geq 0}}\,,
\end{equation}
where
\begin{equation}
    w_{1} =
    \begin{pmatrix}
        2\\
        2\\
        0
    \end{pmatrix}
    \,,\quad
    e_{1} =
    \begin{pmatrix}
        0\\
        0\\
        1
    \end{pmatrix}
    \,,\quad
    p_{1} =
    \begin{pmatrix}
        3\\
        0\\
        2
    \end{pmatrix}
    \,,\quad
    q_{1} =
    \begin{pmatrix}
        0\\
        3\\
        2
    \end{pmatrix}
    \,,
\end{equation}
see \cref{sec:fundamental-Mori-cone-infdih} for more details. The walls of the fundamental Mori cone do contain \mbox{non-trivial} classes with $\mathrm{Stab}_{\infdih}(d) = \mathbb{Z}_{2}$, e.g., $d = (3,0,2)$ with $n_{d} = 66$. Classes $d \in \partial\mathcal{M}_{f}$ with trivial stabilizer also exist, e.g., $d = (0,1,1)$ with $n_{d} = 144$. The intersection of the mirror hyperplanes, which corresponds to the ray along the cusp direction, is also populated by non-trivial classes, like $d = (1,1,0)$ with $n_{d} = 36$, which are stabilized by the entire group. As for the hyperbolic example, due to the permutation symmetry among the first two rows, the full symmetry group is $G = \infdih \rtimes \mathbb{Z}_{2}$.

The CICY 6971 has three inequivalent genus-one fibrations $\pi_{i}: X_{6971} \rightarrow \mathbb{P}^{2}$, where the base is given by the $i$-th row in the configuration matrix and the genus-one fiber by the remaining two rows. One can check that the fibration $\pi_{3}: X_{6971} \rightarrow \mathbb{P}^{2}$ is compatible with the Coxeter action, which acts fiberwise on the genus-one curve, making the appearance of Jacobi theta functions natural. Indeed, all parabolic dihedral models in the database exhibit at least one Coxeter-compatible genus-one fibration.\footnote{The converse is not true and examples of genus-one fibered Calabi-Yau threefolds without an iso-flop Coxeter symmetry exist, e.g., CICY 7884. Note that the existence of a genus-one fibration still leads to a repeating pattern for the GV invariants in the vertical/base-degree zero sector \cite{Toda2012StabilityConditionsCurveCounting,Bridgeland:2024ecw,Cota:2019cjx} and to a Jacobi-form structure for the coefficients of the fixed-base-degree expansion of the topological string partition function \cite{Huang:2015sta,Cota:2019cjx,Schimannek:2021pau}.} The Coxeter-compatible fibration structures can be richer than in the above example, including simultaneously preserved genus-one and K3 fibrations. One such model is the CICY 4151 family described by the configuration matrix
\begin{equation}
    X_{4151} \in
    \begingroup
    \delimiterfactor=1000
    \begin{bNiceArray}{c|ccccc}
        \mathbb{P}^{1} & 1 & 1 & 0 & 0 & 0\\
        \mathbb{P}^{1} & 1 & 0 & 1 & 0 & 0\\
        \mathbb{P}^{2} & 0 & 1 & 1 & 1 & 0\\
        \mathbb{P}^{2} & 1 & 0 & 0 & 0 & 2\\
        \mathbb{P}^{2} & 0 & 0 & 0 & 2 & 1
    \end{bNiceArray}
    ^{5,25}_{-40\rlap{\,,}}
    \endgroup
\end{equation}
where the last two rows correspond to a pair of iso-flop walls leading to a parabolic dihedral symmetry. The projections to each of the first two rows of the configuration matrix yield two inequivalent Coxeter-compatible K3 fibrations. Each of these has a nested genus-one fibration whose fiber is spanned by the two iso-flop rows.

\subsection[Finite dihedral group \texorpdfstring{$\mathrm{I}_{2}(m)$}{I2(m)}: elliptic representation]{Finite dihedral group \texorpdfstring{$\boldsymbol{\mathrm{I}_{2}(m)}$}{I2(m)}: elliptic representation}
\label{sec:dihedral-elliptic-representations}

The elliptic case is the simplest, since it deals with finite groups and, hence, orbit sums that only have a finite number of terms. As discussed in \cite{Lukas:2022crp}, the restrictions $m_{1}, m_{2} \in \mathbb{Z}_{\geq 0}$ and $m_{1}m_{2} < 4$ lead to the following reduced set of possibilities:\footnote{Note that in our exhaustive search we encounter models with a non-direct-product structure and Coxeter group $W = \mathrm{I}_{2}(2)$, cf.\ \cite{Lukas:2022crp}. A concrete example is CICY 3901.}
\begin{itemize}
    \item $\mathrm{I_{2}(2)} \cong \mathrm{A}_{1} \times \mathrm{A}_{1} \cong \mathbb{Z}_{2} \times \mathbb{Z}_{2}$,
    \item $\mathrm{I_{2}(3)} \cong \mathrm{A}_{2} \cong \mathbb{Z}_{3} \rtimes \mathbb{Z}_{2}$,
    \item $\mathrm{I_{2}(4)} \cong \mathrm{B}_{2} \cong \mathbb{Z}_{4} \rtimes \mathbb{Z}_{2}$, and
    \item $\mathrm{I_{2}(6)} \cong \mathrm{G}_{2} \cong \mathbb{Z}_{6} \rtimes \mathbb{Z}_{2}$.
\end{itemize}
These correspond to the finite crystallographic dihedral groups, the only possibility by construction, see \cref{sec:iso-flop-representations}.

Once more, the computation of $\psi_{ld}^{\mathrm{I}_{2}(m)}(T)$ requires us to distinguish between three types of curve classes:
\begin{itemize}
    \item $d \notin \mathrm{Fix}(\mathcal{M}_{1}) \cup \mathrm{Fix}(\mathcal{M}_{2})$, for which $\mathrm{Stab}_{\mathrm{I}_{2}(m)}(d) = 1$;
    \item $d \in ( \mathrm{Fix}(\mathcal{M}_{1}) \cup \mathrm{Fix}(\mathcal{M}_{2}) ) \setminus ( \mathrm{Fix}(\mathcal{M}_{1}) \cap \mathrm{Fix}(\mathcal{M}_{2}) )$, for which $\mathrm{Stab}_{\mathrm{I}_{2}(m)}(d) = \mathbb{Z}_{2}$; and
    \item $d \in \mathrm{Fix}(\mathcal{M}_{1}) \cap \mathrm{Fix}(\mathcal{M}_{2})$, for which $\mathrm{Stab}_{\mathrm{I}_{2}(m)}(d) = \mathrm{I}_{2}(m)$.
\end{itemize}
Unlike for the hyperbolic case and similarly to the parabolic one, all three types are realized by non-trivial curve classes, see the details collected in \cref{sec:dihedral-stabilizer,sec:fundamental-Mori-cone-infdih}. Again, all classes $d \in \mathrm{int}(\mathcal{M}_{f})$ have trivial stabilizer while the converse statement does not hold.

Assume first that $d \notin \mathrm{Fix}(\mathcal{M}_{1}) \cup \mathrm{Fix}(\mathcal{M}_{2})$ with trivial stabilizer. Due to the finite nature of the group, the raw orbit sum contains a finite number of terms and is given by
\begin{equation}
    \psi_{ld}^{\mathrm{I}_{2}(m)}(T) = \sum_{k=0}^{m-1} \left[ e^{2\pi i l \left\langle \mathcal{Q}^{k} d, T \right\rangle} + e^{2\pi i l \left\langle \mathcal{Q}^{k} \mathcal{S} d, T \right\rangle} \right]\,.
\label{eq:elliptic-dih-raw-sum}
\end{equation}
In concrete examples, this yields a simple sum of exponentials. To enable easier comparisons with the previous cases and display a spectral dual expression, we nonetheless treat this sum as we did for the infinite order cases. Since the treatment is analogous to the hyperbolic case after exchanging hyperbolic geometry objects for their elliptic analogues, we keep the discussion brief.

First, since the elliptic representation is always split, we can express $\mathcal{Q}^{k}$ and $\mathcal{Q}^{k}\mathcal{S}$ as in \eqref{eq:Qk-QkS-hyperbolic} and untwist the coordinates by the extension shift vector $b(d_{\perp})$ defined in \eqref{eq:extension-shift-vector}. Using the orbit sum seeds $y(d)$ and $y'(d)$ from \eqref{eq:hyperbolic-orbit-sum-seeds} and the extension shift scalar $\mathcal{C} \left( d_{\perp}, T \right)$ from \eqref{eq:hyperbolic-extension-shift-scalar}, we can separate the extension data and write
\begin{equation}
    \psi_{d,l}^{\mathrm{I}_{2}(m)}(T) = e^{2\pi i l \mathcal{C}(d_{\perp},T)} \sum_{k = 0}^{m-1} \left[ e^{2\pi i l \left\langle Q^{k}y(d), T_{\parallel} \right\rangle} + e^{2\pi i l \left\langle Q^{k}y'(d), T_{\parallel} \right\rangle} \right]\,.
\label{eq:elliptic-dih-raw-sum-untwisted}
\end{equation}
Both terms in \eqref{eq:elliptic-dih-raw-sum-untwisted} have the structure
\begin{equation}
    S_{m} \left(l, z, T_{\parallel} \right) \coloneqq \sum_{k = 0}^{m-1} e^{2\pi i l \left\langle Q^{k} z, T_{\parallel} \right\rangle}\,,
\end{equation}
but with different orbit sum seeds. The departure from the hyperbolic case stems from the elliptic block $\big( V_{0}, \left. \rho \right|_{V_{0}} \big)$, where we now have the geometric representation of $\mathrm{I}_{2}(m)$ acting on the real span of its simple roots \cite{Humphreys1990ReflectionGroups}. The invariant bilinear form $\Sigma$ defined in \eqref{eq:invariant-bilinear-form} is now positive-definite (focusing on the cases $m \in \{3,4,6\}$, since it vanishes for $m=2$), with $M_{1}$ and $M_{2}$ acting as Euclidean reflections. The natural way to rewrite the eigenvalues \eqref{eq:Q-eigenvalues} is in terms of a rotation angle as
\begin{equation}
    v \coloneqq \frac{\pi}{m} \Rightarrow \lambda_{\pm} = e^{\pm 2v i}\,,\quad \cos(2v) = \frac{m_{1}m_{2}-2}{2}\,,\quad \sin(2v) = \frac{\sqrt{m_{1}m_{2}(m_{1}m_{2} - 4)}}{2i}\,.
\end{equation}
The exponent of the orbit sums can be rewritten in the form
\begin{equation}
    \big\langle Q^{k} z, T_{\parallel} \big\rangle = \mathcal{A}\left(z,T_{\parallel}\right) e^{2kiv} + \mathcal{B}\left(z,T_{\parallel}\right) e^{-2kiv}\,,
\label{eq:QkzTparallel-elliptic}
\end{equation}
with $\mathcal{A}\left(z,T_{\parallel}\right)$ and $\mathcal{B}\left(z,T_{\parallel}\right)$ as defined in \eqref{eq:AB-definitions}. Defining\footnote{Unlike in the hyperbolic case, the choice of a logarithm branch results in harmless shifts inside cosines, meaning that we do not need to track it.}
\begin{subequations}
\begin{align}
    \Omega' \left( l, z, T_{\parallel} \right) &\coloneqq 4\pi l \sqrt{\mathcal{A}\left(z,T_{\parallel}\right) \mathcal{B}\left(z,T_{\parallel}\right)}\,,\\
    \delta \left( z, T_{\parallel} \right) &\coloneqq \frac{1}{2i} \log \left( \frac{\mathcal{A}\left(z,T_{\parallel}\right)}{\mathcal{B}\left(z,T_{\parallel}\right)} \right) \mod \pi\,,
\end{align}
\end{subequations}
we can write
\begin{equation}
    2\pi i l \big\langle Q^{k} z, T_{\parallel} \big\rangle = i \Omega' \left( l, z, T_{\parallel} \right) \cos\left(2kv + \delta\left( z, T_{\parallel} \right)\right)\,.
\end{equation}
Using the Jacobi-Anger expansion
\begin{equation}
    e^{i \Delta \cos(\alpha)} = \sum_{m \in \mathbb{Z}} i^{m} J_{m}(\Delta) e^{im\alpha}\,,
\end{equation}
where $J_{m}$ is the ordinary Bessel function of the first kind, and the fact that
\begin{equation}
    \sum_{k=0}^{m-1} e^{2\pi i \frac{k r}{m}} =
    \begin{cases}
        m\,, &m \mid r\,,\\
        0\,, &m \nmid r\,,
    \end{cases}
\end{equation}
we arrive at the expression
\begin{equation}
\begin{split}
    S_{m} \left( l,z,T_{\parallel} \right) &= m \sum_{r \in \mathbb{Z}} i^{mr} J_{mr} \left( \Omega'\left(l,z,T_{\parallel}\right) \right) e^{imr\delta\left( z,T_{\parallel} \right)}\\ &= m \left[ J_{0}\left( \Omega'\left(l,z,T_{\parallel}\right) \right) + 2\sum_{r=1}^{\infty} i^{mr} J_{mr}\left( \Omega'\left(l,z,T_{\parallel}\right) \right) \cos\left( mr\delta \left( z,T_{\parallel} \right) \right) \right]\,,
\end{split}
\label{eq:elliptic-S-sum-Bessel}%
\end{equation}
which is analogous to the hyperbolic expression \eqref{eq:hyperbolic-S-sum-Bessel}.

Employing now the fact that
\begin{equation}
    \Omega'\left(l,y(d),T_{\parallel}\right) = \Omega'\left(l,y'(d),T_{\parallel}\right)
\end{equation}
and defining
\begingroup
\allowdisplaybreaks
\begin{subequations}
\begin{align}
    \zeta' \left( T_{\parallel} \right) &\coloneqq \frac{\delta \left( y(d), T_{\parallel} \right) + \delta \left( y'(d), T_{\parallel} \right)}{2} = \frac{1}{2i} \log \left( \frac{\left\langle v_{+}, T_{\parallel} \right\rangle}{\left\langle v_{-}, T_{\parallel} \right\rangle} \right)\,,\\
    \eta'(d) &\coloneqq \frac{\delta \left( y(d), T_{\parallel} \right) - \delta \left( y'(d), T_{\parallel} \right)}{2} = \frac{1}{2i} \log \left( \frac{\alpha_{y(d)}}{\beta_{y(d)}} \right)\,,
\end{align}
\end{subequations}
\endgroup
we obtain, suppressing the explicit dependencies on the curve class and K\"ahler data,
\begin{equation}
    \psi_{ld}^{\mathrm{I}_{2}(m)}(T) = e^{2\pi i l \mathcal{C}} 2m \left[ J_{0}(\Omega') + 2\sum_{r=1}^{\infty} i^{mr} J_{mr}(\Omega') \cos(mr\zeta') \cos(mr\eta') \right]\,.
\label{eq:elliptic-dih-Bessel-sum}
\end{equation}

For $d \in ( \mathrm{Fix}(\mathcal{M}_{1}) \cup \mathrm{Fix}(\mathcal{M}_{2}) ) \setminus ( \mathrm{Fix}(\mathcal{M}_{1}) \cap \mathrm{Fix}(\mathcal{M}_{2}) )$ we have \mbox{$\mathrm{Stab}_{\mathrm{I}_{2}(m)}(d) = \mathbb{Z}_{2}$}, meaning that the above expression overcounts the contributions to the orbit sum. The correct expression is then
\begin{equation}
    \psi_{ld}^{\mathrm{I}_{2}(m)}(T) = e^{2\pi i l \mathcal{C}} m \left[ J_{0}\left( \Omega' \right) + 2\sum_{r=1}^{\infty} i^{mr} J_{mr}\left( \Omega' \right) \cos\left( mr\delta \right) \right]\,.
\label{eq:elliptic-dih-Bessel-sum-Z2-stabilizer}
\end{equation}
Finally, for $d \in \mathrm{Fix}(\mathcal{M}_{1}) \cap \mathrm{Fix}(\mathcal{M}_{2})$ the entire group acts as the stabilizer and we obtain
\begin{equation}
    \psi_{ld}^{\mathrm{I}_{2}(m)}(T) = e^{2\pi i l \left\langle d, T \right\rangle}\,.
\end{equation}
The reason non-trivial classes can be stabilized by the full group is that, for the elliptic case, the extension shift vector $b\left( d_{\perp} \right)$ displaces the axis of the Euclidean rotation to the interior of the positive quadrant.

As we said earlier, in the elliptic case the raw sum contains finitely many exponential terms. The above rewriting is a plane-wave expansion in terms of cylindrical harmonics, which turns out to be more complicated than the original expression. It sharply localizes around the first Bessel-modes in the deep interior of the moduli space, as expected. One reason to perform the rewriting is that it makes the geometrical analysis of \cref{sec:harmonic-analysis} more natural. Additionally, we know from Glaeser’s theorem \cite{Glaeser1962FonctionsComposees,Barbancon2019FiniteReflection} that for a finite reflection group $W$ acting orthogonally on $V \cong \mathbb{R}^{n}$, any $W$-invariant function of class $C^{\infty}$ can be expressed in terms of functions of class $C^{\infty}$ of the basic invariants of $W$. Moreover, from the Chevalley–Shephard–Todd theorem \cite{ShephardTodd1954FiniteUnitary,Chevalley1955InvariantsReflections} we know that the algebra $\mathbb{C}[V]^{W}$ of $W$-invariants is a polynomial ring and, hence, finitely generated. The resummed expressions for $\psi_{ld}^{\mathrm{I}_{2}(m)}$ above are precisely such a rewriting of the orbit sums as a function of the basic polynomial invariants. We explicitly show this for the example below.

\subsubsection{Six-parameter example}
\label{sec:elliptic-example-I2(4)}

The CICY Coxeter Database contains 28 elliptic $\mathrm{I}_{2}(4)$ models. An illustrative example, with $(h^{1,1},h^{2,1}) = (6,30)$, is the CICY 5528 family, described by the configuration matrix
\begin{equation}
    X_{5528} \in
    \begingroup
    \delimiterfactor=1000
    \begin{bNiceArray}{c|cccccc}
        \mathbb{P}^{1} & 0 & 0 & 0 & 0 & 0 & 2\\
        \mathbb{P}^{4} & 0 & 1 & 1 & 1 & 1 & 1\\
        \mathbb{P}^{1} & 1 & 1 & 0 & 0 & 0 & 0\\
        \mathbb{P}^{1} & 1 & 0 & 1 & 0 & 0 & 0\\
        \mathbb{P}^{1} & 1 & 0 & 0 & 1 & 0 & 0\\
        \mathbb{P}^{1} & 1 & 0 & 0 & 0 & 1 & 0
    \end{bNiceArray}
    ^{6,30}_{-48\rlap{\,.}}
    \endgroup
\end{equation}
Its triple intersection numbers are
\begin{equation}
\begin{split}
    &\kappa _{122} = \kappa _{234} = \kappa _{235} = \kappa _{236} = \kappa _{245} = \kappa _{246} = \kappa _{256} = 4\,,\\
    &\kappa _{123} = \kappa _{124} = \kappa _{125} = \kappa _{126} = 3\,,\qquad \kappa _{223} = \kappa _{224} = \kappa _{225} = \kappa _{226} = 6\,,\\
    &\kappa _{134} = \kappa _{135} = \kappa _{136} = \kappa _{145} = \kappa _{146} = \kappa _{156} = \kappa _{345} = \kappa _{346} = \kappa _{356} = \kappa _{456} = 2\,,\\
    &\kappa _{222} = 8\,,\qquad k_{ijk} = 0\,,\quad \text{for all other } i,j,k \in \{1, \dotsc, 6\}\,.
\end{split}
\end{equation}
The first row of the configuration matrix is of Type 2 and hence leads to an iso-flop wall in the K\"ahler moduli space. The last four rows are of Type 1b and fail the $u$-vector test~\eqref{eq:uvec-definition}, which is a necessary condition for the flop being isomorphic; hence, they correspond to non-isomorphic flop walls. The second row, also of Type 1b, satisfies the $u$-vector test, making it a candidate for an iso-flop wall. Assuming that this is indeed the case, the two reflection generators read
\begin{equation}
    \mathcal{M}_{1} =
    \begin{pNiceArray}{cc|cccc}
        -1 & 1 & 0 & 0 & 0 & 0\\
        0 & 1 & 0 & 0 & 0 & 0\\
        \hline
        0 & 0 & 1 & 0 & 0 & 0\\
        0 & 0 & 0 & 1 & 0 & 0\\
        0 & 0 & 0 & 0 & 1 & 0\\
        0 & 0 & 0 & 0 & 0 & 1
    \end{pNiceArray}
    \,,\qquad
    \mathcal{M}_{2} =
    \begin{pNiceArray}{cc|cccc}
        1 & 0 & 0 & 0 & 0 & 0\\
        2 & -1 & 1 & 1 & 1 & 1\\
        \hline
        0 & 0 & 1 & 0 & 0 & 0\\
        0 & 0 & 0 & 1 & 0 & 0\\
        0 & 0 & 0 & 0 & 1 & 0\\
        0 & 0 & 0 & 0 & 0 & 1
    \end{pNiceArray}
    \,.
\end{equation}

Let us now confirm that the flop across the second row wall is an iso-flop. As discussed in \cref{sec:review-isomorphic-flops,sec:CICY-Coxeter-Database}, the necessary $u$-vector condition checks whether the triple intersection data across the wall is compatible with an iso-flop, while the Hodge numbers $h^{1,1}(X)$ and $h^{2,1}(X)$ remain invariant under the flop; if the change in $c_{2}(X)$ across the wall is such that $c_{2}(X) = \phi^{*}(c_{2}(X'))$, the isomorphism in the smooth category is proven, according to Wall's theorem \cite{Wall1966ClassificationV}. Indeed, from the configuration matrix, one obtains by adjunction
\begin{equation}
    \left( c_{2}(X) \cdot D_{i} \right)_{1 \leq i \leq 6} = (24, 56, 24, 24, 24, 24)^{T}\,.
\end{equation}
The genus-zero GV invariants along the Mori cone wall dual to the K\"ahler basis generator associated with the second configuration matrix row are
\begin{equation}
    n_{(0,1,0,0,0,0)} = 16\,,\qquad n_{(0,d_{2},0,0,0,0)} = 0\,,\quad d_{2} > 1\,,
\end{equation}
leading via \eqref{eq:c2-flop-change} to
\begin{equation}
    \left( c_{2}(X') \cdot D'_{i} \right)_{1 \leq i \leq 6} - \left( c_{2}(X) \cdot D_{i} \right)_{1 \leq i \leq 6} = (0, 32, 0, 0, 0, 0)^{T}\,.
\end{equation}
Hence, we confirm that
\begin{equation}
    \left( c_{2}(X') \cdot D'_{i} \right)_{1 \leq i \leq 6} = \mathcal{M}_{2} \cdot \left( c_{2}(X) \cdot D_{i} \right)_{1 \leq i \leq 6}
\end{equation}
and we obtain an elliptic representation of the $\mathrm{I}_{2}(4)$ Coxeter group.

Given the simple reflection matrices, the model corresponds to the data
\begin{equation}
    \begin{aligned}
        m_{1} &= 1\,,\\
        m_{2} &= 2\,,
    \end{aligned}
    \qquad
    \begin{aligned}
        u'_{1} &= (0,0,0,0)\,,\\
        u'_{2} &= (1,1,1,1)\,,
    \end{aligned}
    \qquad \lambda_{\pm} = \pm i\,,\qquad v = \frac{\pi}{4}\,.
\end{equation}
This leads to
\begingroup
\allowdisplaybreaks
\begin{subequations}
\begin{align}
    \mathcal{C} \left( d_{\perp}, T \right) &= (d_{3}+d_{4}+d_{5}+d_{6})\left( \frac{T_{1}}{2} + T_{2} \right) + d_{3}T_{3} + d_{4}T_{4} + d_{5}T_{5} + d_{6}T_{6}\,,\\
    \Omega'\left( l,y(d),T_{\parallel} \right) &= \sqrt{2} \pi  l \sqrt{
    \begin{aligned}
    &i (T_{1}+(1-i) T_{2}) (T_{1}+(1+i) T_{2})\\
    &\quad \cdot(-2 i d_{1}+(1+i) d_{2}-d_{3}-d_{4}-d_{5}-d_{6})\,,\\
    &\quad \cdot(2d_{1}-(1+i) d_{2}+i (d_{3}+d_{4}+d_{5}+d_{6}))\,,
    \end{aligned}
    }\\
    \delta\left( y(d),T_{\parallel} \right) &= -\frac{1}{2} i \log \left(\frac{(T_{1}+(1-i) T_{2}) (-2 i d_{1}+(1+i) d_{2}-d_{3}-d_{4}-d_{5}-d_{6})}{(T_{1}+(1+i) T_{2})
   (2 d_{1}-(1+i) d_{2}+i (d_{3}+d_{4}+d_{5}+d_{6}))}\right)\,,\\
    \zeta'\left( T_{\parallel} \right) &= -\frac{1}{2} i \log \left(\frac{2 T_{2}+(1+i) T_{1}}{2 T_{2}+(1-i) T_{1}}\right)\\
    \eta'(d) &= -\frac{1}{2} i \log \left(1+\frac{2 i (2 d_{1}-d_{2})}{-2 i d_{1}-(1-i) d_{2}+d_{3}+d_{4}+d_{5}+d_{6}}\right)\,,
\end{align}
\end{subequations}
\endgroup
from which we can compute $\psi_{ld}^{\mathrm{I}_{2}(4)}(T)$ using \eqref{eq:elliptic-dih-Bessel-sum} for curve classes $d \notin \mathrm{Fix}(\mathcal{M}_{1}) \cup \mathrm{Fix}(\mathcal{M}_{2})$ or \eqref{eq:elliptic-dih-Bessel-sum-Z2-stabilizer} for $d \in ( \mathrm{Fix}(\mathcal{M}_{1}) \cup \mathrm{Fix}(\mathcal{M}_{2}) ) \setminus ( \mathrm{Fix}(\mathcal{M}_{1}) \cap \mathrm{Fix}(\mathcal{M}_{2}) )$.

The fundamental Mori cone is
\begin{equation}
    \mathcal{M}_{f} = \left\{ d \in \mathbb{R}^{h}_{\geq 0} \mathrel{\Big|} 2d_{1} - d_{2} \geq 0\,,\quad 2d_{2} - 2d_{1} \geq d_{3} + d_{4} + d_{5} + d_{6} \geq d_{1} \right\}\,,
\end{equation}
which can be written in terms of generators as
\begin{equation}
    \mathcal{M}_{f} = \left\langle w_{1},\dotsc,w_{4},p_{1},\dotsc,p_{4},q_{1},\dotsc,q_{4} \right\rangle_{\mathbb{R}_{\geq 0}}\,,
\end{equation}
where
\begin{equation}
    w_{a'} =
    \begin{pmatrix}
        1\\
        2\\
        2e_{a'}
    \end{pmatrix}
    \,,\quad p_{a'} =
    \begin{pmatrix}
        2\\
        3\\
        2e_{a'}
    \end{pmatrix}
    \,,\quad q_{a'} =
    \begin{pmatrix}
        1\\
        2\\
        e_{a'}
    \end{pmatrix}
    \,,\quad a'\in\{1,\dotsc,4\}
\end{equation}
and $\{ e_{a'} \}_{a' \in \{ 1,\dotsc,4 \}}$ is a canonical basis of $\mathbb{R}^{4}$, see \cref{sec:fundamental-Mori-cone-infdih} for further details. An example of a non-trivial curve class on a wall of the fundamental Mori cone such that $\mathrm{Stab}_{\mathrm{I}_{2}(4)}(d) = \mathbb{Z}_{2}$ is $d = (1,2,1,0,0,0)$, with $n_{d} = 12$. At the intersection of the two mirror hyperplanes we have, e.g., $d = (1,2,1,1,0,0)$ with $n_{d} = 72$, which is stabilized by the entire group. An example of a curve class $d \in \partial\mathcal{M}_{f}$ with trivial stabilizer is $d = (3,5,0,1,1,1)$, with $n_{d} = 56$. As for the previous examples, taking the permutation symmetries of the configuration matrix rows into account, the full symmetry group for this model is $G = \mathrm{I}_{2}(4) \rtimes S_{4}$.

The ring of polynomial invariants for $\mathrm{I}_{2}(4)$ acting on the real span of its simple roots $V_{0} \cong \mathbb{R}^{2}$, i.e., in the geometrical representation, is generated by
\begin{equation}
    \mathbb{C}[V_{0}]^{\mathrm{I}_{2}(m)} = \left\langle p_{2}(x_{1},x_{2}), p_{4}(x_{1},x_{2}) \right\rangle\,,
\end{equation}
where the generators can be chosen to be\footnote{A basis of basic polynomial invariants can be easily obtained using CHEVIE within GAP3. We make a different choice of basis that leads to more compact expressions.}
\begin{subequations}
\begin{align}
    p_{2}(x_{1},x_{2}) &\coloneqq 2 x_{1}^2-2 x_{1} x_{2}+x_{2}^2\,,\\ p_{4}(x_{1},x_{2}) &\coloneqq 16 x_{1}^4-32 x_{1}^3 x_{2}+16 x_{1} x_{2}^3-4 x_{2}^4\,.
\end{align}
\end{subequations}
For the representation acting on the coroots we similarly have
\begin{equation}
    \mathbb{C}[\hat{V}_{0}]^{\mathrm{I}_{2}(m)} = \left\langle \hat{p}_{2}(\hat{x}_{1},\hat{x}_{2}), \hat{p}_{4}(\hat{x}_{1},\hat{x}_{2}) \right\rangle\,,
\end{equation}
with
\begin{subequations}
\begin{align}
    \hat{p}_{2}(\hat{x}_{1},\hat{x}_{2}) &\coloneqq \hat{x}_{1}^2+2 \hat{x}_{1} \hat{x}_{2}+2 \hat{x}_{2}^2\,,\\ \hat{p}_{4}(\hat{x}_{1},\hat{x}_{2}) &\coloneqq \hat{x}_{1}^4+4 \hat{x}_{1}^3 \hat{x}_{2}-8 \hat{x}_{1} \hat{x}_{2}^3-4 \hat{x}_{2}^4\,.
\end{align}
\end{subequations}
One can then check that, e.g., \eqref{eq:elliptic-dih-Bessel-sum} can be rewritten as
\begin{equation}
    \psi_{ld}^{\mathrm{I}_{2}(m)}(T) = 8 e^{2\pi i l \mathcal{C}} \left[ J_{0}(\Omega') + 2\sum_{r=1}^{\infty} J_{4r}(\Omega') T_{r}(\sigma_{T}) T_{r}(\sigma_{y}) \right]\,,
\end{equation}
where $T_{r}$ are the Chebyshev polynomials of the first kind and
\begin{subequations}
\begin{align}
    \Omega'\left( l,y(d),T_{\parallel} \right) &= 2\pi l \sqrt{p_{2}(y(d)_{1},y(d)_{2}) \hat{p}_{2}(T_{1},T_{2})}\,,\\
    \sigma_{T}\left( T_{\parallel} \right) &= -\frac{\hat{p}_{4}(T_{1},T_{2})}{\hat{p}_{2}(T_{1},T_{2})^{2}} = \cos\left(4\zeta'\big( T_{\parallel} \big)\right)\,,\\
    \sigma_{y}(y(d)) &= \frac{p_{4}(y(d)_{1},y(d)_{2})}{4p_{2}(y(d)_{1},y(d)_{2})^{2}} = \cos\left( 4\eta'\big( d \big) \right)\,.
\end{align}
\end{subequations}

%% file: sections/harmonic-analysis.tex
\section{Harmonic analysis of the Gromov-Witten expansion}
\label{sec:harmonic-analysis}

In \cref{sec:dihedral-Coxeter-prepotentials} we have analyzed the $\mathrm{I}_{2}(m)$-invariant functions $\psi_{ld}^{\mathrm{I}_{2}(m)}(T)$ featured in the Gromov-Witten expansion of the prepotential for models displaying a dihedral Coxeter symmetry generated by iso-flops. From their defining expressions as sums over Coxeter orbits, we arrived at more natural\,---\,from the point of view of the symmetries\,---\,objects involving special functions depending on manifestly $\mathrm{I}_{2}(m)$-invariant arguments (possibly up to irrelevant phase shifts) for the hyperbolic and elliptic cases or invariant products of automorphy factors for the parabolic case. Moreover, these sums in terms of special functions sharply localize around the first few contributions when the relevant K\"ahler moduli take values in the interior of the moduli space, a behavior that is opposite to that of the raw Gromov-Witten orbit sums. The situation is mathematically analogous to the complementary localization properties of the spectral dual expansions in the harmonic analysis of waves.

To gain an intuitive understanding of what features are shared among the three dihedral cases and provide a common geometric origin for their resummed expansions, we push this analogy further. This will also inform us about the type of Coxeter-invariant functions $\psi_{ld}^{W}(T)$ that we should expect for the general cases in the CICY Coxeter Database.

The following picture emerges: Heuristically, we can understand the Gromov-Witten expansion as a superposition of plane waves ``propagating" on the moduli space. The plane wave superposition satisfies a massive wave equation in terms of a Laplace-Beltrami operator appropriate to the geometry. This operator is the natural second-order differential operator respecting the isometries of a space. For models with an iso-flop Coxeter symmetry, we can construct a Coxeter-invariant metric on the moduli space realizing the iso-flop transformations as isometries. The resummed expressions for $\psi_{ld}^{W}(T)$ then correspond to an expansion in terms of the eigenfunctions of the resulting Laplace-Beltrami operator, which explains the appearance of special functions adapted to the Coxeter symmetries. In this section, we recover the expressions for $\psi_{ld}^{\mathrm{I}_{2}(m)}(T)$ as a combination of Bessel and Jacobi theta functions from this point of view. The expectation for the general case is that the invariant Gromov-Witten expansion can be expressed in terms of the pullback of the Laplace-Beltrami eigenfunctions defined in a Coxeter-quotient of the moduli space.

\subsection{Coxeter-invariant moduli space geometry}
\label{sec:Coxeter-invariant-moduli-space-geometry}

The Mori and K\"ahler representations of $\mathrm{I}_{2}(m)$ were studied in detail in \cref{sec:dihedral-Mori-Kahler-representations}. One useful result, exploited in \cref{sec:dihedral-hyperbolic-representations,sec:dihedral-elliptic-representations}, is that the extension \eqref{eq:dihedral-rep-short-exact-sequence} is always split for the hyperbolic and elliptic cases. This allows us to define untwisted coordinates (dependent on the extension part of the chosen curve class) for which only a two-dimensional block $\left( V_{0}, \left. \rho \right|_{V_{0}} \right)$ of the representation contains non-trivial information. Below we study the geometry of $\left( V_{0}, \left. \rho \right|_{V_{0}} \right)$ for the hyperbolic and elliptic dihedral cases in preparation for the harmonic analysis of the Gromov-Witten expansion. More concretely, we explicitly determine the natural Laplace-Beltrami operator in the Coxeter quotient of the moduli space. The parabolic case\,---\,for which the extension \eqref{eq:dihedral-rep-short-exact-sequence} is non-split\,---\,behaves differently and will be treated separately in \cref{sec:parabolic-degeneration}. Throughout this section we will focus on the Mori representation to better align with the exposition in \cref{sec:dihedral-Coxeter-prepotentials}; analogous conclusions follow for the K\"ahler representation \textit{mutatis mutandis}.

The two-dimensional representation $\left. \rho \right|_{V_{0}}$ acts on the real vector space $V_{0} \cong \mathbb{R}^{2}$, which is interpreted as the untwisted two-dimensional sublattice of the lattice of effective curves tensored with $\mathbb{R}$. From the two generators $\left. \rho \right|_{V_{0}}(r) = Q$ and $\left. \rho \right|_{V_{0}}(s_{1}) = S$ it is straightforward to check that the action of $\mathrm{I}_{2}(m)$ leaves the symmetric bilinear form
\begin{equation}
    \Sigma =
    \begin{pNiceArray}{cc}[cell-space-limits=2.5pt]
        m_{2} & -\frac{m_{1}m_{2}}{2}\\
        -\frac{m_{1}m_{2}}{2} & m_{1}
    \end{pNiceArray}
\label{eq:dihedral-invariant-bilinear-form}
\end{equation}
invariant, as discussed in \cref{sec:dihedral-hyperbolic-representations}. For the elliptic and parabolic cases, $\Sigma$ is congruent (by a normalization of the simple root lengths) to the Tits form associated to the standard geometric representation of the group, see, for example, \cite{Bourbaki2006GroupesAlgebresLie456,Humphreys1990ReflectionGroups}.\footnote{For $\infdih$ the Tits form convention corresponds to the degenerate bilinear form. The hyperbolic case is a different realization of the same abstract group.} From
\begin{equation}
    \det(\Sigma) = \frac{m_{1}m_{2}}{4}(4-m_{1}m_{2})
\end{equation}
we see that $\Sigma$ behaves differently for the three dihedral cases:
\begin{itemize}
    \item Elliptic $\Rightarrow 0 < m_{1}m_{2} < 4$: Here $\Sigma$ is a non-degenerate symmetric bilinear form with Euclidean signature $(2,0)$.

    \item Parabolic $\Rightarrow m_{1}m_{2} = 4$: Here $\Sigma$ is degenerate; it can be regarded as a degeneration of the hyperbolic geometry.

    \item Hyperbolic $\Rightarrow m_{1}m_{2} > 4$: Here $\Sigma$ is a non-degenerate symmetric bilinear form with Lorentzian signature $(1,1)$.
\end{itemize}

The hyperbolic and elliptic cases are geometrically very similar. Let us treat the former in detail and briefly comment on the latter afterwards. The invariant symmetric bilinear form can be interpreted as a metric, with $\Sigma$ its matrix expression
\begin{equation}
    ds^{2} = \Sigma_{ij}\, dy^{i} dy^{j}
\end{equation}
in the untwisted coordinates. Invariance under the group action and the fact that the eigenvalues of $Q v_{\pm} = \lambda_{\pm}v_{\pm}$ are $\lambda_{\pm} \neq 0$ immediately imply that $\Sigma(v_{\pm},v_{\pm}) = 0$. Hence, the eigenvectors of the boost generator span the null directions of the Lorentzian metric.

Let us define the positive light cone\footnote{Note that, while the restricted Mori cone is naturally closed, the hyperbolic orbit-sum seeds satisfy $y(d), y'(d) \in C^{+}$ for $d \in \mathcal{M}_{f}$ with $\mathrm{Stab}_{\infdih}(d) = 1$, see \cref{sec:AB-properties}. The open positive light cone in the untwisted \(V_0\)-coordinates is, therefore, the appropriate object for the present analysis. Moreover, taking it open also avoids coordinate pathologies when $\alpha = 0$ or $\beta = 0$.}
\begin{equation}
    C^{+} \coloneqq \left\{ y = \alpha v_{+} + \beta v_{-} \in V_{0} \mathrel{|} \alpha, \beta >0 \right\}\,.
\end{equation}
The generators of the null directions can be normalized such that $Sv_{\pm} = v_{\mp}$, see \eqref{eq:hyperbolic-Q-eigenvectors}. Given the multiplicative nature of the boost action $Q: \big(\alpha,\beta\big) \mapsto \big(e^{2u}\alpha,e^{-2u}\beta\big)$, we can find coordinates better suited to describe the quotient by the Coxeter group. In particular, consider the radial and angular coordinates
\begin{equation}
    \rho \coloneqq \sqrt{\alpha\beta} \in \mathbb{R}_{>0}\,,\qquad \sigma = \frac{1}{2} \log\left( \frac{\beta}{\alpha} \right) \in \mathbb{R}\,,
\end{equation}
respectively. The action of $Q$ on these is
\begin{equation}
    Q: (\rho,\sigma) \mapsto (\rho,\sigma-2u)\,,
\end{equation}
meaning that they are good coordinates on the (covering space of the) quotient cylinder
\begin{equation}
    C^{+}/\langle Q \rangle \cong \mathbb{R}_{>0} \times S^{1}\,.
\end{equation}
The generator $S$ acts as a reflection $S: \sigma \mapsto -\sigma$ on the circle coordinate and, as a consequence, the full quotient geometry is the orbifold
\begin{equation}
    C^{+}/\langle Q,S \rangle \cong \mathbb{R}_{>0} \times S^{1}/\mathbb{Z}_{2}\,.
\end{equation}

The invariant bilinear form $\Sigma$ is determined by the dihedral group action up to an overall scale. Hence, to simplify the end result, we will drop global constant Jacobian factors as we perform changes of coordinates by invoking this scale ambiguity. Going to the quotient cylinder coordinates we find
\begin{equation}
    ds^{2} = \Sigma_{ij}\, dy^{i} dy^{j} \sim -d\alpha\, d\beta \sim -d\rho^{2} + \rho^{2} d\sigma^{2}\,.
\end{equation}
To later connect with the orbit sums, it will be useful to consider the alternative cylinder coordinates\footnote{We will abuse notation and refer to the covering space and quotient geometry coordinates by the same symbols, the latter obtained from the former by simply a choice of branch.}
\begin{equation}
    \Omega_{R} \coloneqq 4\pi l \rho\,,\qquad \theta_{R} \coloneqq \frac{\pi}{u}\sigma\,,
\end{equation}
for which we obtain the final (rescaled) metric
\begin{equation}
    ds^{2} = -d\Omega_{R}^{2} + \left( \frac{u}{\pi} \right)^{2} \Omega_{R}^{2}\, d\theta_{R}^{2}\,.
\end{equation}
The canonical second-order differential operator respecting the isometries of the metric, i.e., the action of the Coxeter group, is the Laplace-Beltrami operator
\begin{equation}
    \Delta_{g} \coloneqq \frac{\mathrm{sgn}(g)}{\sqrt{|g|}} \partial_{i} \left( \sqrt{|g|} g^{ij} \partial_{j}f \right)\,,
\end{equation}
which results in
\begin{equation}
    \Delta_{\Sigma}^{R} = \partial_{\Omega_{R}}^{2} + \frac{1}{\Omega_{R}} \partial_{\Omega_{R}} - \left( \frac{\pi}{u} \right)^{2} \frac{1}{\Omega_{R}^{2}} \partial_{\theta_{R}}^{2}\,.
\label{eq:hyperbolic-real-Laplace-Beltrami}
\end{equation}

Let us now relate the quotient cylinder geometry to the orbit sums $S\left(l,z,T_{\parallel}\right)$ defined in \eqref{eq:hyperbolic-one-seed-orbit-sums}. Writing the orbit sum seed in the light cone basis we obtain
\begin{equation}
    z = \alpha_{z}v_{+} + \beta_{z}v_{-}\,.
\end{equation}
The relevant object for $S\left(l,z,T_{\parallel}\right)$ is the pairing
\begin{equation}
    \big\langle Q^{k} z, T_{\parallel} \big\rangle = \mathcal{A}\left(z,T_{\parallel}\right) e^{2ku} + \mathcal{B}\left(z,T_{\parallel}\right) e^{-2ku}\,,
\end{equation}
where the coefficients were defined in \eqref{eq:AB-definitions} but can alternatively be written as
\begin{subequations}
\begin{align}
    \mathcal{A}\left( z,T_{\parallel} \right) &= \alpha_{z} \langle v_{+}, T_{\parallel} \rangle\,,\\
    \mathcal{B}\left( z,T_{\parallel} \right) &= \beta_{z} \langle v_{-}, T_{\parallel} \rangle\,.
\end{align}
\end{subequations}
We therefore see that $\mathcal{A}\left( z,T_{\parallel} \right)$ and $\mathcal{B}\left( z,T_{\parallel} \right)$ are a $T_{\parallel}$-dependent diagonal rescaling of the $(\alpha_{z},\beta_{z})$ coordinates. Importantly, $\langle v_{+}, T_{\parallel} \rangle$ and $\langle v_{-}, T_{\parallel} \rangle$ are complex in general, meaning that we are not simply rescaling the coordinate chart of the real quotient cylinder. The relation to the arguments of the resummed expression \eqref{eq:hyperbolic-S-sum-Bessel} for $S\left(l,z,T_{\parallel}\right)$
is
\begin{equation}
    \Omega \left( l,z,T_{\parallel} \right) = \kappa\left(T_{\parallel}\right) \Omega_{R}(z)\,,\qquad \theta\left( z, T_{\parallel} \right) = \theta_{R}(z) + \theta_{0} \left( T_{\parallel} \right)\,,
\label{eq:Delta-theta-complex-real}
\end{equation}
where
\begin{equation}
    \kappa \left( T_{\parallel} \right) \coloneqq \sqrt{-\langle v_{+}, T_{\parallel} \rangle \langle v_{-}, T_{\parallel} \rangle}\,,\qquad \theta_{0} \left( T_{\parallel} \right) \coloneqq \frac{\pi}{2u} \log \left(\frac{\langle v_{-}, T_{\parallel} \rangle}{\langle v_{+}, T_{\parallel} \rangle} \right)\,,
\end{equation}
on the fixed branch $\mathrm{Im}(\mathcal{A}(z,T_{\parallel})),\, \mathrm{Im}(\mathcal{B}(z,T_{\parallel})) > 0$. Hence, the appropriate geometry corresponds to an embedding of the real quotient cylinder as a $T_{\parallel}$-dependent slice of the complexified quotient cylinder.

To make this more precise, consider the complexified cone
\begin{equation}
    C^{\mathbb{C}} \coloneqq \left\{ y = A v_{+} + B v_{-} \in V_{0} \otimes \mathbb{C} \mathrel{|} A, B \in \mathbb{C}^{*} \right\}\,.
\end{equation}
A choice of coordinates adapted to the complex quotient cylinder $C^{\mathbb{C}}/\langle Q \rangle$ is
\begin{equation}
    \Omega \coloneqq 4\pi l \sqrt{-AB}\,,\qquad \theta \coloneqq \frac{\pi}{2u} \log \left( \frac{B}{A} \right)\,,
\end{equation}
whose definition is analogous to that of the arguments \eqref{eq:hyperbolic-Delta-theta-definition} of the resummed $S\left(l,z,T_{\parallel}\right)$. The complexified representation $\left. \rho^{\mathbb{C}} \right|_{V_{0} \otimes \mathbb{C}}$ preserves the $\mathbb{C}$-bilinear extension of $\Sigma$, with the same formal expression
\begin{equation}
    ds^{2} = -d\Omega^{2} + \left( \frac{u}{\pi} \right)^{2} \Omega^{2}\, d\theta^{2}
\end{equation}
leading to the corresponding complexified Laplace-Beltrami operator
\begin{equation}
    \Delta_{\Sigma} = \partial_{\Omega}^{2} + \frac{1}{\Omega} \partial_{\Omega} - \left( \frac{\pi}{u} \right)^{2} \frac{1}{\Omega^{2}} \partial_{\theta}^{2}\,.
\label{eq:hyperbolic-complex-Laplace-Beltrami}
\end{equation}

The pairing $\big\langle Q^{k} z, T_{\parallel} \big\rangle$ defines a $T_{\parallel}$-dependent $Q$-equivariant map from the real quotient cylinder to a real slice of the complex quotient cylinder. Explicitly, it picks
\begin{equation}
    \begin{split}
        F_{T_{\parallel}}: C^{+} &\longrightarrow C^{\mathbb{C}}\\
        (\alpha,\beta) &\longmapsto \left(\alpha \langle v_{+}, T_{\parallel} \rangle, \beta \langle v_{-}, T_{\parallel} \rangle\right)\,,
    \end{split}
\label{eq:complex-quotient-embedding-Mori-alpha-beta}%
\end{equation}
for fixed $T_{\parallel}$ or, equivalently,
\begin{equation}
    \begin{split}
        F_{T_{\parallel}}: C^{+} &\longrightarrow C^{\mathbb{C}}\\
        \left( \Omega_{R}, \theta_{R} \right) &\longrightarrow \left( \kappa\left(T_{\parallel}\right) \Omega_{R}, \theta_{R} + \theta_{0} \left( T_{\parallel} \right) \right)\,.
    \end{split}
\label{eq:complex-quotient-embedding-Mori}%
\end{equation}
When evaluated at an orbit sum seed $z \in C^{+}$ this yields
\begin{equation}
    F_{T_{\parallel}}(z) = \left(\mathcal{A}\left(z, T_{\parallel} \right), \mathcal{B}\left( z, T_{\parallel} \right)\right)\qquad \text{or}\qquad F_{T_{\parallel}}(z) = \left( \Omega \left( l,z,T_{\parallel} \right), \theta\left( z, T_{\parallel} \right) \right)\,.
\end{equation}
Its geometric meaning becomes clearer in the flat Lorentzian embedding. Choosing Rindler coordinates
\begin{equation}
    X_{R}\coloneqq \Omega_{R} \cosh \left( \frac{u}{\pi} \theta_{R} \right)\,,\qquad Y_{R} \coloneqq \Omega_{R} \sinh \left( \frac{u}{\pi} \theta_{R} \right)\,,
\end{equation}
on the covering space of the real quotient cylinder we find
\begin{equation}
    ds^{2} = -d\Omega_{R}^{2} + \left( \frac{u}{\pi} \right)^{2} \Omega_{R}^{2}\, d\theta_{R}^{2} = -dX^{2}_{R} + dY^{2}_{R}\,.
\end{equation}
Similarly, defining complexified Rindler coordinates\footnote{As a curiosity, we note that the Unruh temperature associated with fixed $|\Omega|$ Rindler trajectories in this space asymptotically vanishes in the large volume region $|\Omega\left( l,z,T_{\parallel} \right)| \rightarrow \infty$ and diverges in the interior of the moduli space $|\Omega\left( l,z,T_{\parallel} \right)| \rightarrow 0$.}
\begin{equation}
    X \coloneqq \Omega \cosh \left( \frac{u}{\pi} \theta \right)\,,\qquad Y \coloneqq \Omega \sinh \left( \frac{u}{\pi} \theta \right)\,,
\end{equation}
we obtain
\begin{equation}
    ds^{2} = -d\Omega^{2} + \left( \frac{u}{\pi} \right)^{2} \Omega^{2}\, d\theta^{2} = -dX^{2} + dY^{2}\,.
\end{equation}
The $T_{\parallel}$-dependent $Q$-equivariant map corresponds then to a combination of a boost and a dilation
\begin{equation}
    \begin{pNiceArray}{c}[cell-space-limits=2.5pt]
        X\\
        Y
    \end{pNiceArray}
    = \kappa\left( T_{\parallel} \right)
    \begin{pNiceArray}{cc}[cell-space-limits=2.5pt]
        \cosh \left( \frac{u}{\pi} \theta_{0} \left( T_{\parallel} \right) \right) & \sinh \left( \frac{u}{\pi} \theta_{0} \left( T_{\parallel} \right) \right)\\
        \sinh \left( \frac{u}{\pi} \theta_{0} \left( T_{\parallel} \right) \right) & \cosh \left( \frac{u}{\pi} \theta_{0} \left( T_{\parallel} \right) \right)
    \end{pNiceArray}
    \begin{pNiceArray}{c}[cell-space-limits=2.5pt]
        X_{R}\\
        Y_{R}
    \end{pNiceArray}
    \,,
\label{eq:hyperbolic-real-slice-boost}
\end{equation}
twisting the real slice inside the complexified space. Hence, changing $T_{\parallel}$ only affects the embedding of the real slice; the harmonic analysis will be performed on the complexified quotient cylinder and the $T_{\parallel}$-dependence only appears through the pullback of the resulting eigenfunctions to the moduli space.

The same results can be obtained\,---\,slightly more efficiently, in fact\,---\,working with the K\"ahler representation. Working at fixed curve class, the real quotient cylinder geometry corresponds then to the real K\"ahler moduli space and the complex quotient cylinder directly arises once we consider the complexified K\"ahler representation. We have chosen to present the material using the Mori representation for better alignment with \cref{sec:dihedral-Coxeter-prepotentials}. The K\"ahler representation point of view would have avoided the discussion of the embedding of the real slice into the complexified space between \eqref{eq:complex-quotient-embedding-Mori-alpha-beta} and \eqref{eq:hyperbolic-real-slice-boost}, but still requires introducing a $z$-dependent evaluation map $E_{z,l}$ defined by the pairing, which we use in \cref{sec:Gromov-Witten-plane-waves}. The Laplace-Beltrami operator obtained for the complex quotient cylinder is identical in both approaches and so are the final orbit sum expressions that we will obtain upon performing the pullback by $E_{z,l}$ to the K\"ahler moduli space.\footnote{It is also possible to compute the pullback to the restricted Mori cone using $F_{T_{\parallel}}$ instead, in which case the real slice embedding would be more relevant, but we seek to obtain a Laplace-Beltrami operator with respect to the independent variables of the prepotential, which are the K\"ahler moduli.}

Having studied the hyperbolic case in detail, let us turn our attention to its elliptic counterpart. Its treatment is analogous and, to avoid unnecessary repetition, we only report the main differences. A prime distinguishing fact between the hyperbolic and elliptic cases is that the M\"obius transformation $\left. \rho \right|_{V_{0}}(r) = Q$, when regarded as acting on the upper half-plane $\overline{\mathbb{H}}$, represents a hyperbolic translation with translation proportional to $u$ for the former and an elliptic rotation with rotation angle proportional to $v$ for the latter. The eigenvalues of $Q$ are $\lambda_{\pm} = e^{\pm2u}$ in the hyperbolic case and $\lambda_{\pm} = e^{\pm 2vi}$ for the elliptic case. At the level of the $V_{0}$ plane geometry, we can transform the hyperbolic geometry into the elliptic one via the substitution $u \mapsto iv$.\footnote{This works at the level of the $V_{0}$ plane geometry, quotients, metric and Laplace-Beltrami operator. The resummed expressions \eqref{eq:hyperbolic-S-sum-Bessel} and \eqref{eq:elliptic-S-sum-Bessel} do not match exactly under this substitution, due to a change in the boundary conditions. We will comment on this fact in \cref{sec:Gromov-Witten-plane-waves}.} The natural quotient geometry by the action of $Q$ is $(V_{0} \setminus \{0\})/\langle Q \rangle$, i.e., the smooth locus of a cone orbifold of the plane. The further quotient by the action of $S$ turns the circle direction into an interval. The resulting (rescaled) metric in appropriate coordinates is
\begin{equation}
    ds^{2} = d{\Omega'_{R}}^{2} + \left( \frac{v}{\pi} \right)^{2} {\Omega'_{R}}^{2} d\delta^{2}_{R}\,,
\end{equation}
leading to the final Laplace-Beltrami operator
\begin{equation}
    \Delta'_{\Sigma} = \partial_{\Omega'}^{2} + \frac{1}{\Omega'} \partial_{\Omega'} + \left( \frac{\pi}{v} \right)^{2} \frac{1}{{\Omega'}^{2}} \partial_{\delta}^{2}\,.
\end{equation}

The parabolic case, on the other hand, can be viewed as a $u \rightarrow 0$ degeneration of the hyperbolic case. The two null directions collapse into a single line stabilized by the action of $Q$ and the metric $\Sigma$ becomes degenerate. This prevents us from deriving a well-defined Laplace-Beltrami operator through the exact same sequence of steps. We treat the parabolic geometry separately in \cref{sec:parabolic-degeneration}.

\subsection{Gromov-Witten plane waves and Bessel eigenmodes}
\label{sec:Gromov-Witten-plane-waves}

An instanton contribution to the Gromov-Witten expansion has the form
\begin{equation}
    e^{2\pi i l \langle d, T \rangle} = e^{-2\pi l \langle d, t \rangle} e^{2\pi i l \langle d, b \rangle}\,.
\end{equation}
This can be interpreted as a plane wave in which the axions $\{ b^{i} \}_{i \in \{ 1, \dotsc, h \}}$ in the $B$-field decomposition $B = b^{i}D_{i} \in H^{2}(X,\mathbb{R})/H^{2}(X,\mathbb{Z})$ are angular variables on a real torus and the curve class $d \in \mathcal{M}$ acts as an integral wavevector via the natural pairing. The real Kähler moduli term acts as an exponential damping, ensuring that the wave dies off at large volume. Meanwhile, the $l$-fold cover term corresponding to the class $ld$ acts as a higher harmonic of the basic wavevector $d$. Since the standard Laplace-Beltrami operator on the flat torus yields
\begin{equation}
    \Delta_{b} e^{2\pi i l \langle d, b \rangle} = -(2\pi)^{2} l^{2} |d|^{2} e^{2\pi i l \langle d, b \rangle}\,,
\end{equation}
we see that the Gromov-Witten expansion is a decomposition of the instanton prepotential in terms of irreducible Fourier modes of the shift symmetry group of the axion torus.

In the study of the $\infdih$-invariant functions $\psi_{ld}^{\infdih}(T)$ for the hyperbolic case, we saw around \eqref{eq:hyperbolic-kernel} that the orbit sum corresponds to a (multiplicative) lattice evaluation of the continuous kernel
\begin{equation}
    f \left( z,T_{\parallel} \,\middle|\, x \right) \coloneqq e^{2\pi i l \left( \mathcal{A} \left( z,T_{\parallel} \right)x + \mathcal{B}\left( z,T_{\parallel} \right)x^{-1} \right)}\,,\quad x \in \mathbb{R}_{>0}\,.
\end{equation}
The exponent can be rewritten, using $x = e^{\tau}$, as
\begin{equation}
    \mathcal{A} \left( z,T_{\parallel} \right)x + \mathcal{B}\left( z,T_{\parallel} \right)x^{-1} = 2\sqrt{\mathcal{A} \left( z,T_{\parallel} \right) \mathcal{B}\left( z,T_{\parallel} \right)} \cosh\left( \tau - \frac{1}{2} \log \left( \frac{\mathcal{B}\left( z,T_{\parallel} \right)}{\mathcal{A} \left( z,T_{\parallel} \right)} \right) \right)\,,
\end{equation}
Taking the K\"ahler-side analogue of \eqref{eq:complex-quotient-embedding-Mori}, we define, for a fixed curve class and harmonic, the evaluation map
\begin{equation}
    \begin{split}
        E_{z,l} : \mathrm{int}\left( \hat{\pi}_{0} \left( \mathcal{K} \right) \right) &\longrightarrow C^{\mathbb{C}}\\
        T_{\parallel} &\longmapsto \left( \Omega \left( l,z,T_{\parallel} \right), \theta\left( z, T_{\parallel} \right) \right)\,.
    \end{split}
\end{equation}
In the coordinates of the covering space of the complex quotient cylinder, the continuous kernel becomes
\begin{equation}
    \Phi_{\tau} \left( \Omega,\theta \right) \coloneqq e^{-\Omega \cosh \left(\tau - \frac{u}{\pi} \theta \right)}\,,
\end{equation}
such that
\begin{equation}
    f \left( z,T_{\parallel} \,\middle|\, x \right) = \Phi_{\tau} \left( E_{z,l}\left( T_{\parallel} \right) \right)\,.
\end{equation}
Singling out the elementary kernel
\begin{equation}
    \Phi \left( \Omega,\theta \right) \coloneqq e^{-\Omega \cosh \left(\frac{u}{\pi} \theta \right)}\,,
\end{equation}
the lattice evaluation corresponds to summing over the boost action $Q^{k}: \theta \mapsto\theta - 2\pi k$, i.e., to the orbit sum
\begin{equation}
    \tilde{S}(\Omega,\theta) \coloneqq \sum_{k \in \mathbb{Z}} \Phi(\Omega,\theta - 2\pi k)\,.
\label{eq:quotient-cylinder-waves-superposition}
\end{equation}
This function descends under $\pi: C^{\mathbb{C}} \rightarrow C^{\mathbb{C}}/\langle Q \rangle$ to a function $S_{Q}$ on the complex quotient cylinder. The orbit sum of instanton contributions is then
\begin{equation}
    S \left( l,z,T_{\parallel} \right) = E_{z,l}^{*} \tilde{S} \left( T_{\parallel} \right) = (\pi \circ E_{z,l})^{*} S_{Q} \left( T_{\parallel} \right)\,.
\end{equation}

Note that the elementary kernel that we lattice evaluate and sum over to construct $\psi_{ld}^{\infdih}(T)$ is an eigenfunction of the Laplace-Beltrami operator on the covering space of the complex quotient cylinder, solving the Helmholtz equation\footnote{The absolute magnitude of the eigenvalue does not hold any special meaning beyond the fact that it is non-zero, due to the scaling ambiguity for the invariant bilinear form. Its sign is a matter of convention, depending on whether we include the metric signature in the definition of the Laplace-Beltrami operator.}
\begin{equation}
    \Delta_{\Sigma} \Phi \left( \Omega,\theta \right) = \Phi \left( \Omega,\theta \right)\,.
\end{equation}
In other words, we are summing over waves on the covering space of the complex quotient cylinder. The instanton contributions are obtained by taking the pullback of these waves to the moduli space, i.e., evaluating $E_{z,l}^{*}\Phi\left( T_{\parallel} \right) = \Phi\left( \Omega\left( l,z,T_{\parallel} \right), \theta\left( z,T_{\parallel} \right) \right)$. Since the Laplace-Beltrami operator was constructed out of the invariant bilinear form, it respects the Coxeter transformations as isometries of the metric. This means that the transformed eigenfunctions satisfy the same wave equation with identical eigenvalues. Due to the linearity of the lattice evaluation, we can exchange the Laplace-Beltrami operator and the infinite sum,\footnote{This can be done formally, but also holds analytically for our region of interest. In the branch $\mathrm{Im}(\mathcal{A}(z,T_{\parallel})),\, \mathrm{Im}(\mathcal{B}(z,T_{\parallel})) > 0$ the evaluated terms $Q^{k}\Phi(\Omega,\theta) = \Phi(\Omega, \theta-2\pi k) = e^{-\Omega \cosh\left( \frac{u}{\pi} \left( \theta - 2\pi k \right) \right)}$ decay super-exponentially as $|k| \rightarrow \infty$. Through a Weierstrass M-test, this implies uniform and absolute convergence for compact subsets.} from which we conclude that
\begin{equation}
    \Delta_{\Sigma} \tilde{S}(\Omega,\theta) = \tilde{S}(\Omega,\theta)\,.
\end{equation}
Separating the multicover factor from the radial variable $\Omega_{1} \coloneqq \Omega/l$ the equation reads
\begin{equation}
    \Delta_{\Sigma} \tilde{S}(l\Omega_{1},\theta) = l^{2}\tilde{S}(l\Omega_{1},\theta)\,,
\end{equation}
where $\Delta_{\Sigma}$ is the same formal expression now written in the $(\Omega_{1},\theta)$ coordinates. From this we see that the higher-harmonics of a curve class contribution do indeed vary at shorter radial scales.

We have decided to perform the harmonic analysis in the complex quotient cylinder, since it leads to simpler expressions. It is also possible to compute the pullback of the metric to the moduli space first, which gives\footnote{Note that this metric has Lorentzian signature, while both the Weil–Petersson and Hodge metrics have Euclidean signature.}
\begin{equation}
    ds^{2} = 16\pi^{2}l^{2}\alpha_{z}\beta_{z} \left( \frac{m_{1}}{m_{2}} \left( dT^{1} \right)^{2} + m_{1}dT^{1}dT^{2} + \left( dT^{2} \right)^{2} \right)\,.
\label{eq:hyperbolic-metric-pullback}
\end{equation}
Using the expression \eqref{eq:hyperbolic-S-sum-Bessel}, one can then check that $S \left( l,z,T_{\parallel} \right)$ is an eigenfunction of the Laplace-Beltrami operator
\begin{equation}
    E_{z,l}^{*}\Delta_{\Sigma} = \frac{1}{4\pi^{2}l^{2}\alpha_{z}\beta_{z}m_{1}(m_{1}m_{2}-4)} \left( m_{2}\partial_{T^{1}}^{2} - m_{1}m_{2} \partial_{T^{1}}\partial_{T^{2}} + m_{1}\partial_{T^{2}}^{2} \right)\,.
\label{eq:hyperbolic-LB-pullback}
\end{equation}
The metric \eqref{eq:hyperbolic-metric-pullback} and, therefore, the Laplace-Beltrami operator \eqref{eq:hyperbolic-LB-pullback} can be directly obtained from the invariant bilinear form fixed by the complexified K\"ahler representation, which is more efficient if one only seeks this final result. Going through the analysis of the complex quotient geometry and realizing them as pullbacks instead, however, makes the geometric role of $\Omega \left( l,z,T_{\parallel} \right)$ and $\theta\left( z, T_{\parallel} \right)$ transparent and the origin of the Bessel functions clear, as we now explain.

Let us come back to the superposition of Gromov-Witten waves. The expression \eqref{eq:quotient-cylinder-waves-superposition} is a sum of eigenfunctions of the Laplace-Beltrami operator on the covering space of the complex quotient cylinder, but not of eigenfunctions on the cylinder itself, since the elementary kernel $\Phi(\Omega,\theta)$ is not $2\pi$-periodic in $\theta$ and does not descend to the quotient. The superposition as a whole is periodic and descends, as mentioned earlier. Hence, we can decompose it in terms of eigenfunctions of the Laplace-Beltrami operator on the complex quotient cylinder as
\begin{equation}
    S_{Q}(\Omega,\theta) = \sum_{m \in \mathbb{Z}} \phi_{m}(\Omega,\theta)\,.
\end{equation}
Taking into account the $\theta$-periodicity, we can separate the radial and angular parts of each eigenmode $\phi_{m}(\Omega,\theta)$ as
\begin{equation}
    \phi_{m}(\Omega,\theta) = R_{m}(\Omega) e^{im\theta}\,.
\end{equation}
Inserting this into the wave equation
\begin{equation}
    \Delta_{\Sigma} \phi_{m}(\Omega,\theta) = \phi_{m}(\Omega,\theta)\,,
\end{equation}
we find, for the radial piece,
\begin{equation}
    \Omega^{2}\frac{d^{2}R_{m}}{d\Omega^{2}} + \Omega \frac{dR_{m}}{d\Omega} - (\Omega^{2} + \nu_{m}^{2}) R_{m} = 0\,,\qquad \nu_{m} \coloneqq \frac{i \pi m}{u}\,.
\end{equation}
This is the modified Bessel equation, whose general solution is
\begin{equation}
    R_{m}(\Omega) = c_{m,1} I_{\nu_{m}}(\Omega) + c_{m,2} K_{\nu_{m}}(\Omega)\,,
\end{equation}
where $I_{\nu_{m}}(\Omega)$ and $K_{\nu_{m}}(\Omega)$ are the modified Bessel functions of the first and second kind, respectively. These have opposite asymptotic exponential behavior as $|\Omega| \rightarrow \infty$, which selects the modified Bessel function of the second kind as the physical solution, reflecting the fact that instanton corrections become exponentially suppressed at large volume. We thus get
\begin{equation}
    S_{Q}(\Omega,\theta) =  \sum_{m \in \mathbb{Z}} c_{m,2} K_{\frac{i \pi m}{u}}(\Omega) e^{im\theta}\,,
\end{equation}
which is the structure found in \eqref{eq:hyperbolic-S-sum-Bessel}. To determine that $c_{m,2} = 1/u$ exactly, one can compute the Fourier coefficients $R_{m}(\Omega)$ of the Fourier expansion on the $\theta$-circle, at which point the calculation is a shorter incarnation of the Mellin-Poisson analysis of \cref{sec:dihedral-hyperbolic-representations}. By setting $z = y(d)$, using the action of $S$ and re-inserting the $e^{2\pi i l \mathcal{C}}$ spectator prefactor that arises after untwisting, we recover the final result \eqref{eq:hyperbolic-infdih-Bessel-sum} for $\psi_{ld}^{\infdih}(T)$ with $d \in \mathcal{M}_{f}$. The structure of the result becomes immediately obvious once the wave equation for the Gromov-Witten expansion is used.

For the Euclidean case, one similarly obtains
\begin{equation}
    {\Omega'}^{2}\frac{d^{2}R_{r}}{d{\Omega'}^{2}} + \Omega' \frac{dR_{r}}{d\Omega'} + ({\Omega'}^{2} - \nu_{r}^{2}) R_{r} = 0\,,\qquad \nu_{r} \coloneqq \frac{\pi r}{v} = mr\,,
\end{equation}
with general solution
\begin{equation}
    R_{r}(\Omega') = c_{r,1} J_{\nu_{r}}(\Omega') + c_{r,2}Y_{\nu_{r}}(\Omega')\,,
\end{equation}
where $J_{\nu_{r}}(\Omega')$ and $Y_{\nu_{r}}(\Omega')$ are the ordinary Bessel functions of the first and second kind, respectively. Since the kernel that we lattice evaluate is analytic at the tip of the cone of the Euclidean orbifold, we need to select the ordinary Bessel functions of the first kind as the physical solution.

Note that, at the level of the local quotient geometry and Laplace-Beltrami operator, the hyperbolic and elliptic cases can be related to each other through the substitution $(u,\Omega,\theta) \mapsto (iv,i\Omega',-\delta)$. The same is not true of the resummed expressions \eqref{eq:hyperbolic-S-sum-Bessel} and \eqref{eq:elliptic-S-sum-Bessel}. The reason is that the physical orbit sums drop part of the Bessel equation solution space due to their asymptotic behavior or regularity conditions, as we have just seen. The aforementioned analytical continuation, however, changes which combination is selected by the boundary conditions, since these are not preserved under the continuation.

As we observed in \cref{sec:dihedral-Coxeter-prepotentials}, the two expressions for $\psi_{ld}^{\mathrm{I}_{2}(m)}(T)$ have complementary convergence properties. This fact is familiar from the Fourier analysis of waves, where a function that is localized in one choice of harmonic basis, such as a plane-wave expansion, may be highly delocalized in another separated basis, such as a spherical-wave expansion, and vice versa. The same phenomenon occurs for the superposition of Gromov-Witten waves in the instanton expansion. The raw orbit sum
\begin{equation}
    S\left(l,z,T_{\parallel}\right) = \sum_{k \in \mathbb{Z}} \Phi\left(\Omega\left(l,z,T_{\parallel}\right),\theta\left(z,T_{\parallel}\right) - 2\pi k\right)
\end{equation}
consists of a superposition of translates of a kernel along a circle direction. It is sharply localized at large volume around the leading instanton contribution and admits efficient numerical evaluation in that regime, while being flatly distributed in the interior of the moduli space. The spectral dual expression
\begin{equation}
    S\left(l,z,T_{\parallel}\right) = \frac{1}{u} \sum_{m \in \mathbb{Z}} K_{\frac{i \pi m}{u}}\left(\Omega\left(l,z,T_{\parallel}\right)\right) e^{im\theta\left(z,T_{\parallel}\right)}\,,
\end{equation}
is flatly distributed at large volume, but sharply localized around the leading Bessel-mode when the relevant K\"ahler moduli are small. Hence, the resummation is not merely a closed form in terms of variables adapted to the Coxeter symmetries, but a reorganization of the function adapted to a regime where the orbit sum representation converges poorly.

\subsection{The parabolic degeneration}
\label{sec:parabolic-degeneration}

While the hyperbolic and elliptic cases behave very similarly and can be treated in an analogous manner, the parabolic case requires a separate analysis. The former cases lead to $\mathrm{I}_{2}(m)$-invariant functions expressed in terms of Bessel functions of the ordinary or modified type, while the latter involves Jacobi theta functions instead. This can, once again, be understood naturally from an analysis of the relevant Laplace-Beltrami operator constructed from the Coxeter-invariant bilinear form.

First, consider the parabolic geometry determined by $\Sigma$, as defined in \eqref{eq:dihedral-invariant-bilinear-form}. We briefly discussed in \cref{sec:Coxeter-invariant-moduli-space-geometry} that it can be understood as a $u \rightarrow 0$ degeneration of the hyperbolic one. Interpreting $Q$ as a M\"obius transformation, the hyperbolic translation shrinks to zero length and the two fixed points on $\partial\mathbb{H}$ coalesce into one. From the point of view of the $V_{0}$ plane geometry, the two null directions fixed by $Q$ and spanning $C^{+}$ degenerate into a cusp direction and the metric $\Sigma$ degenerates. This prevents us from defining a two-dimensional Laplace-Beltrami operator in the way we did above. At the level of the orbit sum, the logarithmically spaced lattice of the hyperbolic case is deformed into an additive lattice, as is appropriate for Jacobi theta functions.

The parabolic representation of $\infdih$, however, is the non-split extension in \eqref{eq:dihedral-rep-short-exact-sequence}. This means that untwisted coordinates that would allow us to reduce the relevant dynamics to two dimensions do not exist. Computing the invariant symmetric bilinear form directly for the entire parabolic representation acting on $V = V_{0} \oplus V_{\perp}$ we find the block structure
\begin{equation}
    \Gamma =
    \begin{pNiceArray}{c|l}[cell-space-limits=5pt]
		A  & B\\
		\hline
		B^{T} & C
	\end{pNiceArray}
    \quad \text{such that} \quad \mathcal{M}_{i}^{T} \Gamma \mathcal{M}_{i} = \Gamma\,,\quad \forall i \in \{1,2\}\,,
\end{equation}
defined up to an overall scale. The condition implies that $A \propto \Sigma$, while the block structure of the generators imposes no constraint on $C = C^{T}$, which we choose to be the identity. Solving the constraint for $B$ we arrive at
\begin{equation}
    \Gamma =
    \begin{pNiceArray}{c|c}[cell-space-limits=5pt]
		\varepsilon \Sigma  & \varepsilon \begin{pmatrix} -\frac{m_{2}}{2}u'_{1}\\[2.5pt] -\frac{m_{1}}{2}u'_{2} \end{pmatrix}\\
		\hline
		\varepsilon \begin{pmatrix} -\frac{m_{2}}{2}u'_{1}\\[2.5pt] -\frac{m_{1}}{2}u'_{2} \end{pmatrix}^{T} & \mathds{1}_{h-2}
	\end{pNiceArray}
    \,,
\end{equation}
where $\varepsilon$ is a relative scale between the choice of $C=\mathds{1}_{h-2}$ and the other blocks. Letting $B_{\varepsilon} \coloneqq \varepsilon^{-1}B$, the determinant is given by
\begin{equation}
    \det(\Gamma) = \det\left(\varepsilon \Sigma - \varepsilon^{2} B_{\varepsilon}B_{\varepsilon}^{T}\right)\,.
\end{equation}
One can check that for the cusp direction generator $\delta_{1}$ we have
\begin{equation}
    \Sigma \cdot \delta_{1} = 0\qquad \text{and}\qquad \delta_{1}^{T} B_{\varepsilon} = -2 \left( u'_{1} + (m_{1}/2) u'_{2} \right)\,.
\end{equation}
The latter expression is the same as \eqref{eq:parabolic-unipotent-index-2} and hence never vanishes due to the non-split nature of the representation. Then, using
\begin{equation}
    \det\left(\varepsilon \Sigma - \varepsilon^{2} B_{\varepsilon}B_{\varepsilon}^{T}\right) = -\varepsilon^{3} \mathrm{tr}\left(\mathrm{adj}(\Sigma)B_{\varepsilon}B_{\varepsilon}^{T}\right) + \mathcal{O}\left(\varepsilon^{4}\right) = - \frac{\varepsilon^{3}}{m_{1}} \left\lVert B_{\varepsilon}^{T} \delta_{1} \right\rVert^{2} + \mathcal{O}(\varepsilon^{4})\,,
\end{equation}
we see that the invariant symmetric bilinear form $\Gamma$ is non-degenerate for a generic choice of $\varepsilon$. Generically, we can choose $\varepsilon = 1$.\footnote{An example where this would not be possible would be a model with $m_{1} = m_{2} = 2$, $u'_{1} = (2,0,\dotsc)$ and $u'_{2} = (0,2,\dotsc)$, which is fine-tuned to make the choice $\varepsilon = 1$ degenerate due to a cancellation with the $\mathcal{O}(\varepsilon^{4})$ terms. Such a cancellation occurs for a total of four parabolic infinite dihedral models in the database, namely the CICYs 7740, 7764, 7828 and 7864. Making the alternative choice $\varepsilon = 2$ is sufficient to obtain a non-degenerate $\Gamma$ for these models.} Moreover, a choice of $\varepsilon$ such that the signature of $\Gamma$ is $(h-1,1)$ is always possible. In any case, the concrete choice of $\varepsilon$ made will not change the rest of our analysis beyond normalization factors.

With the non-degenerate invariant symmetric bilinear form at hand, we can construct a Laplace-Beltrami operator and discuss the harmonic analysis of parabolic models. First, let us note that although untwisting to a two-dimensional geometric problem is not possible, the relevant parabolic dynamics occurs for a three-dimensional subspace $V_{3} \subset V$. This is related to the fact that $\mathcal{Q}$ is unipotent of index 3. Indeed, choose
\begin{equation}
    \overline{n} \in V\quad \text{such that}\quad \mathcal{N}^{2}\overline{n} \neq 0
\end{equation}
and define a Jordan chain of length 3
\begin{equation}
\begin{tikzcd}
    \overline{n} \arrow[r] \arrow[r, "\mathcal{N}"] & \overline{s} \coloneqq \mathcal{N}\overline{n} \arrow[r, "\mathcal{N}"] & \overline{\delta} \coloneqq \mathcal{N}^{2}\overline{n} \arrow[r, "\mathcal{N}"] & 0\,.
\end{tikzcd}
\end{equation}
Here $\overline{\delta}$ is a generator of the cusp direction $\delta$. Define the $\mathcal{Q}$-invariant span
\begin{equation}
    V_{3} \coloneqq \left\langle \overline{n}, \overline{s}, \overline{\delta} \right\rangle_{\mathbb{R}}\,.
\end{equation}
Since the condition selecting $\overline{n}$ is unaltered by rescalings and shifts $\overline{n} \mapsto \overline{n} + \lambda \overline{\delta}$, we can choose a shifted $\breve{n}$ such that $\langle \breve{n}, \breve{n} \rangle = 0$. Moreover, defining the shifted vector $\breve{s} \coloneqq \overline{s} - \overline{\delta}/2$ and possibly rescaling the Jordan chain, the generators correspond to a null frame for the Lorentzian restriction of $\Gamma$, since
\begin{equation}
    \big\langle \overline{\delta}, \overline{\delta} \big\rangle = \big\langle \breve{n}, \breve{n} \big\rangle = \big\langle \breve{s}, \overline{\delta} \big\rangle = \big\langle \breve{s}, \breve{n} \big\rangle =  0\,,\qquad \big\langle \overline{\delta}, \breve{n} \big\rangle = 1\,,\qquad \big\langle \breve{s}, \breve{s} \big\rangle = -1\,.
\end{equation}
The subspace $V_{3}$ is also $\mathcal{S}$-invariant, which can be clearly seen by starting the construction with an $\overline{n}$ such that $\left. \overline{n} \right|_{V_{0}} = 0$. The direction $\langle \overline{n} \rangle_{\mathbb{R}}$ corresponds to a choice of \mbox{rank-one} subspace of $V_{\perp}$ generating the length-three Jordan block. In parabolic models with $h^{1,1}(X) > 3$, the quotient $V/V_{3}$ is acted on trivially by the group; equivalently, after shifting a complement of $\langle \overline{n}\rangle_{\mathbb{R}}$ in $V_{\perp}$ by suitable $V_{0}$-components, one can choose a \mbox{$\Gamma$-orthogonal} spectator subspace $V_{\mathrm{sp}} \subset V$ on which the group acts trivially, meaning that we can regard $V$ as $V = V_{3} \oplus V_{\mathrm{sp}}$ and perform the harmonic analysis focusing on $V_{3}$. Similarly to the spectator directions in the hyperbolic and elliptic cases, $V_{\mathrm{sp}}$ will contribute an additive Laplace-Beltrami factor from a flat geometric piece; all interesting features are captured by its complement.

Using the above null frame for $V_{3}$, any vector can be uniquely written as
\begin{equation}
    \vec{x} = x^{+} \breve{n} + x \breve{s} + x^{-}\overline{\delta} \in V_{3}\,.
\end{equation}
In these coordinates, the induced metric $\left. \Gamma \right|_{V_{3}}$ is
\begin{equation}
    ds^{2} = 2dx^{+}dx^{-} - dx^{2}
\end{equation}
and the action of $\mathcal{Q}^{k}$ shifts them by
\begin{equation}
    \mathcal{Q}^{k}: (x^{+},x,x^{-}) \longmapsto \left( x^{+}, x + kx^{+}, x^{-} + kx + \frac{k^{2}}{2}x^{+} \right)\,.
\end{equation}
This transformation can be untwisted in $V_{3}^{+} \coloneqq \{ x \in V_{3} \mid x^{+} > 0 \}$ through a quadratic change of coordinates
\begin{equation}
    u \coloneqq x^{+}\,,\qquad v \coloneqq \frac{x}{x^{+}}\,,\qquad w \coloneqq x^{-} - \frac{x^{2}}{2x^{+}}\,,\qquad \text{for}\quad x^{+} > 0\,,
\end{equation}
which leads to the simpler $\mathcal{Q}$-action
\begin{equation}
    \mathcal{Q}: (u,v,w) \longmapsto (u,v+1,w)
\label{eq:parabolic-Q-action-uvw}
\end{equation}
and to $\mathcal{S}$ acting as a (possibly affine) reflection with displacement $v_{*}$
\begin{equation}
    \mathcal{S}: (u,v,w) \mapsto (u,v_{*} - v,w)\,.
\end{equation}
Hence, the quotient by the $\mathcal{Q}$-action leads to
\begin{equation}
    V_{3}^{+}/\langle \mathcal{Q} \rangle \cong \mathbb{R}_{>0} \times S^{1} \times \mathbb{R}
\label{eq:parabolic-real-quotient}
\end{equation}
and, further taking the quotient by the $\mathcal{S}$ action, to
\begin{equation}
    V_{3}^{+}/\langle \mathcal{Q}, \mathcal{S} \rangle \cong \mathbb{R}_{>0} \times S^{1}/\mathbb{Z}_{2} \times \mathbb{R}\,.
\end{equation}

In the untwisted coordinates, the metric $\left. \Gamma \right|_{V_{3}^{+}}$ can be written as
\begin{equation}
    ds^{2} = 2du\, dw - u^{2} dv^{2}\,,
\end{equation}
from which we obtain the Laplace-Beltrami operator
\begin{equation}
    \Delta_{\Gamma}^{R} = 2 \partial_{u} \partial_{w} + \frac{1}{u}\partial_{w} - \frac{1}{u^{2}}\partial_{v}^{2}\,.
\label{eq:parabolic-uvw-LB-operator}
\end{equation}
The above operator is defined for the real quotient geometry natural in the geometric analysis of the Mori representation. To relate it to the Laplace-Beltrami operator relevant to our problem, we need to work on a complex version of the quotient geometry and its covering space, as we did for the hyperbolic and elliptic examples in \cref{sec:Coxeter-invariant-moduli-space-geometry}.

First, let us relate the real quotient geometry to the quantities entering the parabolic orbit sum $S \left( l,z,d_{\perp},T_{\parallel} \right)$. These are $z_{\vartheta}\left( l,z,d_{\perp}, T_{\parallel} \right)$ and $\tau\left( l,d_{\perp}, T_{\parallel} \right)$, defined in \eqref{eq:parabolic-z-tau-definition}. From \eqref{eq:parabolic-Qkd-QkSd} note that
\begin{equation}
    \mathcal{Q}^{\mathrm{aff}}z \coloneqq (\mathcal{Q}(z,d_{\perp}))_{\parallel} = z + \alpha(z, d_{\perp})\,,\qquad \alpha(z, d_{\perp}) = Nz + Ud_{\perp}\,,
\end{equation}
and
\begin{equation}
    N\alpha(z ,d_{\perp}) = \beta(d_{\perp})\,.
\end{equation}
Since $\mathrm{Im}(N)$ is the cusp direction, every $Nz$ is proportional to $\beta(d_{\perp})$. Thus, for $\beta(d_{\perp}) \neq 0$ we can define a unique real number $v(z,d_{\perp})$ such that
\begin{equation}
    Nz = v(z,d_{\perp}) \beta(d_{\perp})\,.
\label{eq:v-z-relation}
\end{equation}
Then the relation
\begin{equation}
    N\big(\mathcal{Q}^{\mathrm{aff}}z\big) = (v(z,d_{\perp})+1)\beta(d_{\perp})
\end{equation}
enforces the transformation
\begin{equation}
    v\big(\mathcal{Q}^{\mathrm{aff}}z,d_{\perp}\big) = v(z,d_{\perp}) + 1\,.
\end{equation}
It therefore follows that
\begin{equation}
    z_{\vartheta}\left( l,z,d_{\perp}, T_{\parallel} \right) = z_{0}\left( l,d_{\perp}, T_{\parallel} \right) + \tau\left( l,d_{\perp}, T_{\parallel} \right) v(z,d_{\perp})
\end{equation}
with
\begin{equation}
    z_{0}\left( l,d_{\perp},T_{\parallel} \right) \coloneqq l \left\langle Ud_{\perp} - \frac{\beta(d_{\perp})}{2}, T_{\parallel} \right\rangle\,.
\end{equation}
Here $z_{0}\left( l,d_{\perp},T_{\parallel} \right)$ is the part of $z_{\vartheta}\left( l,z,d_{\perp}, T_{\parallel} \right)$ that does not transform under $\mathcal{Q}$, since it only depends on $d_{\perp}$ and $T_{\parallel}$. For the upcoming geometric discussion, it will be useful to define 
\begin{equation}
    \zeta_{3}\left( l,z,d_{\perp}, T_{\parallel} \right) \coloneqq z_{\vartheta}\left( l,z,d_{\perp}, T_{\parallel} \right) - z_{0}\left( l,d_{\perp},T_{\parallel} \right)\,,
\end{equation}
where the invariant part has been removed. While $z_{\vartheta}\left( l,z,d_{\perp}, T_{\parallel} \right)$ is the physical \mbox{object} appearing as an argument of the Jacobi theta function, $\zeta_{3}\left( l,z,d_{\perp}, T_{\parallel} \right)$ is better adapted to the present parabolic quotient geometry discussion. Given the transformation property \mbox{$Q: (z_{\vartheta}, \tau) \mapsto (z_{\vartheta} + \tau, \tau)$} and the fact that both $\tau$ and $u$ control the quadratic growth in the $\mathcal{Q}$-orbit, it is natural to identify
\begin{equation}
    \tau\left( l,d_{\perp}, T_{\parallel} \right) = \kappa \left( l,T_{\parallel} \right) u (d_{\perp})\,,\qquad \kappa \left( l,T_{\parallel} \right) \coloneqq l \big\langle \overline{\delta}, T_{\parallel} \big\rangle\,.
\end{equation}
Since both $\beta(d_{\perp})$ and $\overline{\delta}$ lie on the cusp direction, $u(d_{\perp})$ acts as the unique scalar such that $\beta(d_{\perp}) = u(d_{\perp})\overline{\delta}$.

To conclude the identification between the real quotient geometry variables $(u,v,w)$ and the quantities appearing in the resummed expression
\begin{equation}
    e^{2\pi il \left\langle d, T \right\rangle} S \left( z,d_{\perp},T_{\parallel} \right) = e^{2\pi il \left\langle d, T \right\rangle} \vartheta\left( z_{\vartheta}, \tau \right)\,,
\end{equation}
we still need to determine the role of $w$. Having argued above that $(u,v)$ are related to the Jacobi theta arguments $(z_{\vartheta},\tau)$, we can imagine $w$ to be related in some fashion to
\begin{equation}
    s(l,d,T) \coloneqq l\langle d, T \rangle\,.
\end{equation}
Following the split $V = V_{3} \oplus V_{\mathrm{sp}}$, it is natural to similarly consider
\begin{equation}
    d = d_{3} + d_{\mathrm{sp}}\,,\qquad T = T_{3} + T_{\mathrm{sp}}\,,\qquad s_{3}(l,d_{3},T_{3}) \coloneqq l\langle d_{3},T_{3} \rangle\,.
\end{equation}
This splits the prefactor into $e^{2\pi i s} = e^{2\pi i s_{3}} e^{2\pi i l \langle d_{\mathrm{sp}}, T_{\mathrm{sp}} \rangle}$, with the second piece corresponding to the spectator plane waves associated to the Laplace-Beltrami operator from $V_{\mathrm{sp}}$, which we can ignore in the following. The $e^{2\pi i s_{3}}$ factor transforms non-trivially under $\mathcal{Q}$ and cancels the automorphy factor picked up by $\vartheta\left( z_{\vartheta}, \tau \right)$ under the same transformation, see the discussion in \cref{sec:dihedral-parabolic-representations}. In the $(u,v,w)$ coordinates, the combination
\begin{equation}
    \sigma_{3}\left(l,z,d_{\perp},T_{\parallel}\right) = \kappa \left(l,T_{\parallel}\right) x^{-}(z,d_{\perp}) = \kappa \left(l,T_{\parallel}\right) \left( w(z,d_{\perp}) + \frac{1}{2}u(d_{\perp})v(z,d_{\perp})^{2} \right)
\end{equation}
transforms as
\begin{equation}
    \mathcal{Q}: \sigma_{3}\left(l,z,d_{\perp},T_{\parallel}\right) \longmapsto \sigma_{3}\left(l,z,d_{\perp},T_{\parallel}\right) + \zeta_{3}\left( l,z,d_{\perp}, T_{\parallel} \right) + \frac{1}{2}\tau\left( l,d_{\perp}, T_{\parallel} \right)\,,
\end{equation}
the same transformation as for the exponent of the aforementioned factor up to a constant negative shift involving $z_{0}\left( l,d_{\perp},T_{\parallel} \right)$.

As in the hyperbolic and elliptic cases, we identified a $T_{\parallel}$-dependent $\mathcal{Q}$-equivariant embedding of the real quotient geometry that appears naturally in the Mori representation into the complex quotient geometry in which the harmonic analysis will take place. Indeed, consider the complex covering space $V_{3}^{\mathbb{C}} \coloneqq V_{3} \otimes \mathbb{C}$ with coordinates $(\tau,\zeta_{3},\sigma_{3})$ and equipped with the metric
\begin{equation}
    ds^{2} = 2 d\tau\, d\sigma_{3} - d\zeta_{3}^{2}\,.
\label{eq:parabolic-complex-metric}
\end{equation}
This metric is preserved by the affine $\mathcal{Q}$-action
\begin{equation}
    \mathcal{Q}: (\tau,\zeta_{3},\sigma_{3}) \longmapsto \left( \tau, \zeta_{3} + \tau, \sigma_{3} + \zeta_{3} + \frac{1}{2}\tau \right)\,,
\label{eq:parabolic-Q-action-tauzetasigma}
\end{equation}
by which we can quotient to obtain $V_{3}^{\mathbb{C}}/\langle \mathcal{Q} \rangle$, the complex analogue of \eqref{eq:parabolic-real-quotient}. The equivalent of \eqref{eq:complex-quotient-embedding-Mori}, i.e., the embedding of the real quotient geometry into the complex one, can be achieved through
\begin{equation}
    \begin{split}
        F_{T_{\parallel}}: V_{3}^{+} &\longrightarrow V_{3}^{\mathbb{C}}\\
        (u,v,w) &\longmapsto \left( \kappa\left( l,T_{\parallel} \right) u, \kappa\left( l,T_{\parallel} \right) u v, \kappa\left( l,T_{\parallel} \right) \left( w + \frac{1}{2}uv^{2} \right) \right)\,.
    \end{split}
\end{equation}
Indeed, this map makes \eqref{eq:parabolic-Q-action-uvw} and \eqref{eq:parabolic-Q-action-tauzetasigma} compatible while satisfying
\begin{equation}
    F_{T_{\parallel}}^{*} \left( 2 d\tau\, d\sigma_{3} - d\zeta_{3}^{2} \right) = \kappa\left( l,T_{\parallel} \right)^{2} \left( 2du\, dw - u^{2} dv^{2} \right)\,.
\end{equation}
The resulting Laplace-Beltrami operator reads
\begin{equation}
    \Delta_{\Gamma} = 2 \partial_{\tau} \partial_{\sigma_{3}} - \partial_{\zeta_{3}}^{2}\,.
\end{equation}

With the relevant geometry at hand, we can now move to the harmonic analysis of the Gromov-Witten expansion using $\Delta_{\Gamma}$. The parabolic orbit sums correspond to an additive lattice evaluation of the continuous kernel
\begin{equation}
    f\left( l,z,d_{\perp},T_{\parallel} \,\middle|\, x \right) \coloneqq e^{2\pi i s_{3}(l,d_{3},T_{3})} e^{\pi i x^{2} \tau \left( l,d_{\perp}, T_{\parallel} \right) + 2\pi i x z_{\vartheta} \left( l,z,d_{\perp}, T_{\parallel} \right)}\,,\quad x \in \mathbb{R}\,.
\end{equation}
In the coordinates of the covering space of the complex quotient geometry, the natural continuous kernel is
\begin{equation}
    \Phi_{x}(\tau,\zeta_{3},\sigma_{3}) \coloneqq e^{2\pi i \sigma_{3}} e^{\pi i x^{2} \tau + 2\pi i x \zeta_{3}}\,,\quad x \in \mathbb{R}\,.
\end{equation}
Using the elementary kernel
\begin{equation}
    \Phi(\tau,\zeta_{3},\sigma_{3}) \coloneqq e^{2\pi i \sigma_{3}}\,,
\end{equation}
the lattice evaluation corresponds to summing over the $\mathcal{Q}$-action, i.e.,
\begin{equation}
    \tilde{\Theta}(\tau,\zeta_{3},\sigma_{3}) \coloneqq \sum_{k \in \mathbb{Z}} \Phi_{k}(\tau,\zeta_{3},\sigma_{3}) = \sum_{k \in \mathbb{Z}} \mathcal{Q}^{k} \Phi(\tau,\zeta_{3},\sigma_{3}) = e^{2\pi i \sigma_{3}} \vartheta(\zeta_{3},\tau)\,.
\label{eq:parabolic-lattice-sampling}
\end{equation}

To connect this geometrically natural object to the orbit sums appearing in \cref{sec:dihedral-parabolic-representations}, we need to perform a twisted pullback to the moduli space, as occurred for the hyperbolic and elliptic cases. Consider the $\mathcal{Q}$-equivariant map
\begin{equation}
    \begin{split}
        A_{a}: V_{3}^{\mathbb{C}} &\longrightarrow V_{3}^{\mathbb{C}}\\
        (\tau,\zeta_{3},\sigma_{3}) &\longmapsto \left( \tau,\zeta_{3} + a, \sigma_{3} + a\frac{\zeta_{3}}{\tau} \right)\,,
    \end{split}
    \qquad a \in \mathbb{C}\,,\quad \tau \neq 0\,.
\end{equation}
One can check that it is not an isometry of the metric \eqref{eq:parabolic-complex-metric}, meaning that the Laplace-Beltrami operator $\Delta_{\Gamma}$ picks up additional factors under the pullback. Nonetheless, since $A_{a} \circ \mathcal{Q} = \mathcal{Q} \circ A_{a}$ we can use it to produce new $\mathcal{Q}$-invariant representatives
\begin{equation}
    \tilde{\Theta}_{a}(\tau,\zeta_{3},\sigma_{3}) \coloneqq A_{a}^{*} \tilde{\Theta}(\tau,\zeta_{3},\sigma_{3}) = e^{2\pi i \left( \sigma_{3} + a\frac{\zeta_{3}}{\tau} \right)} \vartheta(\zeta_{3} + a, \tau)\,.
\end{equation}

To understand the role of the $A_{a}$-twist, recall that the Jacobi theta function is a section of a line bundle on the quotient geometry, cf.\ the discussion of automorphy factors at the end of \cref{sec:dihedral-parabolic-representations}. Indeed, considering the space
\begin{equation}
    \mathcal{B} \coloneqq \left\{ (\tau,\zeta_{3}) \in \mathbb{C}^{2} \,\middle|\, \mathrm{Im}(\tau) > 0 \right\}
\end{equation}
and the group action $\mathcal{Q}_{0} \coloneqq (\tau,\zeta_{3}) \rightarrow (\tau,\zeta_{3}+\tau)$, the Jacobi theta function $\vartheta(\zeta_{3},\tau)$ is not a function of $\mathcal{B}/\langle \mathcal{Q}_{0} \rangle$, but a section of the line bundle $\mathcal{L}_{\vartheta}$ over $\mathcal{B}/\langle \mathcal{Q}_{0} \rangle$ defined by the automorphy factor
\begin{equation}
    j_{\mathcal{Q}}(\tau,\zeta_{3}) \coloneqq e^{-\pi i \tau - 2\pi i \zeta_{3}}\,,
\end{equation}
in other words,
\begin{equation}
    \mathcal{L}_{\vartheta} \coloneqq \left( \mathcal{B} \times \mathbb{C} \right)/\sim\,,\qquad (\tau,\zeta_{3}+\tau,\upsilon) \sim (\tau,\zeta_{3},j_{\mathcal{Q}}(\tau,\zeta_{3})^{-1}\upsilon)\,.
\end{equation}
The enlarged space $V_{3}^{\mathbb{C}}$ with coordinates $(\tau,\zeta_{3},\sigma_{3})$ and with the $\mathcal{Q}$-action \eqref{eq:parabolic-Q-action-tauzetasigma} was chosen such that $\eta_{3} \coloneqq e^{2\pi i \sigma_{3}}$ is a trivializing coordinate for the pullback of $\mathcal{L}_{\vartheta}^{-1}$ to the covering space, making
\begin{equation}
    \Theta(\tau,\zeta_{3},\sigma_{3}) = \eta_{3} \vartheta(\zeta_{3},\tau) = e^{2\pi i \sigma_{3}} \vartheta(\zeta_{3},\tau)
\end{equation}
an honest $\mathcal{Q}$-invariant function on $V_{3}^{\mathbb{C}}$. Shifting the Jacobi theta argument to $\vartheta(\zeta_{3} + a,\tau)$, with $a \in \mathbb{C}$, we obtain a section of $\mathcal{L}_{\vartheta} \otimes \mathcal{L}_{a}$ instead, where $\mathcal{L}_{a}$ is a flat line bundle over $\mathcal{B}/\langle \mathcal{Q}_{0} \rangle$ with holonomy $e^{-2\pi i a}$. Hence, $\eta_{3}$ must be twisted in the opposite direction to become a trivializing coordinate of the pullback of $\mathcal{L}_{\vartheta}^{-1} \otimes \mathcal{L}_{a}^{-1}$ to the covering space. This defines the $A_{a}$-twist map used above. Once the twisted bundle $\mathcal{L}_{\vartheta} \otimes \mathcal{L}_{a}$ has been chosen, there still remains the freedom to choose the trivialization of the bundle. This corresponds to the vertical translations
\begin{equation}
    \begin{split}
        B_{\chi}: V_{3}^{\mathbb{C}} &\longrightarrow V_{3}^{\mathbb{C}}\\
        (\tau,\zeta_{3},\sigma_{3}) &\longmapsto (\tau,\zeta_{3},\sigma_{3} + \chi)\,,
    \end{split}
\end{equation}
which are both $\mathcal{Q}$-equivariant and isometries of the metric \eqref{eq:parabolic-complex-metric}. Their effect is to multiply $\eta_{3} = e^{2\pi i \sigma_{3}}$ by a nowhere-vanishing factor $e^{2\pi i \chi}$. Defining $A_{a,\chi} \coloneqq B_{\chi} \circ A_{a}$, the freedom in choosing the $\mathcal{Q}$-invariant representative results in
\begin{equation}
    \tilde{\theta}_{a,\chi}(\tau,\zeta_{3},\sigma_{3}) \coloneqq A_{a,\chi}^{*}\tilde{\Theta}(\tau,\zeta_{3},\sigma_{3}) = e^{2\pi i \left( \sigma_{3} + a\frac{\zeta_{3}}{\tau} + \chi \right)} \vartheta(\zeta_{3} + a, \tau)\,.
\end{equation}
The evaluated coordinates $\sigma_{3}\left( l,z,d_{\perp},T_{\parallel} \right)$ and $\zeta_{3}\left( l,z,d_{\perp},T_{\parallel} \right)$, natural in the complex quotient geometry, differ from the \cref{sec:dihedral-parabolic-representations} arguments $s_{3}\left( l,d_{3},T_{3} \right)$ and $z_{\vartheta}\left( l,z,d_{\perp},T_{\parallel} \right)$ by the $\mathcal{Q}$-independent shift $z_{0}\left( l,d_{\perp},T_{\parallel} \right)$ related to the line bundle twist and a choice of normalization associated with its trivialization.

To directly connect to the orbit sums of \cref{sec:dihedral-parabolic-representations}, define the map
\begin{equation}
    \begin{split}
        \mathscr{A}: V_{3}^{\mathbb{C}} \times \mathbb{C} \times \mathbb{C} &\longrightarrow V_{3}^{\mathbb{C}}\\
        (\tau,\zeta_{3},\sigma_{3};a,\chi) &\longmapsto \left( \tau,\zeta_{3} + a, \sigma_{3} + a\frac{\zeta_{3}}{\tau} + \chi \right)\,,
    \end{split}
    \qquad \tau \neq 0\,,
\end{equation}
which for fixed $a, \chi \in \mathbb{C}$ reduces to the $A_{a,\chi}$-twist. Consider then the universal twisted family of functions
\begin{equation}
    \Theta(\tau,\zeta_{3},\sigma_{3};a,\chi) \coloneqq \mathscr{A}^{*}\tilde{\Theta}(\tau,\zeta_{3},\sigma_{3};a,\chi) = e^{2\pi i \left( \sigma_{3} + a\frac{\zeta_{3}}{\tau} + \chi \right)} \vartheta(\zeta_{3} + a, \tau)\,.
\end{equation}
Due to the $\mathcal{Q}$-equivariant nature of each $A_{a,\chi}$-twist, $\Theta(\tau,\zeta_{3},\sigma_{3};a,\chi)$ is $\mathcal{Q}$-invariant and therefore descends to a family on $V_{3}^{\mathbb{C}}/\langle \mathcal{Q} \rangle \times \mathbb{C} \times \mathbb{C}$. At fixed curve class and harmonic, define the evaluation map
\begin{equation}
    \begin{split}
        E_{z,d_{\perp},l}: \mathrm{int}\left( \hat{\pi}_{3}\left( \mathcal{K} \right) \right) &\longrightarrow V_{3}^{\mathbb{C}}\\
        T_{3} &\longmapsto \left( \tau\left(l,d_{\perp},T_{\parallel}\right), \zeta_{3}\left( l,z,d_{\perp},T_{\parallel} \right), \sigma_{3}\left( l,z,d_{\perp},T_{\parallel} \right) \right)\,,
    \end{split}
\end{equation}
where $\hat{\pi}_{3}: \hat{V}^{\mathbb{C}} \rightarrow \hat{V}_{3}^{\mathbb{C}}$ implements the moduli space split used in this section. To take into account the line bundle twist freedom, let us define an evaluation map to the domain of the universal twisted family, i.e.,
\begin{equation}
    \begin{split}
        \mathscr{E}_{z,d_{\perp},l}: \mathrm{int}\left( \hat{\pi}_{3}\left( \mathcal{K} \right) \right) &\longrightarrow V_{3}^{\mathbb{C}} \times \mathbb{C} \times \mathbb{C}\\
        T_{3} &\longmapsto \left( E_{z,d_{\perp},l}(T_{3}); z_{0}\left( l,d_{\perp},T_{\parallel} \right), \chi(l,z,d_{\perp},T_{3}) \right)\,.
    \end{split}
\end{equation}
The correct trivialization is fixed by
\begin{equation}
    \chi(l,z,d_{\perp},T_{3}) = s_{3}\left(l,d_{3},T_{3}\right) - \sigma_{3}\left( l,z,d_{\perp},T_{\parallel} \right) - z_{0}(l,d_{\perp},T_{\parallel}) \frac{\zeta_{3}\left( l,z,d_{\perp},T_{\parallel} \right)}{\tau\left( l,d_{\perp},T_{\parallel} \right)}\,.
\end{equation}
The orbit sums relevant for $\psi_{ld}^{\infdih}(T)$ are, in view of the above considerations, the pullback of the universal twisted family
\begin{equation}
    e^{2\pi i s_{3}\left(l,d_{3},T_{3}\right)} S\left( l,z,d_{\perp},T_{\parallel} \right) = \mathscr{E}_{z,d_{\perp},l}^{*}\Theta\left(T_{3}\right)\,.
\end{equation}

In the parabolic case, the lattice sum \eqref{eq:parabolic-lattice-sampling} directly yields the Jacobi theta function. Still, we can motivate the result from a harmonic analysis point of view by noting that the elementary kernel satisfies the harmonic equation
\begin{equation}
    \Delta_{\Gamma}\Phi(\tau,\zeta_{3},\sigma_{3}) = 0\,,
\end{equation}
which also holds for the continuous kernel
\begin{equation}
    \Delta_{\Gamma}\Phi_{x}(\tau,\zeta_{3},\sigma_{3}) = 0\,.
\end{equation}
Hence, all $\mathcal{Q}$-images of the elementary kernel solve the same harmonic equation and, due to the linearity of the lattice evaluation, we find that
\begin{equation}
    \Delta_{\Gamma}\tilde{\Theta}(\tau,\zeta_{3},\sigma_{3}) = 0\,.
\end{equation}
Separating the $\sigma_{3}$-dependent overall factor
\begin{equation}
    \tilde{\Theta}(\tau,\zeta_{3},\sigma_{3}) = e^{2\pi i \sigma_{3}} \phi(\zeta_{3},\tau)
\end{equation}
the harmonic equation reduces it to the heat equation in one spatial dimension
\begin{equation}
    \partial_{\tau}\phi(\zeta_{3},\tau) = \frac{1}{4\pi i} \partial^{2}_{\zeta_{3}} \phi(\zeta_{3},\tau).
\end{equation}
The standard heat kernel on a circle is the Jacobi theta function, which can  then be interpreted as a superposition of Fourier modes on a circular spatial dimension.

As for the hyperbolic and elliptic cases, it is possible to first compute the pullback of the metric to the moduli space, obtaining
\begin{equation}
    ds^{2} = 2l^{2} \left\langle \beta(d_{\perp}),dT_{\parallel} \right\rangle \left\langle d_{3}, dT_{3} \right\rangle - l^{2} \left\langle \alpha(z,d_{\perp}) - \frac{1}{2}\beta(d_{\perp}), dT_{\parallel} \right\rangle^{2}\,.
\end{equation}
Acting on $e^{2\pi i s_{3}\left(l,d_{3},T_{3}\right)} S\left( l,z,d_{\perp},T_{\parallel} \right)$ with the Laplace-Beltrami operator\footnote{Note that, although the third component of $d_{3} \in V_{3}$ appears in the denominator, it is non-vanishing for curve classes not stabilized by the entire infinite dihedral group, which is the context of the discussion.}
\begin{equation}
    \begin{split}
        \mathscr{E}_{z,d_{\perp},l}^{*}\mathscr{A}^{*}\Delta_{\Gamma} &= \frac{2}{l^{2}d_{3,3}(\beta_{1}\alpha_{2}-\beta_{2}\alpha_{1})} \left( \alpha_{2} \left( \partial_{T^{1}} - \frac{d_{3,1}}{d_{3,3}}\partial_{T^{3}} \right) - \alpha_{1} \left( \partial_{T^{2}} - \frac{d_{3,2}}{d_{3,3}} \partial_{T^{3}} \right) \right) \partial_{T^{3}}\\
        &\quad -\frac{1}{l^{2}(\beta_{1}\alpha_{2}-\beta_{2}\alpha_{1})^{2}} \left( -\beta_{2} \left( \partial_{T^{1}} - \frac{d_{3,1}}{d_{3,3}}\partial_{T^{3}} \right) + \beta_{1} \left( \partial_{T^{2}} - \frac{d_{3,2}}{d_{3,3}} \partial_{T^{3}} \right) \right)^{2}\,,
    \end{split}
\end{equation}
one observes that the harmonic equation is indeed satisfied. Here we have used $\{\alpha_{1},\alpha_{2}\}$ and $\{\beta_{1},\beta_{2}\}$ to refer to the coordinates of $\alpha(z,d_{\perp}) - \beta(d_{\perp})/2$ and $\beta(d_{\perp})$, respectively.

%% file: sections/general-Coxeter-prepotentials.tex
\section{General Coxeter prepotentials}
\label{sec:general-Coxeter-prepotentials}

From the CICY Coxeter Database discussed in \cref{sec:CICY-Coxeter-Database}, we saw that the most prevalent Coxeter group $W$ in the dataset with $\rank(W) \geq 2$ is the dihedral group $\mathrm{I}_{2}(m)$. Taking it as a concrete working example, we completely characterized the $\mathrm{I}_{2}(m)$-invariant functions that act as building blocks of the prepotential. Moreover, in \cref{sec:harmonic-analysis} we understood that the structure of these is naturally explained by realizing that the Gromov-Witten expansion is a superposition of eigenfunctions of the Laplace-Beltrami operator derived from the Coxeter symmetries.

The obvious next step is to generalize these results to the full set of Coxeter groups featured in the database by constructing their corresponding Laplace-Beltrami operators and performing the pertinent harmonic decomposition of the instanton prepotential. While we leave the complete analysis of the general case for future work, we would still like to outline some steps in this direction below. First, we discuss how the raw orbit sums can be explicitly and efficiently constructed in practice for Coxeter groups. Second, we leverage the dihedral results to offer a resummed expression for the general case in terms of dihedral building blocks by decomposing the problem into the action of dihedral subgroups.

\subsection{Finite-state automata and Coxeter groups}
\label{sec:finite-state-automata-Coxeter-groups}

For simplicity, we focus throughout this section on curve classes $d \in \mathcal{M}_{f}$ with $\mathrm{Stab}_{W}(d) = 1$ to avoid subtleties related to non-trivial stabilizers. Given a Coxeter group $W$, the \mbox{$W$-invariant} functions appearing in the instanton expansion can be constructed through the orbit sum
\begin{equation}
    \psi_{ld}^{W}(T) = \sum_{g \in W} e^{2\pi i l \langle gd, T \rangle}\,.
\end{equation}
For the infinite dihedral group, constructing this orbit sum explicitly is straightforward, as explained in \cref{sec:dihedral-Coxeter-prepotentials}. Using the presentation of $\infdih \cong \mathbb{Z} \rtimes \mathbb{Z}_{2}$ given in \eqref{eq:infdih-non-Coxeter-presentation}, all elements of the group are of the form $r^{k}$ or $r^{k}s_{1}$, with $k \in \mathbb{Z}$, reducing the problem to two simple sums indexed by the integers. Hence, our first task is to find a similarly simple way to express the raw orbit sum for the general case. We will achieve this through the construction of specific finite-state automata.

A finite-state automaton is a set of memory states plus labeled transitions among them. More formally, a (deterministic) finite-state automaton is a tuple $\mathcal{A} = (Q,\Sigma,\delta,q_{0},F)$, where $Q$ is a finite set of states, $\Sigma$ is a finite alphabet, $\delta: Q \times \Sigma \rightarrow Q$ is a transition function, $q_{0} \in Q$ is an initial state and $F \subseteq Q$ is a set of accepting states \cite{HopcroftMotwaniUllman2006Automata}. In the context of group theory, finite-state automata make an appearance through the notion of automatic groups \cite{EpsteinCannonHoltLevyPatersonThurston1992}. These are finitely generated groups equipped with finite-state automata that, roughly speaking, encode information about their Cayley graph. More specifically, automatic groups come equipped with a word acceptor automaton, which accepts a word if it corresponds to an element of the group expressed in the chosen representative language, and a multiplier automaton, that compares the standard spelling of an element of the group and its product with one of the generators. An important feature of automatic groups is that the word problem (checking whether two words represent the same group element) is solvable in quadratic time in terms of the word length. Finitely generated Coxeter groups admit an automatic structure through the Brink-Howlett automaton \cite{BrinkHowlett1993FinitenessProperty}, see also \cite{HohlwegNadeauWilliams2016AutomataReducedWords}, with which it is possible to perform various practical computations \cite{Casselman1994MachineCalculationsWeyl,Casselman1995AutomataCoxeter,Casselman2002ComputationCoxeterI,Casselman2008ComputationCoxeterII}.

Our use of finite-state automata will be more pedestrian, and we will not invoke the more sophisticated results associated with the Brink-Howlett automaton. Rather, we will intuitively construct small finite-state automata, tailored to our examples, that will allow us to sum over all elements of the Coxeter group without overcounting. Instead of developing this abstractly, we illustrate the discussion through a series of examples.

\subsubsection{Infinite dihedral group}
\label{sec:fsa-dihedral}

Let us start by revisiting the infinite dihedral group $\infdih$ from this point of view. The alphabet $\Sigma = \{1,2\}$ of the finite-state automaton stands for the simple reflections. The possible states of the automaton are $Q = \{ q_{0},q_{1},q_{2} \}$, where $q_{0}$ is the initial state corresponding to the empty word (identity element) and $q_{1}$ and $q_{2}$ store in memory whether the last letter appended to the word was $s_{1}$ or $s_{2}$, respectively. The transition function $\delta: Q \times \Sigma \rightarrow Q$ is defined by
\begin{equation}
    \delta(q_{0},1) = q_{1}\,,\qquad \delta(q_{0},2) = q_{2}\,,\qquad \delta(q_{1},2) = q_{2}\,,\qquad \delta(q_{2},1) = q_{1}\,.
\end{equation}
We forbid (leave undefined)  the transitions $\delta(q_{1},1)$ and $\delta(q_{2},2)$, which would lead to adjacent repetitions of the same simple reflection and, hence, non-reduced words. Since for $\infdih$ reduced words are unique per group element, enumerating only reduced words ensures that we do not overcount as we sum over the finite-state automaton. Finally, our accepting states are $F = \{ q_{1}, q_{2} \}$.

While the formal definition of the finite-state automaton is useful, it is much more practical to use representations in terms of graphs, see~\cref{fig:infdih-automaton}. Each node represents a possible state of the automaton. The edges encode how the state of the automaton can be updated given its current state. We abuse notation in the diagrams and directly identify the automaton states with certain group elements. By avoiding self-loops, the automaton traverses the Cayley graph of $\infdih$, shown in \cref{fig:infdih-Cayley}, without repeating elements.
\begin{figure}[t]
	\centering
	\CenteredTikzPair
		{0.6\textwidth}{0.4\textwidth}
		{figures/infinite-dihedral-Cayley}{Cayley graph of $\infdih$.}{fig:infdih-Cayley}
		{figures/infinite-dihedral-automaton}{Finite-state automaton for $\infdih$.}{fig:infdih-automaton}
	\caption{The finite-state automaton allows us to traverse the Cayley graph of $\infdih$ without repeating elements.}
	\label{fig:infdih-Cayley-automaton}
\end{figure}
Hence, summing over all elements of the group once can be done by starting at the identity and simply updating the automaton.

Let us assume that we have initialized the automaton at the identity state $e$ and updated it once to either $s_{1}$ or $s_{2}$. Since we are building the Cayley graph by appending group generators on the right, the automaton should be interpreted to act in the same way. We can encode this at the level of linear algebra in the following way. Define the diagonal matrix
\begin{equation}
    D \coloneqq
    \begin{pmatrix}
        s_{1} & 0\\
        0 & s_{2}
    \end{pmatrix}
\end{equation}
and represent the aforementioned states after one update by the row vectors
\begin{equation}
    s_{1} \longmapsto \vec{s}_{1} \coloneqq (s_{1}, 0) = e_{1}^{T} \cdot D\,,\qquad s_{2} \longmapsto \vec{s}_{2} \coloneqq (0, s_{2}) = e_{2}^{T} \cdot D\,.
\end{equation}
The (partial) transition function of the automaton can then be represented by
\begin{equation}
    A_{1} =
    \begin{pmatrix}
        0 & 0\\
        1 & 0
    \end{pmatrix}
    \,,\qquad
    A_{2} =
    \begin{pmatrix}
        0 & 1\\
        0 & 0
    \end{pmatrix}
    \,,
\end{equation}
acting on row vectors, such that $A_{i} \cdot D$ appends the letter $s_{i}$ on the right. The word $s_{i_{0}}s_{i_{1}} \cdots s_{i_{n}}$ is obtained through $\vec{s}_{i_{0}} \cdot A_{i_{1}} \cdot D \cdots A_{i_{n}} \cdot D \cdot \vec{1}$, where $i_{j} \in \{1,2\}$ and $\vec{1}$ is a column vector of ones. Our automaton only produces reduced words; any concatenation of matrices corresponding to a non-reduced word yields $(0,0)$ as output. Non-zero outputs correspond to traversing one branch of the Cayley graph without backtracking.

Our interest is not in specific words but in summing over all group elements, which for $\infdih$ is equivalent to summing over all reduced words relative to a chosen generating set. Define
\begin{equation}
    A = A_{1} + A_{2} =
    \begin{pmatrix}
        0 & 1\\
        1 & 0
    \end{pmatrix}
    \,,
\end{equation}
which is the (transpose of) the adjacency matrix of the automaton graph with the identity node removed. As an initial state, consider the superposition of states $\vec{s}_{1} + \vec{s}_{2} = \vec{1}^{T} \cdot D$. Then, the formal sum
\begin{equation}
    S_{\infdih} = 1 + \sum_{n \in \mathbb{Z}_{\geq 0}} \vec{1}^{\,T} \cdot D \cdot (A \cdot D)^{n} \cdot \vec{1} = 1 + (s_{1} + s_{2}) + (s_{1}s_{2} + s_{2}s_{1}) + \cdots
\end{equation}
contains all elements of the group exactly once. Treating it like a formal geometric series, we can efficiently express the result as
\begin{equation}
    S_{W} = 1 + \vec{\sigma}^{T} \cdot D \cdot (\mathds{1}-K)^{-1} \cdot \vec{1}\,,\qquad K \coloneqq A \cdot D\,,
\label{eq:dfa-sum}
\end{equation}
with $W = \infdih$. Here the initialization vector is simply $\vec{\sigma} = (1,1)^{T}$, but it can vary in more complicated examples.

\subsubsection{Universal Coxeter groups}
\label{sec:fsa-universal}

The infinite dihedral group is a particular case of the universal Coxeter group
\begin{equation}
    W_{n} \coloneqq  \mathbb{Z}_{2} * \overset{n}{\cdots} * \mathbb{Z}_{2}\,,
\end{equation}
for which every group element has a unique reduced expression. The Cayley graph of $W_{n}$ is a tree graph in which every node has $1$ incoming and $n-1$ outgoing edges, except for the identity which has no incoming edges. We show the Cayley graph of $W_{3}$ in \cref{fig:universal-rank-3-Cayley}.
\begin{figure}[t]
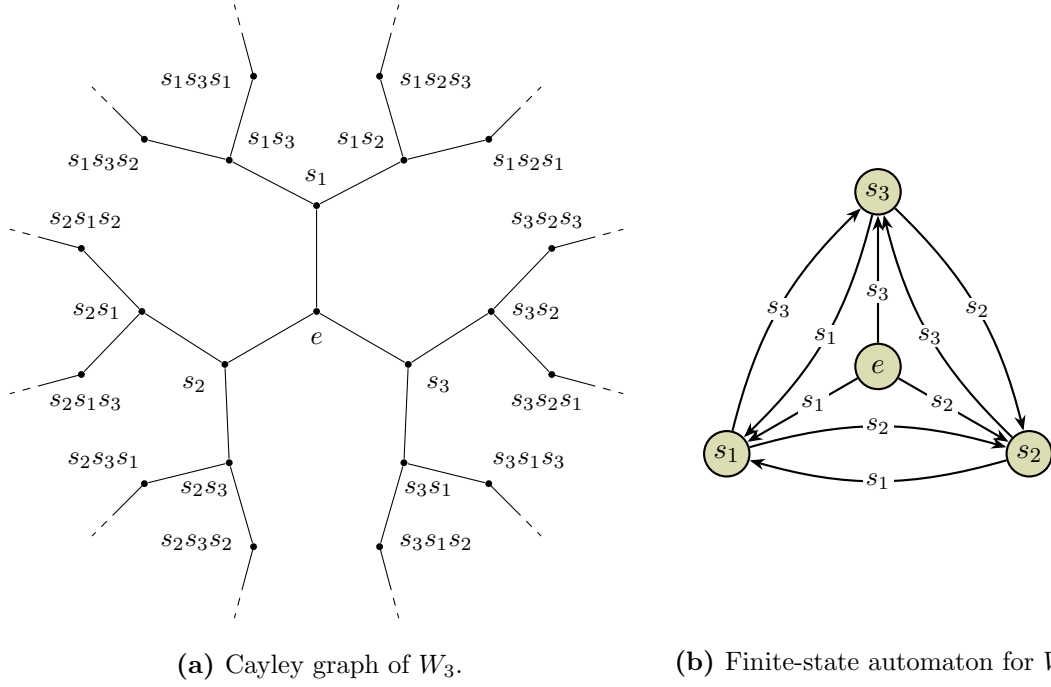

	\centering
	\CenteredTikzPair
		{0.6\textwidth}{0.4\textwidth}
		{figures/universal-rank-3-Cayley}{Cayley graph of $W_{3}$.}{fig:universal-rank-3-Cayley}
		{figures/universal-rank-3-automaton}{Finite-state automaton for $W_{3}$.}{fig:universal-rank-3-automaton}
	\caption{The Cayley graph of the universal Coxeter group $W_{n}$ is a tree graph. Traversing it with a non-backtracking finite-state automaton produces each group element once.}
	\label{fig:universal-rank-3-Cayley-automaton}
\end{figure}
The finite-state automaton and its updating rules are a direct generalization of those for $\infdih$, see \cref{fig:universal-rank-3-automaton} for the $W_{3}$ automaton.

The formal sum over all elements of the group is obtained by allowing the finite-state automaton to traverse all paths in the Cayley graph starting from the identity without backtracking. This leads to $S_{W_{n}}$ as in \eqref{eq:dfa-sum} with
\begin{equation}
    A =
    \begin{pmatrix}
        0 & 1 & 1 & \cdots & 1 \\
        1 & 0 & 1 & \cdots & 1 \\
        1 & 1 & 0 & \ddots & \vdots \\
        \vdots& \vdots& \ddots & \ddots & 1 \\
        1 & 1 & \cdots& 1 & 0
    \end{pmatrix}
    \,,\qquad D =
    \begin{pmatrix}
        s_{1} & 0 & 0 & \cdots & 0 \\
        0 & s_{2} & 0 & \cdots & 0 \\
        0 & 0 & s_{3} & \ddots & \vdots \\
        \vdots& \vdots& \ddots & \ddots & 0 \\
        0 & 0 & \cdots& 0 & s_{n}
    \end{pmatrix}
    \,,
\end{equation}
and the initialization vector $\vec{\sigma} = \vec{1}$.

\subsubsection{General Coxeter groups}
\label{sec:fsa-general}

For a general Coxeter group, two distinct generators are allowed to have a non-trivial relation. As a consequence, the Cayley graph ceases to be a tree graph, featuring cycles that encode these relations. The subgraph covered by updating the finite-state automaton chosen for our purposes must avoid cycles by never traversing certain edges of the Cayley graph. This ensures that a unique reduced word per group element is produced by the finite-state automaton and that \eqref{eq:dfa-sum} does not overcount contributions. To accomplish this, the possible states of the automaton have to be enlarged such that it can store in memory information about which edges of the Cayley graph were last traversed, instead of simply remembering the last one. Let us illustrate this with a couple of simple examples.

Consider the Coxeter group $W_{(2,\infty,\infty)}$ with Coxeter matrix
\begin{equation}
    C =
    \begin{pmatrix}
        1 & 2 & \infty\\
        2 & 1 & \infty\\
        \infty & \infty & 1
    \end{pmatrix}
    \,.
\end{equation}
The non-trivial relation means that two reduced words $s_{1}s_{2}$ and $s_{2}s_{1}$ represent the same group element. In a Coxeter group, the reduced words corresponding to a group element all have the same length \cite{BjornerBrenti2005CombinatoricsCoxeter}. Hence, choosing the canonical representative through lexicographic order is enough to avoid overcounting. This leads to the Cayley graph shown in \cref{fig:three-simple-reflections-with-relations-Cayley}, where the edges depicted in red allow for the same element to be reached by more than one non-backtracking path from the identity.
\begin{figure}[t]
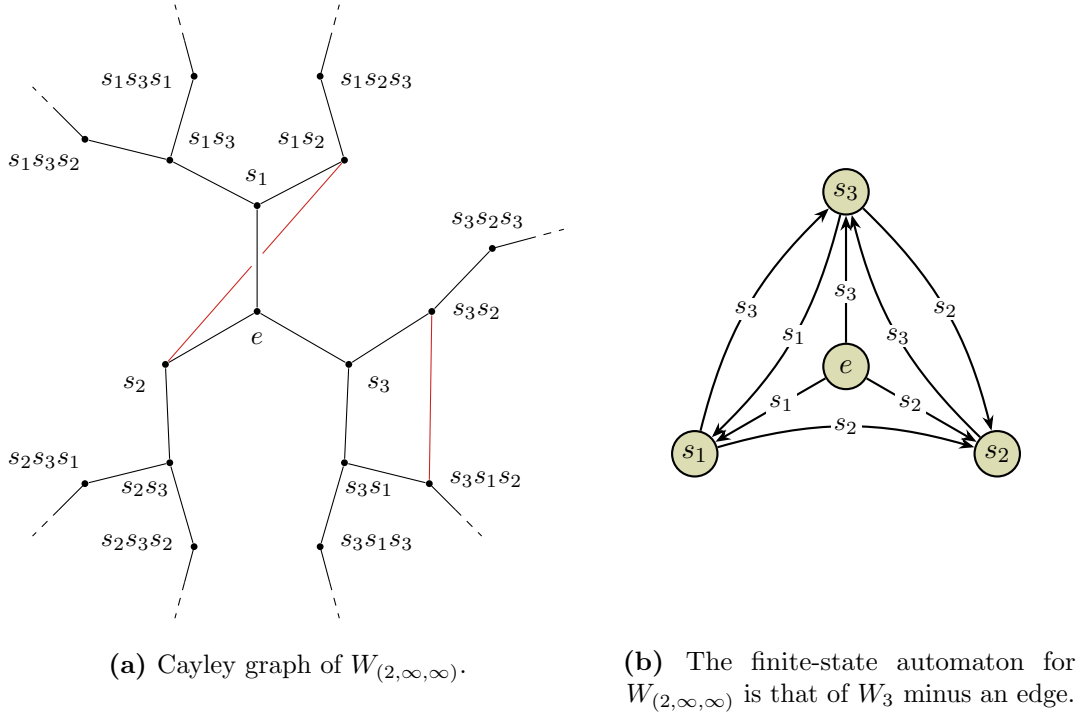

	\centering
	\CenteredTikzPair
		{0.6\textwidth}{0.4\textwidth}
		{figures/three-simple-reflections-with-relations-Cayley}{Cayley graph of $W_{(2,\infty,\infty)}$.}{fig:three-simple-reflections-with-relations-Cayley}
		{figures/three-simple-reflections-with-relations-automaton}{The finite-state automaton for $W_{(2,\infty,\infty)}$ is that of $W_{3}$ minus an edge.}{fig:three-simple-reflections-with-relations-automaton}
	\caption{The chosen non-backtracking finite-state automaton for $W_{(2,\infty,\infty)}$ avoids the Cayley graph edges depicted in red. This prevents the automaton from producing more than one reduced word per element in the enumeration.}
	\label{fig:three-simple-reflections-with-relations-Cayley-automaton}
\end{figure}
Since the non-trivial relation among generators is a commutation relation, the finite-state automaton relevant for our purposes remains simple. It is obtained from that of $W_{3}$ by removing the edge that produces $s_{2}s_{1}$ blocks, see \cref{fig:three-simple-reflections-with-relations-automaton}. Hence, we obtain $S_{W_{(2,\infty,\infty)}}$ from $\eqref{eq:dfa-sum}$ through
\begin{equation}
    A =
    \begin{pmatrix}
        0 & 1 & 1\\
        0 & 0 & 1\\
        1 & 1 & 0
    \end{pmatrix}
    \,,\qquad
    D =
    \begin{pmatrix}
        s_{1} & 0 &0\\
        0 & s_{2} & 0\\
        0 & 0 & s_{3}
    \end{pmatrix}
    \,,
\end{equation}
and the initialization vector $\vec{\sigma} = \vec{1}$. The matrix $A$ is non-symmetric because the automaton graph after removing the identity node has become directed. This is enough in this example.

Higher-order relations not only require non-symmetric $A$ matrices, but also an enlarged space of states for the automaton. To exemplify this, consider the Coxeter group $W_{(3,\infty,\infty)}$ with Coxeter matrix
\begin{equation}
    C =
    \begin{pmatrix}
        1 & 3 & \infty\\
        3 & 1 & \infty\\
        \infty & \infty & 1
    \end{pmatrix}
    \,.
\end{equation}
The first element of the group that can be represented with two different reduced words is $s_{1}s_{2}s_{1} = s_{2}s_{1}s_{2}$ of length $c_{12} = 3$. According to our lexicographic order choice, we must forbid the second one from appearing. Additionally, we must prevent any reduced words containing the alternating pattern of length $c_{12} + 1 = 4$ given by $s_{1}s_{2}s_{1}s_{2}$ or $s_{2}s_{1}s_{2}s_{1}$. The second one cannot exist once we have forbidden the length 3 reduced word $s_{2}s_{1}s_{2}$, but the first one must be explicitly prevented. The Cayley graph of the group is presented in \cref{fig:three-simple-reflections-with-higher-relations-Cayley}.
\begin{figure}[t]
	\centering
	\CenteredTikzPair
		{0.6\textwidth}{0.4\textwidth}
		{figures/three-simple-reflections-with-higher-relations-Cayley}{Cayley graph of $W_{(3,\infty,\infty)}$.}{fig:three-simple-reflections-with-higher-relations-Cayley}
		{figures/three-simple-reflections-with-higher-relations-automaton}{The finite-state automaton for $W_{(3,\infty,\infty)}$, with enlarged memory.}{fig:three-simple-reflections-with-higher-relations-automaton}
	\caption{The finite-state automaton for $W_{(3,\infty,\infty)}$ avoids repeating group elements by remembering a larger number of past steps as it traverses the Cayley graph without backtracking, which requires additional states.}
	\label{fig:three-simple-reflections-with-higher-relations-Cayley-automaton}
\end{figure}
A non-backtracking finite-state automaton that avoids repeating elements for $W_{(3,\infty,\infty)}$ must have enough memory to remember, at least in some directions, the last three letters that were appended to the reduced word that is being constructed. This is achieved by an enlarged set of states with the appropriate transition function, as shown in \cref{fig:three-simple-reflections-with-higher-relations-automaton}. Then $S_{W_{(3,\infty,\infty)}}$ is obtained using the matrices
\begin{equation}
    A =
    \begin{pmatrix}
        0 & 0 & 0 & 0 & 1 & 1\\
        0 & 0 & 0 & 0 & 0 & 1\\
        0 & 0 & 0 & 0 & 0 & 1\\
        0 & 1 & 0 & 0 & 0 & 1\\
        0 & 0 & 1 & 0 & 0 & 1\\
        1 & 0 & 0 & 1 & 0 & 0\\
    \end{pmatrix}
    \,,\qquad D =
    \begin{pmatrix}
        s_{1} & 0 & 0 & 0 & 0 & 0\\
        0 & s_{1} & 0 & 0 & 0 & 0\\
        0 & 0 & s_{1} & 0 & 0 & 0\\
        0 & 0 & 0 & s_{2} & 0 & 0\\
        0 & 0 & 0 & 0 & s_{2} & 0\\
        0 & 0 & 0 & 0 & 0 & s_{3}\\
    \end{pmatrix}
    \,,
\end{equation}
where we have chosen to order the states like $(1,21,121,2,12,3)$. The initialization vector reads $\vec{\sigma} = (1,0,0,1,0,1)^{T}$, which selects the reduced words given by a single generator.

\subsection{Sum over instanton contributions}
\label{sec:fsa-instanton-sum}

Through a series of examples, we have seen that the formal sum $S_{W}$ over all elements of a general Coxeter group $W$ can be expressed in terms of the adjacency matrix of the graph underlying an appropriately chosen finite-state automaton. To make this useful for our purposes, we need to translate this into an orbit sum of instanton contributions to the prepotential.

For each simple reflection $s_{i}$, consider the linear operator
\begin{equation}
\begin{split}
    \phi_{i} : \hat{V}_{\mathbb{C}} &\longrightarrow \hat{V}_{\mathbb{C}}\\
    T &\longmapsto \phi_{i}(T) = \hat{\mathcal{M}}_{i} \cdot T\,.
\end{split}
\end{equation}
Next, define a composition operator acting on functions $f(T) \in C^{\infty}\big( \hat{V}_{\mathbb{C}} \big)$ of the K\"ahler moduli by
\begin{equation}
\begin{split}
    \mathscr{S}_{i} : C^{\infty}\big( \hat{V}_{\mathbb{C}} \big) &\longrightarrow C^{\infty}\big( \hat{V}_{\mathbb{C}}  \big)\\
    f &\longmapsto f \circ \phi_{i}\,,
\end{split}
\end{equation}
for $i \in \{ 1, \dotsc, h \}$. Given a word $w = s_{i_{1}} \cdots s_{i_{n}} \in W$, it acts on such functions by operator composition $\mathscr{S}_{w} \coloneqq \mathscr{S}_{i_{1}} \cdots \mathscr{S}_{i_{n}}$, meaning that
\begin{equation}
    \mathscr{S}_{w}(f)(T) = f(\phi_{i_{n}} \circ \cdots \circ \phi_{i_{1}} (T))\,.
\end{equation}
Hence, acting on a given instanton contribution we obtain
\begin{equation}
    \mathscr{S}_{w} \left( e^{2\pi i l \langle d,T \rangle} \right) = e^{2\pi i l \left\langle d, \hat{\mathcal{M}}_{i_{n}} \cdots \hat{\mathcal{M}}_{i_{1}} T \right\rangle} = e^{2\pi i l \left\langle \mathcal{M}_{i_{1}} \cdots \mathcal{M}_{i_{n}} d,T \right\rangle} = e^{2\pi i l \langle wd, T \rangle}\,.
\end{equation}
To declutter the notation, we will often write $\mathscr{S}_{w}f(T)$ to refer to $\mathscr{S}_{w}(f)(T)$.

With these definitions, upgrade the formal sum $S_{W}$ to a sum $\mathscr{S}_{W}$ in the algebra of composition operators through the substitution $D \mapsto \mathscr{D}$ obtained by $s_{i} \mapsto \mathscr{S}_{i}$. In the resulting expression, juxtapositions are to be interpreted as products of composition operators. Let us denote the $l$-th multicover instanton contribution for a given curve class $d \in \mathcal{M}_{f}$ with trivial stabilizer by
\begin{equation}
\begin{split}
    f_{l} : \hat{V}_{\mathbb{C}} &\longrightarrow \mathbb{C}\\
    T &\longmapsto e^{2\pi i l \langle d, T \rangle}\,.
\end{split}
\end{equation}
The orbit sum of instanton contributions can then be expressed as
\begin{equation}
\begin{split}
    \psi_{ld}^{W}(T) &= \sum_{w \in W} e^{2\pi i l \langle wd,T \rangle}\\
    &= \mathscr{S}_{W} (f_{l})(T) = \left[ 1 + \vec{\sigma}^{T} \cdot \mathscr{D} \cdot (\mathds{1}-\mathscr{K})^{-1} \cdot \vec{1} \right] f_{l}(T)\,,\qquad \mathscr{K} \coloneqq A \cdot \mathscr{D}\,.
\label{eq:orbit-sum-dfa}
\end{split}
\end{equation}
This formula cleanly separates the group combinatorics, encapsulated in $A$, from its action on the K\"ahler moduli space, corresponding to $\mathscr{D}$.

This rewriting of $\psi_{ld}^{W}(T)$ still corresponds to a superposition of plane Gromov-Witten waves, rather than to a resummation into eigenfunctions of the natural Laplace-Beltrami operator associated to the group action as seen in \cref{sec:dihedral-Coxeter-prepotentials,sec:harmonic-analysis}. The obtained expression is, nonetheless, more useful than simply writing $\psi_{ld}^{W}(T) = \sum_{w \in W} e^{2\pi i l \langle wd,T \rangle}$ at an abstract level. First, it replaces an implicit definition in terms of an infinite sum for finite input data related to the finite-state automaton constructed for $W$, namely, the finite matrices $A$ and $\mathscr{D}$. This reduces the treatment of infinite group cases to simpler finite linear algebra objects. Second, once the finite-state automaton is constructed, \eqref{eq:orbit-sum-dfa} automatically runs over the set of canonical reduced words chosen (lexicographically ordered, in our case) without the need to worry about overcounting group elements. Third, finite truncations of $(\mathds{1}-\mathscr{K})^{-1} = \sum_{n \in \mathbb{Z}_{\geq 0}} \mathscr{K}^{n}$ directly correspond to partial sums up to a given word length (or path length traversed by the automaton), providing a systematic way to study growth and asymptotics that is not immediately apparent from the abstract sum. Finally, the resolvent expression \eqref{eq:orbit-sum-dfa} will allow us to perform certain compressions of the orbit sum, as we now explore.

\subsection{Reduction to dihedral blocks}
\label{fsa-dihred}

Having reworked the orbit sum into a finite-state linear algebra expression, the natural next step would be to resum it into a spectral dual decomposition. Analyzing the quotient moduli space geometry by the action of the Coxeter group and building the corresponding Laplace-Beltrami operator would inform us about the type of eigenfunctions that are natural for each model. This is an interesting problem that we hope to revisit in future work. More immediately, we take an alternative perspective that allows us to directly leverage the results from \cref{sec:dihedral-Coxeter-prepotentials} for the general case by performing a dihedral block decomposition of the finite-state automata.

Given a Coxeter group $(W,S)$, each pair of simple reflections generates a distinguished dihedral subgroup $\mathrm{I}_{2}^{ij}(m) \coloneqq \langle s_{i}, s_{j} \rangle \leq W$ corresponding to an edge of the Coxeter diagram (when $m=2$ the edge is conventionally omitted). Forgetting about the larger Coxeter symmetry and focusing on a single subgroup of the aforementioned type, the $\rank(W) = 2$ results derived in preceding sections apply immediately. Regarding $W$ as a collection of $\mathrm{I}_{2}^{ij}(m)$ subgroups plus the (possibly infinite) alternation phenomenon between rank-2 channels $\langle s_{i}, s_{j} \rangle$, it is reasonable to expect that we can decompose the general problem into a series of dihedral blocks that we already know how to treat and their global mixing, which we still need to explicitly understand. The resolvent formula for the orbit sum obtained from the finite-state automaton is well-suited for this problem: after separating the dihedral blocks and resumming them, the alternation between channels is captured by the remainder of the original resolvent formula. We develop this approach below. As above, we illustrate the general method using $W_{3}$, $W_{(2,\infty,\infty)}$ and $W_{(3,\infty,\infty)}$ as examples.

\subsubsection[Dihedral reduction of \texorpdfstring{${W_{3}}$}{W3}]{Dihedral reduction of \texorpdfstring{$\boldsymbol{W_{3}}$}{W3}}
\label{sec:fsa-dihred-W3}

The full orbit sum for the instanton contribution corresponding to the $l$-th multicover of the curve class $d \in \mathcal{M}_{f}$ with trivial stabilizer is given by \eqref{eq:orbit-sum-dfa}. We recall that for $W_{3}$ the input data is
\begin{equation}
    A =
    \begin{pmatrix}
        0 & 1 & 1\\
        1 & 0 & 1\\
        1 & 1 & 0
    \end{pmatrix}
    \,,\qquad
    \mathscr{D} =
    \begin{pmatrix}
        \mathscr{S}_{1} & 0 & 0\\
        0 & \mathscr{S}_{2} & 0\\
        0 & 0 & \mathscr{S}_{3}
    \end{pmatrix}
    \,,\qquad
    \vec{\sigma} =
    \begin{pmatrix}
        1\\
        1\\
        1
    \end{pmatrix}
    \,.
\end{equation}
For the purpose of separating the dihedral contributions in the orbit sum, it is convenient to refine the finite-state automaton for $W_{3}$ such that it can remember the last two letters appended to the word. This allows it to produce the same enumeration of elements of the group while being able to detect when an alternation of letters within an $\mathrm{I}_{2}^{ij}(\infty)$ subgroup ends. Focusing on the newly added states, the construction is simply the line graph of the original finite-state automaton. Indeed, the allowed transitions among the new states $\{ 12,21,13,31,23,32 \}$ are encoded, using the same conventions as earlier, in the matrices
\begin{equation}
    A^{(2)} =
    \begin{pmatrix}
        0 & 1 & 0 & 0 & 1 & 0\\
        1 & 0 & 1 & 0 & 0 & 0\\
        0 & 0 & 0 & 1 & 0 & 1\\
        1 & 0 & 1 & 0 & 0 & 0\\
        0 & 0 & 0 & 1 & 0 & 1\\
        0 & 1 & 0 & 0 & 1 & 0
    \end{pmatrix}
    \,,\qquad
    \mathscr{D}^{(2)} =
    \begin{pmatrix}
        \mathscr{S}_{2} & 0 & 0 & 0 & 0 & 0\\
        0 & \mathscr{S}_{1} & 0 & 0 & 0 & 0\\
        0 & 0 & \mathscr{S}_{3} & 0 & 0 & 0\\
        0 & 0 & 0 & \mathscr{S}_{1} & 0 & 0\\
        0 & 0 & 0 & 0 & \mathscr{S}_{3} & 0\\
        0 & 0 & 0 & 0 & 0 & \mathscr{S}_{2}
    \end{pmatrix}
    \,,
\end{equation}
with starting length-2 state
\begin{equation}
    \vec{\sigma}_{2} =
    \begin{pmatrix}
        \mathscr{S}_{1}\mathscr{S}_{2}\\
        \mathscr{S}_{2}\mathscr{S}_{1}\\
        \mathscr{S}_{1}\mathscr{S}_{3}\\
        \mathscr{S}_{3}\mathscr{S}_{1}\\
        \mathscr{S}_{2}\mathscr{S}_{3}\\
        \mathscr{S}_{3}\mathscr{S}_{2}
    \end{pmatrix}
    \,.
\end{equation}
The full orbit sum is then given by
\begin{equation}
    \psi_{ld}^{W_{3}}(T) = \left[ 1 + (\mathscr{S}_{1} + \mathscr{S}_{2} + \mathscr{S}_{3}) + \left( \vec{\sigma}_{2}^{T} \cdot \left( \mathds{1} - K^{(2)} \right)^{-1} \cdot \vec{1} \right) \right] f_{l}(T)\,,\qquad K^{(2)} \coloneqq A^{(2)}\mathscr{D}^{(2)}\,,
\end{equation}
where we have separated the length-0 and length-1 words already.

Let us classify the edges of the line graph  into stay edges $ij \rightarrow ji$ that stay within a dihedral subgroup $\mathrm{I}_{2}^{ij}(\infty)$ and switch edges $ij \rightarrow jk$ that introduce the third letter of $W_{3}$. According to this split, we can decompose the automaton matrix into
\begin{equation}
    A^{(2)} = A^{(2)}_{\mathrm{st}} + A^{(2)}_{\mathrm{sw}}\,,\qquad
    A^{(2)}_{\mathrm{st}} =
    \begin{pmatrix}
        0 & 1 & 0 & 0 & 0 & 0\\
        1 & 0 & 0 & 0 & 0 & 0\\
        0 & 0 & 0 & 1 & 0 & 0\\
        0 & 0 & 1 & 0 & 0 & 0\\
        0 & 0 & 0 & 0 & 0 & 1\\
        0 & 0 & 0 & 0 & 1 & 0
    \end{pmatrix}
    \,,\qquad A^{(2)}_{\mathrm{sw}} =
    \begin{pmatrix}
        0 & 0 & 0 & 0 & 1 & 0\\
        0 & 0 & 1 & 0 & 0 & 0\\
        0 & 0 & 0 & 0 & 0 & 1\\
        1 & 0 & 0 & 0 & 0 & 0\\
        0 & 0 & 0 & 1 & 0 & 0\\
        0 & 1 & 0 & 0 & 0 & 0
    \end{pmatrix}
    \,,
\end{equation}
with $K_{\mathrm{st}}$ and $K_{\mathrm{sw}}$ defined in the obvious way. Given the block structure of $A^{(2)}_{\mathrm{st}}$, the resolvent of $K_{\mathrm{st}}$ has the form
\begin{equation}
    (\mathds{1} - K_{\mathrm{st}})^{-1} = \left( \mathds{1} - K^{(12)}_{\mathrm{st}} \right)^{-1} \oplus \left( \mathds{1} - K^{(13)}_{\mathrm{st}} \right)^{-1} \oplus \left( \mathds{1} - K^{(23)}_{\mathrm{st}} \right)^{-1}\,,
\end{equation}
where
\begin{equation}
    \left( \mathds{1} - K^{(ij)}_{\mathrm{st}} \right)^{-1} =
    \begin{pmatrix}
        (1 - \mathscr{S}_{i}\mathscr{S}_{j})^{-1} & \mathscr{S}_{i}(1-\mathscr{S}_{j}\mathscr{S}_{i})^{-1}\\
        \mathscr{S}_{j}(1-\mathscr{S}_{i}\mathscr{S}_{j})^{-1} & (1 - \mathscr{S}_{j}\mathscr{S}_{i})^{-1}
    \end{pmatrix}
    \,.
\label{eq:infdih-resolvent-block}
\end{equation}
Letting the resolvent act on the Gromov-Witten terms we obtain the stay function vector
\begin{equation}
    H(T) \coloneqq \left[ (\mathds{1} - K_{\mathrm{st}})^{-1} \cdot \vec{1} \right] f_{l}(T) =
    \begin{pmatrix}
        H_{12}(T)\\
        H_{21}(T)\\
        H_{13}(T)\\
        H_{31}(T)\\
        H_{23}(T)\\
        H_{32}(T)\\
    \end{pmatrix}
    \,,
\end{equation}
where the entries are
\begin{equation}
\begin{split}
    H_{ij}(T) &\coloneqq \left[ (1 - \mathscr{S}_{i}\mathscr{S}_{j})^{-1} + \mathscr{S}_{i}(1 - \mathscr{S}_{j}\mathscr{S}_{i})^{-1} \right]f_{l}(T)\\
    &\phantom{\vcentcolon}= \sum_{k \in \mathbb{Z}_{\geq 0}} (\mathscr{S}_{i}\mathscr{S}_{j})^{k} f_{l}(T) + \sum_{k \in \mathbb{Z}_{\geq 0}} \mathscr{S}_{i}( \mathscr{S}_{j}\mathscr{S}_{i} )^{k} f_{l}(T)\,.
\end{split}
\label{eq:stay-function-vector-infdih-entries}%
\end{equation}
Each function $H_{ij}(T)$ is built from one-sided versions of the orbit sums studied in \cref{sec:dihedral-Coxeter-prepotentials}.

From the factorization
\begin{equation}
    \mathds{1} - K^{(2)} = (\mathds{1} - K_{\mathrm{st}})(\mathds{1} - M)\,,\qquad M \coloneqq (\mathds{1} - K_{\mathrm{st}})^{-1}K_{\mathrm{sw}}\,,
\end{equation}
it follows that the resolvent can be written as
\begin{equation}
    \left( \mathds{1} - K^{(2)} \right)^{-1} = (\mathds{1} - M)^{-1} (\mathds{1} - K_{\mathrm{st}})^{-1}\,.
\label{eq:resolvent-inverse-factorization}
\end{equation}
Hence, the dihedral reduction of the orbit sum yields
\begin{equation}
    \psi_{ld}^{W_{3}}(T) = f_{l}(T) + \sum_{i=1}^{3} (\mathscr{S}_{i}f_{l})(T) + \vec{\sigma}_{2}^{T} \cdot (\mathds{1} - M)^{-1} \cdot H(T)\,,
\label{eq:dihedral-reduction-orbit-sum-W3}
\end{equation}
where we see the combinations of special functions studied in \cref{sec:dihedral-Coxeter-prepotentials} appear through $H(T)$ and then summed over including the mixing between dihedral channels.

\subsubsection[Dihedral reduction of \texorpdfstring{${W_{(2,\infty,\infty)}}$}{W(2,Inf,Inf)}]{Dihedral reduction of \texorpdfstring{$\boldsymbol{W_{(2,\infty,\infty)}}$}{W(2,Inf,Inf)}}
\label{sec:fsa-dihred-W2Inf}

Continuing through the list of selected examples, let us consider $W_{(2,\infty,\infty)}$ next. We recall that the difference with respect to $W_{3}$ is the relation $s_{1}s_{2} = s_{2}s_{1}$. As a consequence, we saw in \cref{sec:fsa-general} that the corresponding finite-state automaton lacks the $2 \rightarrow 1$ edge to avoid overcounting group elements, see \cref{fig:three-simple-reflections-with-relations-Cayley-automaton}. Following the same procedure as above, we separate the length-0 and length-1 words and define an automaton to generate the remaining ones. This automaton has as nodes the edges of the original one, meaning that the set of states is now only enlarged by $\{ 12, 13, 31, 23, 32 \}$. The allowed transitions are described by
\begin{equation}
    A^{(2)} =
    \begin{pmatrix}
        0 & 0 & 0 & 1 & 0\\
        0 & 0 & 1 & 0 & 1\\
        1 & 1 & 0 & 0 & 0\\
        0 & 0 & 1 & 0 & 1\\
        0 & 0 & 0 & 1 & 0
    \end{pmatrix}
    \,,\qquad \mathscr{D}^{(2)} =
    \begin{pmatrix}
        \mathscr{S}_{2} & 0 & 0 & 0 & 0\\
        0 & \mathscr{S}_{3} & 0 & 0 & 0\\
        0 & 0 & \mathscr{S}_{1} & 0 & 0\\
        0 & 0 & 0 & \mathscr{S}_{3} & 0\\
        0 & 0 & 0 & 0 & \mathscr{S}_{2}
    \end{pmatrix}
    \,,
\end{equation}
with starting length-2 state
\begin{equation}
    \vec{\sigma}_{2} =
    \begin{pmatrix}
       \mathscr{S}_{1} \mathscr{S}_{2}\\
       \mathscr{S}_{1} \mathscr{S}_{3}\\
       \mathscr{S}_{3} \mathscr{S}_{1}\\
       \mathscr{S}_{2} \mathscr{S}_{3}\\
       \mathscr{S}_{3} \mathscr{S}_{2}
    \end{pmatrix}
    \,.
\end{equation}

Performing the same split between stay and switch edges as above, we can decompose the $A^{(2)}$ matrix into
\begin{equation}
    A^{(2)}_{\mathrm{st}} =
    \begin{pmatrix}
        0 & 0 & 0 & 0 & 0\\
        0 & 0 & 1 & 0 & 0\\
        0 & 1 & 0 & 0 & 0\\
        0 & 0 & 0 & 0 & 1\\
        0 & 0 & 0 & 1 & 0
    \end{pmatrix}
    \,,\qquad A^{(2)}_{\mathrm{sw}} =
    \begin{pmatrix}
        0 & 0 & 0 & 1 & 0\\
        0 & 0 & 0 & 0 & 1\\
        1 & 0 & 0 & 0 & 0\\
        0 & 0 & 1 & 0 & 0\\
        0 & 0 & 0 & 0 & 0
    \end{pmatrix}
    \,,
\end{equation}
with $K_{\mathrm{st}}$ and $K_{\mathrm{sw}}$ defined accordingly. The block decomposition of the resolvent of $K_{\mathrm{st}}$ has then the form
\begin{equation}
    (\mathds{1} - K_{\mathrm{st}})^{-1} = \left( \mathds{1} \right) \oplus \left( \mathds{1} - K^{(13)}_{\mathrm{st}} \right)^{-1} \oplus \left( \mathds{1} - K^{(23)}_{\mathrm{st}} \right)^{-1}\,,
\end{equation}
where we observe that the first block is one-dimensional and trivial while the other two are of the same form as~\eqref{eq:infdih-resolvent-block}. The trivial block appears because the $s_{1}s_{2} = s_{2}s_{1}$ relation allows for no dihedral oscillations in the $12$-channel in our chosen canonical language, forcing a switch as soon as the $12$ state is reached. The stay function vector
\begin{equation}
    H(T) \coloneqq \left[ (\mathds{1} - K_{\mathrm{st}})^{-1} \cdot \vec{1} \right] (f_{l})(T) =
    \begin{pmatrix}
        H_{12}(T)\\
        H_{13}(T)\\
        H_{31}(T)\\
        H_{23}(T)\\
        H_{32}(T)\\
    \end{pmatrix}
\end{equation}
has entries defined as in \eqref{eq:stay-function-vector-infdih-entries} with the exception of $H_{12}(T)$, for which we simply have
\begin{equation}
    H_{12}(T) \coloneqq f_{l}(T)\,.
\end{equation}

Using the same factorization of the resolvent as for $W_{3}$ in~\eqref{eq:resolvent-inverse-factorization}, we obtain an expression analogous to \eqref{eq:dihedral-reduction-orbit-sum-W3} given by
\begin{equation}
    \psi_{ld}^{W_{(2,\infty,\infty)}}(T) = f_{l}(T) + \sum_{i=1}^{3} (\mathscr{S}_{i}f_{l})(T) + \vec{\sigma}_{2}^{T} \cdot (\mathds{1} - M)^{-1} \cdot H(T)\,.
\end{equation}
The infinite dihedral blocks $\mathrm{I}_{2}^{13}(\infty)$ and $\mathrm{I}_{2}^{23}(\infty)$ are represented in $H(T)$ through the one-sided versions of the orbit sums studied in \cref{sec:dihedral-Coxeter-prepotentials}, which are then summed over including the mixing between dihedral channels. The $\mathrm{I}_{2}^{12}(2)$ block only has one length-2 reduced word in our lexicographic choice of canonical language. Hence, the corresponding entry in $H(T)$ is simply the orbit sum seed.

\subsubsection[Dihedral reduction of \texorpdfstring{${W_{(3,\infty,\infty)}}$}{W(3,Inf,Inf)}]{Dihedral reduction of \texorpdfstring{$\boldsymbol{W_{(3,\infty,\infty)}}$}{W(3,Inf,Inf)}}
\label{sec:fsa-dihred-W3Inf}

Finally, let us treat the $W_{(3,\infty,\infty)}$ example. Here the relation distinguishing it from $W_{3}$ is of higher order than in the preceding case, namely $(s_{1}s_{2})^{3} = 1$. This meant that already the original finite-state automaton required an enlarged memory to remember certain paths beyond the last appended letter, see \cref{fig:three-simple-reflections-with-higher-relations-Cayley-automaton}. Upon separating the length-0 and length-1 words, the automaton generating the remaining ones needs to remember, in order to distinguish dihedral oscillations and switches, not only the last two appended letters but also additional information to ensure that no group elements are overcounted. This leads to the additional set of states $\{ 12, 21, 121, 13, 31, 23, 32 \}$. The allowed transitions are encoded in the matrices
\begin{equation}
    A^{(2)} =
    \begin{pmatrix}
        0 & 0 & 1 & 0 & 0 & 1 & 0\\
        0 & 0 & 0 & 1 & 0 & 0 & 0\\
        0 & 0 & 0 & 1 & 0 & 0 & 0\\
        0 & 0 & 0 & 0 & 1 & 0 & 1\\
        1 & 0 & 0 & 1 & 0 & 0 & 0\\
        0 & 0 & 0 & 0 & 1 & 0 & 1\\
        0 & 1 & 0 & 0 & 0 & 1 & 0
    \end{pmatrix}
    \,,\qquad \mathscr{D}^{(2)} =
    \begin{pmatrix}
        \mathscr{S}_{2} & 0 & 0 & 0 & 0 & 0 & 0\\
        0 & \mathscr{S}_{1} & 0 & 0 & 0 & 0 & 0\\
        0 & 0 & \mathscr{S}_{1} & 0 & 0 & 0 & 0\\
        0 & 0 & 0 & \mathscr{S}_{3} & 0 & 0 & 0\\
        0 & 0 & 0 & 0 & \mathscr{S}_{1} & 0 & 0\\
        0 & 0 & 0 & 0 & 0 & \mathscr{S}_{3} & 0\\
        0 & 0 & 0 & 0 & 0 & 0 & \mathscr{S}_{2}
    \end{pmatrix}
    \,,
\end{equation}
with starting length-2 state
\begin{equation}
    \vec{\sigma}_{2} =
    \begin{pmatrix}
        \mathscr{S}_{1} \mathscr{S}_{2}\\
        \mathscr{S}_{2} \mathscr{S}_{1}\\
        0\\
        \mathscr{S}_{1} \mathscr{S}_{3}\\
        \mathscr{S}_{3} \mathscr{S}_{1}\\
        \mathscr{S}_{2} \mathscr{S}_{3}\\
        \mathscr{S}_{3} \mathscr{S}_{2}
    \end{pmatrix}
    \,.
\end{equation}

Once more, we decompose $A^{(2)}$ into stay and switch edges to obtain
\begin{equation}
    A^{(2)}_{\mathrm{st}} =
    \begin{pmatrix}
        0 & 0 & 1 & 0 & 0 & 0 & 0\\
        0 & 0 & 0 & 0 & 0 & 0 & 0\\
        0 & 0 & 0 & 0 & 0 & 0 & 0\\
        0 & 0 & 0 & 0 & 1 & 0 & 0\\
        0 & 0 & 0 & 1 & 0 & 0 & 0\\
        0 & 0 & 0 & 0 & 0 & 0 & 1\\
        0 & 0 & 0 & 0 & 0 & 1 & 0
    \end{pmatrix}
    \,,\qquad A^{(2)}_{\mathrm{sw}} =
    \begin{pmatrix}
        0 & 0 & 0 & 0 & 0 & 1 & 0\\
        0 & 0 & 0 & 1 & 0 & 0 & 0\\
        0 & 0 & 0 & 1 & 0 & 0 & 0\\
        0 & 0 & 0 & 0 & 0 & 0 & 1\\
        1 & 0 & 0 & 0 & 0 & 0 & 0\\
        0 & 0 & 0 & 0 & 1 & 0 & 0\\
        0 & 1 & 0 & 0 & 0 & 0 & 0
    \end{pmatrix}
    \,,
\end{equation}
with the decomposition descending to $K_{\mathrm{st}}$ and $K_{\mathrm{sw}}$. The resolvent of $K_{\mathrm{st}}$ has again three non-trivial blocks
\begin{equation}
    (\mathds{1} - K_{\mathrm{st}})^{-1} = \left( \mathds{1} - K^{(12)}_{\mathrm{st}} \right)^{-1} \oplus \left( \mathds{1} - K^{(13)}_{\mathrm{st}} \right)^{-1} \oplus \left( \mathds{1} - K^{(23)}_{\mathrm{st}} \right)^{-1}\,.
\end{equation}
The latter two are of the same form as~\eqref{eq:infdih-resolvent-block}, while the first one is simply
\begin{equation}
    \left( \mathds{1} - K^{(12)}_{\mathrm{st}} \right)^{-1} = \mathds{1} + K_{\mathrm{st}}^{(12)} =
    \begin{pmatrix}
        1 & 0 & \mathscr{S}_{1}\\
        0 & 1 & 0\\
        0 & 0 & 1
    \end{pmatrix}
    \,,
\end{equation}
due to the nilpotent nature of $K^{(12)}_{\mathrm{st}}$. The stay function vector
\begin{equation}
    H(T) \coloneqq \left[ (\mathds{1} - K_{\mathrm{st}})^{-1} \cdot \vec{1} \right] (f_{l})(T) =
    \begin{pmatrix}
        H_{12}(T)\\
        H_{21}(T)\\
        H_{121}(T)\\
        H_{13}(T)\\
        H_{31}(T)\\
        H_{23}(T)\\
        H_{32}(T)\\
    \end{pmatrix}
\end{equation}
contains more varied entries this time. The infinite dihedral blocks provide $H_{13}(T)$, $H_{31}(T)$, $H_{23}(T)$ and $H_{32}(T)$, defined as in \eqref{eq:stay-function-vector-infdih-entries}. Those corresponding to the $\mathrm{I}_{2}^{12}(3)$ subgroup are orbit sum seeds and one additional image under the group action, namely
\begin{equation}
    H_{12}(T) = (1+\mathscr{S}_{1})f_{l}(T)\,,\qquad H_{21}(T) = f_{l}(T)\,,\qquad H_{121}(T) = f_{l}(T)\,.
\end{equation}
As the order of the relations between generators increases, small dihedral oscillations start to be possible within the corresponding channels, leading to additional terms in the stay function vector.

Yet again, employing the factorization \eqref{eq:resolvent-inverse-factorization} we are led to the orbit sum expression
\begin{equation}
    \psi_{ld}^{W_{(3,\infty,\infty)}}(T) = f_{l}(T) + \sum_{i=1}^{3} (\mathscr{S}_{i}f_{l})(T) + \vec{\sigma}_{2}^{T} \cdot (\mathds{1} - M)^{-1} \cdot H(T)\,,
\end{equation}
with the appropriate definitions for the $W_{(3,\infty,\infty)}$ example. The parabolic and hyperbolic infinite dihedral blocks appear explicitly through one-sided versions of the combinations of special functions discussed in \cref{sec:dihedral-Coxeter-prepotentials}. Elliptic dihedral blocks, on the other hand, lead to finite sums of exponentials that can be rewritten in terms of ordinary Bessel functions of the first kind as in \cref{sec:dihedral-elliptic-representations}.

\subsubsection{Unilateral dihedral orbit sums}
\label{sec:fsa-dihred-unilateral-blocks}

The dihedral reduction of the orbit sums $\psi_{ld}^{W}(T) = \sum_{w \in W} e^{2\pi i l \langle wd,T \rangle}$ allows us to express the result in terms of what we have called the stay function vector $H(T)$. This object is the result of explicitly applying the stay resolvent to the orbit sum seed, meaning that it produces objects naturally associated with the dihedral subgroups corresponding to the edges of the Coxeter diagram of $W$. The mixing between the pure dihedral channels appears through a second resolvent acting on $H(T)$, which encodes the remaining transitions allowed by the refined \mbox{finite-state} automaton. The entries of $H(T)$ are therefore amenable to the same resummations performed in \cref{sec:dihedral-Coxeter-prepotentials} for the $W = \mathrm{I}_{2}(m)$ case, albeit with some differences.

Let us focus first on the infinite dihedral subgroups $\mathrm{I}_{2}^{ij}(\infty)$. In \cref{sec:dihedral-Coxeter-prepotentials} we encounter full bilateral orbit sums over the group elements\footnote{The sums $S_{\infdih}(l,z,T)$ in this section are not to be confused with the $S \left( l,z,T_{\parallel} \right)$ sums appearing in \cref{sec:dihedral-Coxeter-prepotentials}. The former include certain prefactors present in $\psi_{ld}^{\infdih}(T)$ that were excluded in the definition of the latter.}
\begin{equation}
    S_{\infdih}(l,z,T) \coloneqq \sum_{k \in \mathbb{Z}} e^{2\pi i l \langle \mathcal{Q}^{k} z, T \rangle}\,.
\end{equation}
In contrast, the infinite dihedral entries of $H(T)$ correspond to unilateral orbit sums
\begin{equation}
    S_{\infdih}^{+}(l,z,T) \coloneqq \sum_{k \in \mathbb{Z}_{\geq 0}} e^{2\pi i l \langle \mathcal{Q}^{k} z, T \rangle}\,.
\end{equation}
Indeed, from \eqref{eq:stay-function-vector-infdih-entries} we read off that
\begin{equation}
    H_{ij}(T) = \sum_{k \in \mathbb{Z}_{\geq 0}} \left[ e^{2\pi i l \langle \mathcal{Q}_{ij}^{k} d,T \rangle} + e^{2\pi i l \langle \mathcal{M}_{i} \mathcal{Q}_{ji}^{k} d, T \rangle} \right] = S_{\infdih}^{+}(l,z,T) + S_{\infdih}^{+}(l,\mathcal{S}z,T)\,,
\end{equation}
where we have used the dihedral relation $\mathcal{M}_{i} \mathcal{Q}_{ji}^{k} = \mathcal{Q}_{ij}^{k} \mathcal{M}_{i}$. These occur because the stay resolvent generates alternations of two letters by applying non-negative powers of the rotation of the corresponding dihedral group, as can be seen in \eqref{eq:infdih-resolvent-block}. Let us briefly comment on these monoid orbit sums.

\paragraph{Elliptic dihedral blocks.} In the elliptic $\mathrm{I}_{2}^{ij}(m)$ case, the full orbit sum is finite and the corresponding entries of $H(T)$ are simply truncations of it. The number of terms included grows with $m$. Using the Jacobi-Anger expansion it is possible to express these entries as combinations of ordinary Bessel functions of the first kind as in \cref{sec:dihedral-elliptic-representations}.

\paragraph{Parabolic dihedral blocks.} For the parabolic $\mathrm{I}_{2}^{ij}(\infty)$ case, the full orbit sum leads to Jacobi theta functions
\begin{equation}
    S_{\infdih}(l,z,T) = e^{2\pi i l \langle z,T \rangle} \vartheta(z_{\vartheta},\tau)\,,
\end{equation}
where the coordinates were defined in \cref{sec:dihedral-parabolic-representations}. The unilateral object involves partial theta functions \cite{AndrewsBerndt2009RamanujanLostNotebookII}
\begin{equation}
    \vartheta^{+} (z_{\vartheta},\tau) \coloneqq \sum_{k \in \mathbb{Z}_{\geq 0}} e^{\pi i k^{2} \tau + 2\pi i k z_{\vartheta}}\,,\quad z_{\vartheta} \in \mathbb{C}\,,\quad \mathrm{Im}(\tau) > 0\,,
\end{equation}
instead, yielding
\begin{equation}
    S_{\infdih}^{+}(l,z,T) = e^{2\pi i l \langle z,T \rangle} \vartheta^{+}(z_{\vartheta},\tau)\,.
\end{equation}
This highlights the fact that the dihedral decomposition performed above breaks modularity for parabolic dihedral subgroups. The partial theta function can be expressed in terms of the Jacobi theta function $\vartheta(z_{\vartheta},\tau)$ and a non-modular correction $\Psi(z_{\vartheta},\tau)$ as
\begin{equation}
    \vartheta^{+}(z_{\vartheta},\tau) = \frac{1}{2} \left( \vartheta(z_{\vartheta},\tau) + 1 \right) + \frac{1}{2} \Psi(z_{\vartheta},\tau)\,,
\end{equation}
where the last term is the false Jacobi theta function \cite{Bringmann2021FalseThetaRodgers}
\begin{equation}
    \Psi(z_{\vartheta},\tau) \coloneqq \sum_{k \in \mathbb{Z}} \mathrm{sgn}(k) e^{\pi i k^{2} \tau + 2\pi i k z_{\vartheta}}\,,\quad z_{\vartheta} \in \mathbb{C}\,,\quad \mathrm{Im}(\tau) > 0\,,\qquad \mathrm{sgn}(0) \coloneqq0\,.
\end{equation}
In \cref{sec:parabolic-degeneration}, we related the appearance of Jacobi theta functions to the heat equation in one spatial dimension. The fact that the stay resolvent only inserts positive powers of the rotation element of the dihedral subgroup can be interpreted as solving the equation while forbidding negative winding numbers, thereby breaking modularity.

\paragraph{Hyperbolic dihedral blocks.} In the hyperbolic $\mathrm{I}_{2}^{ij}(\infty)$ case, the orbit sum leads to a combination of modified Bessel functions of the second kind
\begin{equation}
    S_{\infdih} \left( l,z,T \right) = e^{2\pi i l \mathcal{C}} \sum_{k \in \mathbb{Z}} e^{2\pi i l \left( \mathcal{A}q^{k} + \mathcal{B}q^{-k} \right)} = e^{2\pi i l \mathcal{C}}\frac{1}{u} \sum_{m \in \mathbb{Z}} K_{\frac{\pi i m}{u}}\left( \Omega \left( l,z,T_{\parallel} \right) \right) e^{im\theta\left(z,T_{\parallel}\right)}\,.
\end{equation}
The unilateral expression
\begin{equation}
    S_{\infdih}^{+} \left( l,z,T \right) = e^{2\pi i l \mathcal{C}} \sum_{k \in \mathbb{Z}_{\geq 0}} e^{2\pi i l \left( \mathcal{A}q^{k} + \mathcal{B}q^{-k} \right)}
\end{equation}
is a (multiplicative) half-lattice evaluation. Seeking the same decomposition as for the additive lattice result from the parabolic case, we can write the unilateral expression as
\begin{equation}
    S_{\infdih}^{+} \left( l,z,T \right) = \frac{1}{2} \left( S_{\infdih}(l,z,T) + e^{2\pi i l \langle z, T \rangle} \right) + \frac{1}{2}\Lambda(l,z,T)\,.
\end{equation}
The correction term is given by
\begin{equation}
    \Lambda(l,z,T) = e^{2\pi i l \mathcal{C}} \frac{1}{2\pi i} \int_{\mathrm{Re}(s) = c>0} 2K_{s}(\Omega) e^{us} \frac{\sinh((us/\pi)\theta)}{\sinh(us)} ds\,,
\end{equation}
obtained by using the inverse Mellin transform and summing a geometric series, the latter absolutely converging for any $c>0$ choice of contour, which also avoids integrand poles.

%% file: sections/conclusions.tex
\section{Conclusions and future directions}
\label{sec:conclusions}

Iso-flops connect chambers of the extended K\"ahler cone of Calabi-Yau threefolds that describe diffeomorphic families. In the context of 4D $\mathcal{N} = 2$ Type IIA compactifications, chambers connected by such iso-flops lead to equivalent descriptions of the same theory in lower dimensions. As a consequence, the prepotential of Type IIA compactifications on Calabi-Yau threefolds with iso-flops must transform in a very controlled way under these small birational modifications, which severely restricts its functional form. More specifically, the iso-flops act as simple reflections that generate a Coxeter group $W$ acting on the curve classes or, dually, on the complexified K\"ahler moduli. The instanton contributions associated to non-flopping curves must then organize into Coxeter-invariant functions $\psi_{ld}^{W}(T)$.

In our introductory remarks, we highlighted two questions that immediately arise in view of the above observations:
\begin{enumerate}
    \item What are the possible Coxeter groups generated by iso-flops in the K\"ahler moduli space of Calabi-Yau threefolds?

    \item What are the properties of the Coxeter-invariant functions $\psi_{ld}^{W}(T)$ acting as building blocks of the 4D $\mathcal{N} = 2$ Type IIA prepotential?
\end{enumerate}
Using the set of K\"ahler-favorable CICYs to carry out explicit computations, we have addressed both points in the body of this work, which we now summarize.

\paragraph{CICY Coxeter Database.} For the 4874 K\"ahler-favorable CICYs, we have answered the first question in \cref{sec:CICY-Coxeter-Database} by performing an exhaustive search for iso-flops in their K\"ahler cone and classifying the Coxeter groups that they generate. A total of 2182 models with $h^{1,1}(X) \leq 11$ exhibit a non-trivial Coxeter symmetry, chosen among 19 different possible Coxeter groups. Out of these families, 590 display a $\rank(W) \geq 2$ Coxeter group; see \cref{tab:CICY-Coxeter-database} for the complete list. The database contains 251 models with an infinite Coxeter symmetry, for which $h^{1,1}(X) \leq 5$. The most prevalent $\rank(W) \geq 2$ case is the dihedral group, with a total of 481 counts, 189 of which correspond to the infinite dihedral case. The CICY Coxeter database, provided as an ancillary file with the arXiv submission of this paper and available as an online version at \cite{AlvarezGarciaCICYCoxeterDatabase}, is a version of the favorable CICY list with iso-flop data appended to it, namely, which rows of the configuration matrix lead to iso-flops, the Coxeter matrix of the groups that they generate and the representation with which they act on the K\"ahler moduli space of the variety.

\paragraph{Dihedral Coxeter prepotentials.} Moving to the second question, the abundance and simplicity of the dihedral group singles it out as an ideal example that showcases the constraining power of infinite Coxeter symmetries. For the three dihedral cases (hyperbolic, parabolic and elliptic), the raw orbit sum definition of the Coxeter-invariant functions can be resummed into
\begingroup
\allowdisplaybreaks
\begin{align}
    \text{H:}\quad &\psi_{ld}^{\infdih}(T) = e^{2\pi i l \mathcal{C}} \frac{2}{u} \left[ K_{0}(\Omega) + 2\sum_{m=1}^{\infty} K_{\frac{\pi i m}{u}}(\Omega) \cos(m\zeta) \cos(m\eta) \right]\,,\\
    \text{P:}\quad &\psi_{ld}^{\infdih}(T) = e^{2\pi il \left\langle d, T \right\rangle} \left[ \vartheta\left( z_{\theta}, \tau \right) + e^{2\pi i\xi} \vartheta\left( z_{\theta}^{S}, \tau \right) \right]\,,\\
    \text{E:}\quad &\psi_{ld}^{\mathrm{I}_{2}(m)}(T) = e^{2\pi i l \mathcal{C}} 2m \left[ J_{0}(\Omega') + 2\sum_{r=1}^{\infty} i^{mr} J_{mr}(\Omega') \cos(mr\zeta') \cos(mr\eta') \right]\,,
\end{align}
\endgroup
where $K_{\nu}$ and $J_{m}$ are Bessel functions and $\vartheta$ is the Jacobi theta function. Their arguments depend on the curve class and the complexified K\"ahler moduli, see \cref{sec:dihedral-Coxeter-prepotentials} for their precise definition and for how the above expressions are modified when the chosen curve class has a non-trivial stabilizer. The resummed presentation of $\psi_{ld}^{\mathrm{I}_{2}(m)}(T)$ is more natural from the point of view of the Coxeter symmetries and showcases how the structure of the prepotential is constrained in the presence of iso-flops. Moreover, the convergence properties of $\psi_{ld}^{\mathrm{I}_{2}(m)}(T)$ in its raw orbit sum presentation and in its resummed form are complementary. The former localizes around a few leading instanton contributions in the large volume regime, but converges very slowly in the interior of the moduli space, where it is flatly distributed among the exponential terms. In contrast, the resummed expressions have a slow convergence rate in the large volume region, but sharply localize around the first few Bessel-modes in the interior of the moduli space, acting as a spectral dual decomposition of the standard Gromov-Witten expansion.

\paragraph{Harmonic analysis of the Gromov-Witten expansion.} The results obtained for the dihedral Coxeter prepotentials through direct computation can be geometrically justified by performing a harmonic analysis of the Gromov-Witten expansion. A worldsheet instanton contribution can be heuristically regarded as a plane wave, the angular coordinate corresponding to the axion, multiplied by an exponential dampening factor at large volume, controlled by the real K\"ahler modulus. The Gromov-Witten expansion is then a superposition of such plane waves in the moduli space. The Coxeter action on the K\"ahler moduli leaves a symmetric bilinear form $\Sigma$ invariant, which can be interpreted as a metric whose isometries contain the Coxeter group. Constructing the Laplace-Beltrami operator $\Delta_{\Sigma}$ associated to this metric, i.e., the canonical second-order operator respecting its isometries, we find that
\begin{equation}
    \Delta_{\Sigma}\psi_{ld}^{W}(T) = \lambda \psi_{ld}^{W}(T)\,,
\end{equation}
for some $\lambda$ that depends on the case under study. In other words, the Coxeter-invariant functions $\psi_{ld}^{W}(T)$ solve the Helmholtz equations for $\Delta_{\Sigma}$. Hence, the resummed expressions for $\psi_{ld}^{W}(T)$ are just a decomposition into eigenfunctions of the Laplace-Beltrami operator associated to the Coxeter action on the moduli space. Indeed, we have explicitly checked that, after a separation of variables, the Helmholtz equation for the hyperbolic and elliptic cases reduces to the modified and ordinary Bessel equation, respectively, while in the parabolic case we recover the heat equation. This makes the appearance of Bessel and Jacobi theta functions in the prepotential expected. For general Coxeter groups we believe that, similarly, the prepotential will accept a decomposition in terms of the eigenfunctions of the appropriate Laplace-Beltrami operator with Coxeter isometries, acting as the spectral dual decomposition for the Gromov-Witten expansion and with complementary convergence properties to the raw orbit sum of exponential worldsheet instanton contributions.

\paragraph{Dihedral block decomposition.} While its complete treatment is left for future work, we can leverage the dihedral results to learn something about the structure of $\psi_{ld}^{W}(T)$ in the general case. First, we need to express the raw orbit sum in a tractable manner while making sure to not overcount curve classes. We approach this by constructing \mbox{finite-state} automata that traverse the Cayley graph of the Coxeter groups without repeating elements. Summing over the automata results in a resolvent-type formula encoding the raw orbit sum in terms of finite linear algebra objects, essentially the adjacency matrix of the automaton. Refining the finite-state automaton such that it can remember at least the last two steps taken in the Cayley graph, we can detect dihedral alternations contained within a dihedral subgroup associated to an edge of the Coxeter graph of the group. These can be resummed to obtain special functions similar to those encountered in the dihedral Coxeter prepotentials. As a result, we can perform a dihedral block decomposition of $\psi_{ld}^{W}(T)$ in terms of a sum over these special functions, which encodes the mixing between dihedral channels. Said sum is presented in terms of a remainder resolvent-type formula.

\vspace{\baselineskip}
To sum up, isomorphic flops generate a Coxeter action on the moduli space, forcing the 4D $\mathcal{N} = 2$ Type IIA prepotential to organize into Coxeter-invariant functions. These are eigenfunctions of an appropriate Laplace-Beltrami operator and admit a decomposition into the natural harmonics adapted to the Coxeter-symmetric geometry. Multicover contributions associated to a given curve class are higher-harmonics of the same basic wavevector. This spectral dual decomposition of the Gromov-Witten expansion involves interesting special functions and convergence properties.

A variety of open questions merit further attention, and we hope to revisit them in future work. The CICY Coxeter Database constructed in this paper focuses on the set of K\"ahler-favorable CICYs. It would be interesting to extend the classification of iso-flop Coxeter symmetries to the Kreuzer-Skarke database or to even more general Calabi-Yau threefolds. It is conceivable that new crystallographic Coxeter groups acting on the moduli space beyond those listed in \cref{sec:CICY-Coxeter-Database} could be found in this way. Moreover, iso-flops are very common for K\"ahler-favorable CICYs; a wider survey would help us understand whether this phenomenon is equally prevalent outside of CICYs.

Throughout the text, we have used the dihedral group as the particular example treated in full detail. This allowed us to gain some intuition for the general case, in the form of the harmonic analysis of the Gromov-Witten expansion. Still within the context of the CICY Coxeter Database, it would be desirable to explicitly construct the Laplace-Beltrami operators corresponding to the other groups and study how the prepotential decomposes into their eigenfunctions.

In addition, we would like to understand whether any physical meaning can be assigned to the Helmholtz equation satisfied by the non-flopping contributions to the prepotential, and whether further constraints can be derived from it. For example, since the \mbox{higher-genus} Gopakumar-Vafa invariants also organize into orbits of the Coxeter group, similar conclusions should hold for the higher-genus partition functions. Furthermore, the stringy tree-level quaternion K\"ahler geometry of the hypermultiplet moduli space follows from the special K\"ahler geometry of the vector multiplet moduli space via the c-map. It is then natural to ask whether iso-flop Coxeter symmetries also provide some additional structure for hypermultiplet-related objects.

Venturing beyond the eight-supercharges setting, a clear target of phenomenological interest is that of 4D $\mathcal{N} = 1$ Type II compactifications descending from Calabi-Yau threefolds with an iso-flop Coxeter symmetry. Considering the action of the iso-flops not only on the moduli but also on the flux data of the compactification should lead to some relations among flux vacua. Moreover, depending on the choice of fluxes, part of the Coxeter symmetry could survive for the $\mathcal{N} = 1$ theory as well. It would be interesting to explore whether the distribution of flux vacua is affected by the presence of iso-flops in the moduli space and the prepotential structure that they enforce. Continuing with the 4D $\mathcal{N} = 1$ SUGRA setting, heterotic compactifications on Calabi-Yau threefolds with iso-flops are also an interesting arena; the need to track the effect of these transitions on the heterotic gauge bundle makes the problem geometrically richer.

Altogether, Calabi-Yau threefolds enjoying iso-flop Coxeter symmetries lead to more constrained low-energy theories upon compactifying string theory on them. We hope to continue investigating how strings look when seen through a kaleidoscope in future work.

%% file: sections/acknowledgements.tex
\subsection*{\texorpdfstring{\textbf{Acknowledgments}}{Acknowledgments}}

We thank Sarah Harrison, Vinicius Nevoa, Sanjay Raman, Cumrun Vafa and Kai Xu for useful discussions. The work of F.\,R.\ is supported by the NSF grants PHY-2210333 and PHY-2019786 (The NSF AI Institute for Artificial Intelligence and Fundamental Interactions). The work of R.\,A.-G.\ and F.\,R.\ is also supported by startup funding from Northeastern University.

%% file: appendices/dihedral-stabilizer.tex
\section{Curve classes with non-trivial \texorpdfstring{$\boldsymbol{\mathrm{I}_{2}(m)}$}{I2(m)} stabilizer}
\label{sec:dihedral-stabilizer}

To avoid overcounting curve classes, the definition of $\psi_{ld}^{\mathrm{I}_{2}(m)}$ takes into account the stabilizer subgroup $\mathrm{Stab}_{\mathrm{I}_{2}(m)}(d)$ fixing a given $d \in \mathcal{M}$. As required to compute the orbit sums of \cref{sec:dihedral-Coxeter-prepotentials}, we describe the stabilizers for the Mori representation of the dihedral group. The fixed loci also feed into the explicit choice of fundamental Mori cone in \cref{sec:fundamental-Mori-cone-infdih}. We employ the notation introduced in \cref{sec:review-isomorphic-flops,sec:dihedral-Coxeter-prepotentials}.

\paragraph{\texorpdfstring{$\boldsymbol{\infdih}$}{I2(Inf)}: hyperbolic representation.} Consider the matrix $\mathcal{Q}^{k_{0}}$ representing an arbitrary rotation, with $k_{0} \in \mathbb{Z}^{*}$. First, note that for $d \in \mathrm{Fix}(\mathcal{Q}^{k_{0}})$
\begin{equation}
    \mathcal{Q}^{k_{0}}d=d \iff Q^{k_{0}}y(d) + b(d_{\perp}) = y(d) + b(d_{\perp}) \iff y(d) = 0\,,
\end{equation}
since the eigenvalues of $Q$ are $\lambda_{\pm} \neq 1$. This implies that $\mathrm{Fix}(\mathcal{Q}^{k_{0}}) = \mathrm{Fix}(\mathcal{Q})$ and that $\mathrm{Fix}(\mathcal{Q}^{k_{0}}) \subseteq \mathrm{Fix}(\mathcal{Q}^{k}\mathcal{S})$, for all $k \in \mathbb{Z}$. Hence, $d \in \mathrm{Fix}(\mathcal{Q}^{k_{0}})$ implies that the stabilizer is $\mathrm{Stab}_{\infdih}(d) = \infdih$. The common fixed-point set of the rotations is therefore
\begin{equation}
    \mathcal{F} \coloneqq \mathrm{Fix}(\mathcal{Q}) = \left\{ \left( b(d_{\perp}); d_{\perp} \right)^{T} \in \mathbb{R}^{h} \,\middle|\, d_{\perp} \in \mathbb{R}^{h-2} \right\}\,.
\end{equation}
The hyperbolic M\"obius transformation $Q$ leaves no non-trivial vectors fixed; the above shows that only the extension data entering $\mathcal{Q}$ can remedy this. From the components of $b(d_{\perp})$ in \eqref{eq:extension-shift-vector}, we see that, for the hyperbolic $m_{1}m_{2} > 4$ case, $\mathcal{M} \cap \mathcal{F} = \{0\}$ unless $h^{1,1}(X) > 2$ and $\deltaone = \deltatwo = 0$. Within the K\"ahler-favorable CICY Coxeter database, all $h^{1,1}(X) > 2$ examples have $(u'_{1})^{a'} + (u'_{2})^{a'} > 0$, for $a' \in \{ 1, \dotsc, h-2 \}$, making such a vanishing of products impossible unless $d_{\perp} = 0$. We conclude that no non-trivial curve classes $d \in \mathcal{M}$ are fixed by rotations.

Next, consider the matrix $\mathcal{Q}^{k_{0}}\mathcal{S}$ representing an arbitrary reflection, with $k_{0} \in \mathbb{Z}$. We observe that
\begin{equation}
    \mathcal{Q}^{k_{0}}\mathcal{S} d = d \iff Q^{k_{0}}S y(d) + b(d_{\perp}) = y(d) + b(d_{\perp})\,.
\end{equation}
In addition to the trivial solution $y(d) = 0$, this is solved by the mirror hyperplane of $Q^{k_{0}}S$, which is spanned by
\begin{equation}
    v_{k_{0}} \coloneqq
    \begin{pmatrix}
        m_{1}(1 + \tanh(u) \tanh(k_{0}u))\\
        2
    \end{pmatrix}
    \,.
\label{eq:hyperbolic-infdih-mirrors}
\end{equation}
Hence, the fixed-point set of an arbitrary reflection is given by
\begin{equation}
    \mathrm{Fix}(\mathcal{Q}^{k_{0}}\mathcal{S}) = \mathcal{F} \oplus \langle (v_{k_{0}};0) \rangle_{\mathbb{R}}\,.
\end{equation}
Those curve classes fixed by two reflections are also fixed by the rotation represented by their product and, therefore, by the full group. As a consequence, the case of interest is that of $d \in \mathrm{Fix}(\mathcal{Q}^{k_{0}}\mathcal{S}) \setminus \mathcal{F}$ for a single $k_{0} \in \mathbb{Z}$, for which we have $\mathrm{Stab}_{\infdih}(d) = \mathbb{Z}_{2}$. The intersection with $\mathcal{M}$ can contain non-trivial curve classes; see \cref{sec:hyperbolic-dihedral-example} for an example.

In summary, for the hyperbolic case the only curve classes in the Mori cone with a non-trivial stabilizer are those sitting on top of the mirror hyperplane of a single reflection and have $\mathbb{Z}_{2}$ stabilizer.

\paragraph{\texorpdfstring{$\boldsymbol{\infdih}$}{I2(Inf)}: parabolic representation.} It will be useful to recall that, in the parabolic case, the M\"obius transformation $Q$ fixes the cusp direction $\delta \coloneqq \langle \delta_{1} \rangle_{\mathbb{R}} = \langle \delta_{2} \rangle_{\mathbb{R}}$, where
\begin{equation}
    \delta_{1} =
    \begin{pmatrix}
        m_{1}\\
        2
    \end{pmatrix}
    \,,\qquad
    \delta_{2} =
    \begin{pmatrix}
        2\\
        m_{2}
    \end{pmatrix}
    \,.
\end{equation}
It corresponds to the coalesced hyperbolic simple reflection mirrors in the parabolic limit, see \cref{sec:parabolic-degeneration}. Note that for the nilpotent piece $N$ of $Q$ we have that $\ker(N) = \mathrm{im}(N) = \delta$. We will also refer to the generators
\begin{equation}
    n_{1} \coloneqq
    \begin{pmatrix}
        -2\\
        m_{1}
    \end{pmatrix}
    \,,\qquad
    n_{2} \coloneqq
    \begin{pmatrix}
        m_{2}\\
        -2
    \end{pmatrix}
    \,,
\end{equation}
of the orthogonal direction $n \coloneqq \langle n_{1} \rangle_{\mathbb{R}} = \langle n_{2} \rangle_{\mathbb{R}}$.

Focus first on the rotation represented by $\mathcal{Q}$. For $d \in \mathrm{Fix}(\mathcal{Q})$ it holds that
\begin{equation}
    \mathcal{Q}d =d \iff N d_{\parallel} + U d_{\perp} = 0 \iff
    \begin{cases}
        \phantom{-}2 d_{1} - m_{1}d_{2} = -\left( \deltaone + m_{1} \deltatwo \right)\,,\\
        -2d_{2} + m_{2}d_{1} = -\deltatwo\,.
    \end{cases}
\end{equation}
The condition implies that $U d_{\perp} \in \delta$, i.e., the fixed-point set is non-trivial whenever the extension data aligns with the cusp direction. This is equivalent to the compatibility condition $m_{2} \deltaone + 2 \deltatwo = 0$ of the overdetermined linear system above. The infinite set of non-trivial solutions is given by
\begin{equation}
    \mathrm{Fix}(\mathcal{Q}) = \left\{ d \in \mathbb{R}^{h} \,\middle|\, d_{\parallel} = -\left( \frac{\deltatwo}{m_{2}}, 0 \right)^{T} + \delta\,,\quad Ud_{\perp} \in \delta \right\}\,.
\end{equation}
Since $u'_{1}$, $u'_{2}$ and $d_{\perp}$ have non-negative components, non-trivial solutions only exist when $\deltaone = \deltatwo = 0$. In fact, all parabolic examples in the K\"ahler-favorable CICY Coxeter database have $(u'_{1})^{a'} + (u'_{2})^{a'} > 0$, for $a' \in \{ 1, \dotsc, h-2 \}$, making the requirement $d_{\perp} = 0$. Hence, the above simplifies to
\begin{equation}
    \mathcal{M} \cap \mathrm{Fix}(\mathcal{Q}) \subseteq \left\{ d \in \mathbb{R}^{h} \,\middle|\, d_{\parallel} \in \delta\,,\quad d_{\perp} = 0 \right\}\,.
\end{equation}
For an arbitrary rotation represented by $\mathcal{Q}^{k_{0}}$, with $k_{0} \in \mathbb{Z}^{*}$, we observe that
\begin{equation}
    \mathcal{Q}^{k_{0}}d = d \iff k_{0}Nd_{\parallel} + \left( k_{0}\mathds{1}_{2} + \binom{k_{0}}{2} N \right) Ud_{\perp} = 0\,.
\end{equation}
Taking the product with an element of $n$, the above implies that $U d_{\perp} \in \delta$. Then $NUd_{\perp} = 0$ and the condition reduces to the one for $\mathcal{Q}d=d$, meaning that $\mathrm{Fix}(\mathcal{Q}^{k_{0}}) = \mathrm{Fix}(\mathcal{Q})$. Since the cusp direction is fixed by both simple reflections $M_{1}$ and $M_{2}$, it follows that for $d \in \mathcal{M} \cap \mathrm{Fix}(\mathcal{Q})$ we have $\mathrm{Stab}_{\infdih}(d) = \infdih$.

Consider now an arbitrary reflection represented by $\mathcal{Q}^{k_{0}}\mathcal{S}$, with $k_{0} \in \mathbb{Z}$. For an element of its fixed-point set $d \in \mathrm{Fix}(\mathcal{Q}^{k_{0}}\mathcal{S})$ we have
\begin{equation}
    \mathcal{Q}^{k_{0}}\mathcal{S}d = d \iff (Q^{k_{0}}S - \mathds{1}_{2})d_{\parallel} + (Q^{k_{0}}U_{1} + R_{k_{0}}U)d_{\perp} = 0\,.
\end{equation}
This gives two equivalent equations that can be written as the condition
\begin{equation}
    \mathrm{Fix}(\mathcal{Q}^{k_{0}}\mathcal{S}) = \left\{ d \in \mathbb{R}^{h} \,\middle|\, \left\langle n_{2}, d_{\parallel} \right\rangle = \frac{1}{2} \left( m_{2}(k_{0}+1)\deltaone + 2k_{0}\deltatwo \right) \right\}\,.
\end{equation}
In other words, for fixed $d_{\perp}$, the mirror hyperplane is an affine line in the $d_{\parallel}$-plane parallel to the cusp direction $\delta$. When $d_{\perp} = 0$, it is precisely the cusp direction and $\left. \mathrm{Fix}(\mathcal{Q}^{k_{0}}\mathcal{S}) \right|_{d_{\perp}=0} = \mathrm{Fix}(\mathcal{Q})$. For $d \in \mathrm{Fix}(\mathcal{Q}^{k_{0}}\mathcal{S}) \setminus \mathrm{Fix}(\mathcal{Q})$ we have $\mathrm{Stab}_{\infdih}(d) = \mathbb{Z}_{2}$. As before, a curve class fixed by two reflections is automatically fixed by the entire group.

To summarize, non-trivial stabilizers for the parabolic case arise for curve classes in the Mori cone that sit on top of the mirror hyperplane of a single reflection, in which case they have a $\mathbb{Z}_{2}$ stabilizer, or for curves classes contained in the cusp direction within the $d_{\parallel}$-plane, in which case they are stabilized by the full $\infdih$ group. Both types of non-trivial stabilizers appear in concrete models, see \cref{sec:parabolic-dihedral-example} for an example.

\paragraph{\texorpdfstring{$\boldsymbol{\mathrm{I}_{2}(m)}$}{I2(m)}: elliptic representation.} Consider an arbitrary rotation represented by $\mathcal{Q}^{k_{0}}$, with $k_{0} \in \mathbb{Z}$ and $k_{0} \neq 0 \mod m$. Arguing analogously to the hyperbolic case, the fixed-point set is given once more by $\mathcal{F} = \mathrm{Fix}(\mathcal{Q}) = \mathrm{Fix}(\mathcal{Q}^{k_{0}})$. The crucial difference is that for the elliptic $m_{1}m_{2} < 4$ case, the components of $b(d_{\perp})$ in \eqref{eq:extension-shift-vector} are non-negative, allowing for non-trivial curve classes $d \in \mathcal{M} \cap \mathcal{F}$ with $\mathrm{Stab}_{\mathrm{I}_{2}(m)}(d) = \mathrm{I}_{2}(m)$.

Next, consider an arbitrary reflection represented by $\mathcal{Q}^{k_{0}}\mathcal{S}$, with $k \in \mathbb{Z}$. Following the same steps as for the hyperbolic case we find that
\begin{equation}
    \mathrm{Fix}(\mathcal{Q}^{k_{0}}\mathcal{S}) = \mathcal{F} \oplus \langle (v_{k_{0}};0) \rangle_{\mathbb{R}}\,,
\end{equation}
with $v_{k_{0}} \in \ker\left(Q^{k_{0}}S - \mathds{1}_{2}\right)$. When $m>2$, a general formula for $v_{k_{0}}$ is
\begin{equation}
    v_{k_{0}} =
    \begin{pmatrix}
        m_{1} \cos((k_{0}+1)v)\\
        2 \cos(v) \cos(k_{0}v)
    \end{pmatrix}
    \,.
\end{equation}
For $d \in \mathrm{Fix}(\mathcal{Q}^{k_{0}}\mathcal{S}) \setminus \mathcal{F}$ for a single $k_{0} \in \mathbb{Z}$, we have $\mathrm{Stab}_{\mathrm{I}_{2}(m)}(d) = \mathbb{Z}_{2}$. As above, a curve class fixed by two reflections is fixed by the full group.

In summary, for the elliptic case we have non-trivial stabilizers for curve classes in the Mori cone located on a mirror hyperplane, with $\mathbb{Z}_{2}$ stabilizer, or those at the intersection of two mirrors, with $\mathrm{I}_{2}(m)$ stabilizer. See \cref{sec:elliptic-example-I2(4)} for concrete examples of both.

%% file: appendices/fundamental-Mori-cone-infdih.tex
\section{Fundamental Mori cone for the \texorpdfstring{$\boldsymbol{\mathrm{I}_{2}(m)}$}{I2(m)} action}
\label{sec:fundamental-Mori-cone-infdih}

As reviewed in \cref{sec:coxeter-prepotentials} and used throughout \cref{sec:dihedral-Coxeter-prepotentials,sec:general-Coxeter-prepotentials}, the contribution of the non-flopping curve classes $d \in \mathcal{M}_{\mathrm{restr}}$ to the instanton prepotential can be rewritten in terms of a sum of $W$-invariant functions $\Psi_{d}^{W}(T)$ running over a fundamental domain $\mathcal{M}_{f} \subseteq \mathcal{M}_{\mathrm{restr}}$ for the $W$-action on the restricted Mori cone. Explicitly characterizing $\mathcal{M}_{f}$ for the dihedral case is rather immediate after the analysis of \cref{sec:dihedral-stabilizer}, as we work out below.

\paragraph{\texorpdfstring{$\boldsymbol{\infdih}$}{I2(Inf)}: hyperbolic representation.} The fundamental Mori cone $\mathcal{M}_{f}$ is given by the positive orthant wedge between the mirror hyperplanes associated to the two simple reflections $\mathcal{M}_{1}$ and $\mathcal{M}_{2}$. This leads to
\begin{equation}
    \mathcal{M}_{f} = \left\{ d \in \mathbb{R}^{h}_{\geq 0} \,\middle|\, 2d_{1} - m_{1}d_{2} \leq \deltaone\,,\quad 2d_{2} - m_{2}d_{1} \leq \deltatwo \right\}\,.
\label{eq:hyperbolic-fundamental-Mori-cone}
\end{equation}
This rational polyhedral cone can be described as
\begin{equation}
    \mathcal{M}_{f} = \left\langle w_{1}, w_{2}, e_{1}, \dotsc, e_{h-2}, p_{1}, \dotsc, p_{h-2}, q_{1}, \dotsc, q_{h-2} \right\rangle_{\mathbb{R}_{\geq 0}}\,,
\end{equation}
where the generating vectors are the following:\footnote{Throughout this section, we use integral generating vectors. Depending on the model, their entries may have common factors that are to be divided out to obtain the primitive generators.}
\begin{enumerate}
    \item Two-dimensional chamber: Whenever $d_{\perp} = 0$, the problem is equivalent to the two-dimensional case. The corresponding rational polyhedral cone is spanned by the $M_{1}$ and $M_{2}$ mirrors along
    \begin{equation}
        w_{1} =
        \begin{pNiceArray}{c}[cell-space-limits=5pt]
            m_{1}\\
            2\\
            \hline
            0
        \end{pNiceArray}
        \,,\qquad w_{2} =
        \begin{pNiceArray}{c}[cell-space-limits=5pt]
            2\\
            m_{2}\\
            \hline
            0
        \end{pNiceArray}
        \,.
    \end{equation}
    This rational cone accumulates via the two-dimensional $\infdih$-action towards the irrational cone defined by the eigenvectors of $Q$. The effect of the extension data $d_{\perp}$ in \eqref{eq:hyperbolic-fundamental-Mori-cone} is to simply shift the vertex of the cone to the intersection of the simple reflection mirrors $b(d_{\perp})$, as is evident from rewriting the inequalities in terms of the untwisted coordinates $y(d)_{1}$ and $y(d)_{2}$ of the orbit sum seed.

    \item Pure spectator rays: A canonical basis $\{ e_{a'} \}_{a' \in \{1, \dotsc, h-2\}}$ of $\mathbb{R}^{h-2}$ cuts the positive orthant along the $d_{\perp}$-directions on which the group action is trivial.

    \item Inequality saturating rays: The vectors
    \begin{equation}
        p_{a'} =
        \begin{pNiceArray}{c}[cell-space-limits=5pt]
            (u'_{1})^{a'}\\
            0\\
            \hline
            2e_{a'}
        \end{pNiceArray}
        \,,\qquad
        q_{a'} =
        \begin{pNiceArray}{c}[cell-space-limits=5pt]
            0\\
            (u'_{2})^{a'}\\
            \hline
            2e_{a'}
        \end{pNiceArray}
        \,,\qquad a'\in \{ 1, \dotsc, h-2 \}\,,
    \end{equation}
    obtained by saturating the inequalities for $d_{2}=0$ and $d_{1}=0$, respectively.
\end{enumerate}
After taking into account the obvious redundancies whenever $(u'_{1})^{a'} = 0$ or $(u'_{2})^{a'} = 0$, the above generating set is minimal.

\paragraph{\texorpdfstring{$\boldsymbol{\infdih}$}{I2(Inf)}: parabolic representation.}

Again, from the positive orthant wedge between the simple reflection mirrors, we find
\begin{equation}
    \mathcal{M}_{f} = \left\{ d \in \mathbb{R}^{h}_{\geq 0} \,\middle|\, \etad + \deltaone \geq 0\,,\quad \etad - \frac{m_{1}}{2} \deltatwo \leq 0 \right\}\,.
\label{eq:parabolic-fundamental-Mori-cone}
\end{equation}
For $d_{\perp} = 0$, this corresponds to the ray along the cusp direction, as expected from the parabolic limit of the hyperbolic case, see \cref{sec:parabolic-degeneration}. The ray thickens into a cusp strip as $d_{\perp}$ becomes non-zero. The rational polyhedral cone can be described as
\begin{equation}
    \mathcal{M}_{f} = \left\langle w_{1}, e_{1}, \dotsc, e_{h-2}, p_{1}, \dotsc, p_{h-2}, q_{1}, \dotsc, q_{h-2} \right\rangle_{\mathbb{R}_{\geq 0}}\,,
\end{equation}
similar to the hyperbolic case above, with one of the two-dimensional chamber rays removed due to the parabolic degeneration $\delta = \langle w_{1} \rangle_{\mathbb{R}} = \langle w_{2} \rangle_{\mathbb{R}}$.

\paragraph{\texorpdfstring{$\boldsymbol{\mathrm{I}_{2}(m)}$}{I2(m)}: elliptic representation.} The fundamental Mori cone for the elliptic case is similar to the one for the hyperbolic case; the positive orthant wedge between the mirror hyperplanes associated to the two simple reflections $\mathcal{M}_{1}$ and $\mathcal{M}_{2}$ leads to the same inequalities but flipped. Here, however, this wedge is not automatically within $\mathcal{M}_{\mathrm{restr}}$. To ensure this, we must impose that the entire finite dihedral orbit of the curve class $d$ remains within $\mathcal{M}$. This results in
\begin{equation}
    \mathcal{M}_{f} = \left\{ d \in \mathbb{R}^{h}_{\geq 0} \,\middle|\, 
    \begin{aligned}
        2d_{1} - m_{1}d_{2} \geq \deltaone\,,\quad &b(d_{\perp}) + Q^{k}y(d) \in \mathbb{R}^{2}_{\geq 0}\,,\\
        2d_{2} - m_{2}d_{1} \geq \deltatwo\,,\quad &b(d_{\perp}) + Q^{k}Sy(d) \in \mathbb{R}^{2}_{\geq 0}\,,\\
        &\forall k \in \{0, \dotsc, m-1\}
    \end{aligned}
    \right\}\,.
\label{eq:elliptic-fundamental-Mori-cone}
\end{equation}
The generating rays for this rational polyhedral cone can be obtained by a standard conversion from the above half-space description to an extremal-ray description. We will not write them here since the minimal list of generating rays depends on $m \in \{ 2,3,4,6 \}$ and on the concrete extension data of the model.

%% file: appendices/AB-properties.tex
\section{Properties of \texorpdfstring{$\boldsymbol{\mathcal{A}\left( z,T_{\parallel}\right),\,\mathcal{B}\left( z,T_{\parallel}\right),\,\zeta\left( T_{\parallel}\right)\;\text{and}\;\eta(d)}$}{A(z,T), B(z,T), zeta(T) and eta(d)}}
\label{sec:AB-properties}

Given the Mori representation $\rho: \infdih \rightarrow \mathrm{GL}(V)$ with $V \cong \mathbb{R}^{h} \supset \mathcal{M}$ and satisfying the extension \eqref{eq:dihedral-rep-short-exact-sequence}, let us denote the projection onto $V_{0}$ by $\pi_{0}: V \rightarrow V_{0}$, \textit{mutatis mutandis} for the K\"ahler representation.

In \cref{sec:dihedral-hyperbolic-representations} we invoked the fact that
\begin{equation}
    \mathrm{Im}\left(\mathcal{A}\left( z,T_{\parallel} \right)\right),\, \mathrm{Im}\left(\mathcal{B}\left(z,T_{\parallel}\right)\right) > 0\quad \textrm{for}\quad z \in C^{+}\,,\quad T_{\parallel} \in \mathrm{int}(\hat{\pi}_{0}(\mathcal{K}))\,,
\end{equation}
together with $y(d),\, y'(d) \in C^{+}$ for $d \in \mathcal{M}_{f}$ with $\mathrm{Stab}_{\infdih}(d) = 1$, to argue for the convergence properties of $f \left( z,T_{\parallel} \,\middle|\, x \right)$ and the applicability of \eqref{eq:gradshteyn-identity}. Let us justify both claims below.

First, using the notation $T^{i} = a^{i} + it^{i}$, $i \in \{1,\dotsc,h\}$, for the complexified K\"ahler moduli, the definitions \eqref{eq:AB-definitions}, the fact that $m_{1}m_{2} > 4$ for the hyperbolic $\infdih$ case and writing $z$ in the eigenbasis $\{v_{+},v_{-}\}$ of $Q$ as
\begin{equation}
    z = \alpha_{z}v_{+} + \beta_{z}v_{-}\,,
\end{equation}
we observe that
\begin{equation}
\mathrm{Im}\left(\mathcal{A}\left( z,T_{\parallel} \right)\right),\, \mathrm{Im}\left(\mathcal{B}\left(z,T_{\parallel}\right)\right) > 0 \Leftrightarrow
    \begin{cases}
        \left\langle \alpha_{z} (\lambda_{+} - \lambda_{-}) v_{+}, t_{\parallel} \right\rangle > 0\,,\\
        \left\langle \beta_{z} (\lambda_{+} - \lambda_{-}) v_{-}, t_{\parallel} \right\rangle > 0\,.
    \end{cases}
\end{equation}
Since $\lambda_{+} - \lambda_{-} > 0$ and $\left\langle v_{\pm}, t_{\parallel} \right\rangle > 0$ for $T_{\parallel} \in \mathrm{int}(\hat{\pi}_{0}(\mathcal{K}))$, the inequalities are satisfied when $\alpha_{z},\,\beta_{z}>0$, i.e., for $z \in C^{+} = \langle v_{+}, v_{-} \rangle_{\mathbb{R}_{>0}}$.

Next, let us show that the orbit sum seeds $y(d),\, y'(d) \in C^{+}$ for the relevant curve classes $d \in \mathcal{M}_{f}$ with $\mathrm{Stab}_{\infdih}(d) = 1$. In untwisted coordinates, the two inequalities in \eqref{eq:hyperbolic-fundamental-Mori-cone} derived from the mirror hyperplane equations read
\begin{equation}
    2y(d)_{1} - m_{1}y(d)_{2} \leq 0\qquad \text{and}\qquad 2y(d)_{2} - m_{2}y(d)_{1} \leq 0\,.
\end{equation}
Note that, if $y(d)_{2} = 0$ these inequalities together with $m_{2} > 0$ force $y(d) = 0$, which contradicts the trivial stabilizer assumption. Additionally, assuming $y(d)_{2} < 0$ leads to the rewriting of the inequalities
\begin{equation}
    \frac{m_{1}}{2} \leq \frac{y(d)_{1}}{y(d)_{2}} \leq \frac{2}{m_{2}}\,,
\end{equation}
which contradicts the hyperbolic case condition $m_{1}m_{2} > 4$. Hence, we can rewrite the inequalities as
\begin{equation}
    \frac{2}{m_{2}} \leq \frac{y(d)_{1}}{y(d)_{2}} \leq \frac{m_{1}}{2}\,.
\end{equation}
Given that $m_{1}m_{2} > 4$, we see that
\begin{equation}
    \mu_{-} < \frac{2}{m_{2}} < \frac{m_{1}}{2} < \mu_{+}\,,
\end{equation}
from which we conclude that $y(d) \in C^{+}$. Since $y'(d) = Sy(d)$ and the action of $S$ on the null direction generators is $Sv_{\pm} = v_{\mp}$, we conclude that $y'(d) \in C^{+}$ as well.

Now let us focus on $\zeta\left( T_{\parallel} \right)$ and $\eta(d)$. These objects were defined in \eqref{eq:phi-psi-definition} in terms of combinations of $\theta \left( z,T_{\parallel} \right)$ and, naively, they both depend on $d$ and $T_{\parallel}$. First, note that by comparing $\left\langle Q^{k} y, T_{\parallel} \right\rangle$ computed in the eigenbasis $\{v_{+},v_{-}\}$ to \eqref{eq:QkzTparallel}, we see that we can write
\begin{subequations}
\begin{align}
    \mathcal{A}\left( y(d),T_{\parallel} \right) &= \alpha_{y(d)} \left\langle v_{+}, T_{\parallel} \right\rangle\,,\\
    \mathcal{B}\left( y(d),T_{\parallel} \right) &= \beta_{y(d)} \left\langle v_{-}, T_{\parallel} \right\rangle\,.
\end{align}
\end{subequations}
Using that $y'(d) = M_{1}y$, it follows that
\begin{subequations}
\begin{align}
    \mathcal{A}\left( y'(d),T_{\parallel} \right) &= \beta_{y(d)} \left\langle v_{+}, T_{\parallel} \right\rangle\,,\\
    \mathcal{B}\left( y'(d),T_{\parallel} \right) &= \alpha_{y(d)} \left\langle v_{-}, T_{\parallel} \right\rangle\,.
\end{align}
\end{subequations}
From the previous discussion, we know that $\alpha_{y(d)},\,\beta_{y(d)},\,\alpha_{y'(d)},\,\beta_{y'(d)} > 0$. Moreover, $\mathrm{{Im}\left( \left\langle v_{\pm}, T_{\parallel} \right\rangle \right)} > 0$ implies that $\arg\left( \left\langle v_{-}, T_{\parallel} \right\rangle/\left\langle v_{+}, T_{\parallel} \right\rangle \right) \in (-\pi,\pi)$. Hence, we can safely apply the principal branch logarithmic identity
\begin{equation}
    \log(a \xi) = \log(a) + \log(\xi)\,,
\end{equation}
for $a \in \mathbb{R}_{>0}$ and $\xi \in \mathbb{C} \setminus (-\infty,0]$, to find the simplified expressions
\begin{equation}
    \zeta\left( T_{\parallel} \right) = \frac{\pi}{2u} \log\left( \frac{\left\langle v_{-}, T_{\parallel} \right\rangle}{\left\langle v_{+}, T_{\parallel} \right\rangle} \right)\,,\qquad \eta(d) = \frac{\pi}{2u} \log\left( \frac{\beta_{y(d)}}{\alpha_{y(d)}} \right)\,,
\end{equation}
where the split dependence on $d$ and $T_{\parallel}$ is manifest. The $y(d)$ coordinate quotient reads
\begin{equation}
    \frac{\beta_{y(d)}}{\alpha_{y(d)}} = \frac{\mu_{+} b_{y(d)} - a_{y(d)}}{a_{y(d)} - \mu_{-}b_{y(d)}}\,,
\end{equation}
with
\begin{subequations}
\begin{align}
    a_{y(d)} &\coloneqq (m_{1}m_{2}-4)d_{1} + 2 \left\langle u'_{1},d_{\perp} \right\rangle + m_{1} \left\langle u'_{2}, d_{\perp} \right\rangle\,,\\
    b_{y(d)} &\coloneqq (m_{1}m_{2}-4)d_{2} + m_{2} \left\langle u'_{1}, d_{\perp} \right\rangle + 2 \left\langle u'_{2},d_{\perp} \right\rangle\,.
\end{align}
\end{subequations}